\documentclass[showpacs,pre,onecolumn]{revtex4-1}
\usepackage{amsmath}
\usepackage{graphicx}
\usepackage{subfigure}

\begin{document}

\title{Solitons supported by localized nonlinearities in periodic media}
\author{Nir Dror and Boris A. Malomed }
\affiliation{Department of Physical Electronics, School of
Electrical Engineering, Faculty of Engineering, Tel Aviv University,
Tel Aviv 69978, Israel}

\begin{abstract}
Nonlinear periodic systems, such as photonic crystals and Bose-Einstein
condensates (BECs) loaded into optical lattices, are often described by the
nonlinear Schr\"{o}dinger/Gross-Pitaevskii equation with a sinusoidal
potential. Here, we consider a model based on such a periodic potential,
with the nonlinearity (attractive or repulsive) concentrated either at a
single point or at a symmetric set of two points, which are represented,
respectively, by a single $\delta $-function or a combination of two $\delta
$-functions. With the attractive or repulsive sign of the nonlinearity, this
model gives rise to ordinary solitons or gap solitons (GSs), which reside,
respectively, in the semi-infinite or finite gaps of the system's linear
spectrum, being pinned to the $\delta $-functions. Physical realizations of
these systems are possible in optics and BEC, using diverse variants of the
nonlinearity management. First, we demonstrate that the single $\delta $%
-function multiplying the nonlinear term supports families of \emph{stable}
regular solitons in the self-attractive case, while a family of solitons
supported by the attractive $\delta $-function in the absence of the
periodic potential is completely unstable. In addition, we show that the $%
\delta $-function can support \emph{stable} GSs in the first finite bandgap
in both the self-attractive and repulsive models. The stability analysis for
the GSs in the second finite bandgap is reported too, for both signs of the
nonlinearity. Alongside the numerical analysis, analytical approximations
are developed for the solitons in the semi-infinite and first two finite
gaps, with the single $\delta $-function positioned at a minimum or maximum
of the periodic potential. In the model with the symmetric set of two $%
\delta $-functions, we study the effect of the \textit{spontaneous symmetry
breaking} of the pinned solitons. Two configurations are considered, with
the $\delta $-functions set symmetrically with respect to the minimum or
maximum of the underlying potential.
\end{abstract}

 \maketitle

\section{Introduction}

\label{sec:Introduction}

In the course of the last decade, a great deal of interest has been drawn to
theoretical and experimental studies of the nonlinear dynamics in systems
with periodic potentials. Physical realizations of this topic are well known
in nonlinear optics \cite{KA,Optics} and Bose-Einstein condensates (BECs).
In optical media, the periodic (lattice) potentials may be created as
permanent or virtual ones (in the latter case, these are photonic lattices
induced in photorefractive crystals \cite{Moti,Optics}). In BECs, similar
potentials can be induced in the form of optical lattices (OLs), i.e.,
interference patterns formed by laser beams shone through the condensate.
The OLs make it possible to study a great variety of dynamical effects in
BECs \cite{Morsch,Maciek}. Similar periodic potentials may also be imposed
by magnetic lattices \cite{magnetic}.

It is commonly known that, in the free space, stable one-dimensional (1D)
solitons exist in optical waveguides and BECs with the attractive cubic
nonlinearity, while in the 2D and 3D geometry the solitons are unstable to
the collapse \cite{Berge}. On the other hand, it has been predicted that 2D
\cite{Baizakov1,Efremidis,Musslimani,Baizakov3} and 3D \cite%
{Baizakov1,Baizakov3} solitons can be stabilized by dint of the
corresponding OLs. Moreover, low-dimensional OLs, i.e., quasi-1D and
quasi-2D lattices in the 2D \cite{Baizakov3} and 3D \cite%
{Baizakov3,Mihalache1} space, respectively, also provide for the
stabilization of fully localized multidimensional solitons. A related
prediction is the existence of stable 2D \cite{Kartashov,Baizakov4} and 3D
\cite{Mihalache2} solitons in models with radial OLs.

In BECs with repulsive interactions between atoms, i.e., the repulsive
intrinsic nonlinearity, solitons cannot exist in free space, but gap
solitons (GSs) may be supported by the OL. The principle behind the
formation of the GS is that the periodic potential can invert the sign of
the effective mass of collective excitations, which may then balance the
repulsive nonlinearity. In the 1D case, several species of GSs are known,
which include fundamental solitons and their two- and three-peak bound
complexes, in the first and second bandgaps \cite{Abdullaev,Kivshar}, and \textit{%
subfundamental} solitons in the second gap (the latter means a twisted
soliton squeezed into a single cell of the potential lattice, whose norm is
smaller than the norm of the fundamental soliton existing at the same value
of the chemical potential) \cite{Nikos,Mayteevarunyoo,JCuevas}. The origin
of the GS families may be traced back to bifurcations generating them from
Bloch waves at edges of the bandgaps \cite{Kivshar,Yang}.

Multidimensional GSs are represented by fundamental solitons \cite%
{MultidimensionalGS1,MultidimensionalGS2,Sakaguchi,Gubeskys2} and gap-type
vortices \cite%
{Gubeskys1,MultidimensionalGS2,Ostrovskaya,Sakaguchi,Mayteevarunyoo2}. In
particular, the simplest gap vortices are composed of four density peaks,
and fall into two different categories: densely packed squares (alias
off-site-centered vortices), in which the center is positioned around a
local maximum of the OL potential \cite%
{MultidimensionalGS2,Ostrovskaya,Mayteevarunyoo2}, and rhombic
(on-site-centered) configurations, featuring a nearly empty lattice cell in
the middle \cite{MultidimensionalGS1,MultidimensionalGS2,Sakaguchi,Gubeskys2}%
. While GS families in 2D are also generated by bifurcations from the
respective Bloch waves \cite{Yang2D,Yang}, there are gap-vortex families
which do not originate from such bifurcations \cite{Yang2Dnonbif}.

In the experiment, an effectively one-dimensional GS, composed of few
hundred atoms, was created in the condensate of $^{87}$Rb \cite{Eiermann}.
In another experiment, which involved a stronger OL, broad confined states
were created \cite{Anker}. They were interpreted as modes intermediate
between the GSs and extended nonlinear Bloch waves \cite{Alexander2}.

In a different framework, solitons can be supported by a spatial modulation
of the local nonlinearity (a particular case of the ``nonlinearity
management" \cite{book}). In terms of solid-state physics, an effective
potential structure induced by means of this method is called a \textit{%
pseudopotential }\cite{Harrison}. Realizations of such structures are
possible in optics and BECs, where they are expected to give rise to many
new effects, see recent review \cite{RMP}. In optical media,
nonlinearity-modulation profiles can be induced by means of nonuniform
distributions of resonant dopants, similar to those in resonantly-absorbing
Bragg reflectors, see original works \cite{RABR} and review \cite%
{RABR-review}. Such dopant-density patterns can be created by means of
available technology \cite{technology}. For matter waves in BECs, similar
nonlinear profiles can be created via the spatial modulation of $a_{s}(x)$,
the local value of the \textit{s}-wave scattering length. Such modulations
can be induced by means of the Feshbach resonance, controlled by a
nonuniform dc magnetic field \cite{Inouye} or resonant optical field \cite%
{Fedichev,Theis}. It was also predicted that the Feshbach resonance may be
controlled by dc electric fields \cite{Marinescu}. Note that the Feshbach
resonance allows one to create a pseudopotential corresponding to a \emph{%
sign-changing} function $a_{s}(x)$ (actually, this is possible in the
condensate of $^{7}$Li \cite{RandyLev}), which implies the spatial
alternation between the attractive and repulsive signs of the nonlinearity.

In the context of the BECs, many works dealt with solitons and related
dynamical states in the framework of 1D pseudopotential structures \cite%
{BECpseudopotential}. Nonlinear structures based on the spatial modulation
of the Kerr coefficient were also considered in optics, giving rise to
similar states \cite{Hao}. Solitons supported by 2D pseudopotentials were
theoretically studied too, with a conclusion that it is much more difficult
to stabilize them than using linear OL potentials \cite{Sakaguchi2}. The
results accumulated in the studies of solitons in nonlinear lattices are
summarized in a recent review \cite{RMP}.

A specific example of such nonlinear pseudopotential settings is a model
where the nonlinearity is concentrated at a single point, which is
represented by a $\delta $-function. A prototypical model of this type was
introduced in Ref. \cite{Azbel}. It may represent a planar linear waveguide
with a narrow nonlinear stripe embedded into it (for the case of the
second-harmonic-generating quadratic nonlinearity, spatial solitons in a
similar setting were considered in Ref. \cite{Canberra}). In terms of the
BECs, localized nonlinearity may be induced through the Feshbach resonance
imposed by a focused laser beam. It should be said that a more realistic
model would also include a uniform background nonlinearity, as it is
difficult to create a setting where it might be completely eliminated.
Nevertheless, it makes sense to consider, as the basic one, the model in
which the nonlinearity is represented solely by the localized terms, as it
is reasonable to assume that the weakly nonlinear background will not lead
to a drastic change of the results. Extending the studies in this direction,
a model of a double-well pseudopotential, based on a symmetric set of two $%
\delta $-functions, or their regularized counterparts, was introduced, with
the purpose of studying the spontaneous symmetry breaking (SSB) of localized
modes \cite{DeltaSSB}. In particular, it is possible to find full analytical
solutions for symmetric, asymmetric and antisymmetric states in the model
with two ideal $\delta $-functions. In this model, where the SSB bifurcation
is of the subcritical type, the symmetric solutions are stable up to the
bifurcation point. Beyond this point, the symmetric states and the emerging
asymmetric states are unstable, as well as all the antisymmetric ones.
Symmetry breaking in a circular nonlinear lattice, with a smooth spatial
modulation of $a_{s}$, was studied, in the framework of both the
Gross-Pitaevskii equation (GPE) and many-body quantum system, in Ref. \cite%
{HP}.

A natural extension of the settings outlined above is a model integrating
the linear OL potential and the nonlinearity concentrated at one or two
points, represented by the respective $\delta $-functions or their
regularized versions. These systems are the subject of the present work. In
particular, we demonstrate that, while cusp-shaped solitons pinned to the $%
\delta $-function multiplying the attractive cubic nonlinearity turn out to
be unstable, the linear periodic potential readily stabilizes them. Another
issue of obvious interest is whether the strongly localized repulsive
nonlinearity may support GSs in finite bandgaps of the OL-induced spectrum
(we demonstrate that this is possible indeed).

The rest of the paper is organized as follows. The model is formulated in
section~\ref{sec:model}. For the system including the single $\delta $%
-function and OL potential, analytical approximations, based on the
perturbation theory, are developed for the pinned modes in the first and
second finite bandgaps, as well as in the semi-infinite gap, in section \ref%
{sec:Approximation}. A comparison with numerical findings is performed too.
Detailed results of the numerical analysis for the existence and, most
important, stability of the pinned modes in the semi-infinite, first and
second gaps are reported in section \ref{sec:SingleDelta}. Both the
attractive and repulsive signs of the $\delta $-functional nonlinearity are
considered, for different positions of the $\delta $-function with respect
to the underlying lattice. In section~\ref{sec:TwoDeltas}, the analysis is
reported for symmetric, antisymmetric and asymmetric modes supported by a
pair of the $\delta $-functions, positioned symmetrically with respect to a
maximum or minimum of the OL potential. The paper is concluded by section~%
\ref{sec:Conclusion}.

\section{The model}

\label{sec:model} The model featuring the cubic nonlinearity represented by
the $\delta $-function was introduced in Ref. \cite{Azbel}, in the context
of tunneling of interacting particles through a junction:
\begin{equation}
i\psi _{t}+\frac{1}{2}\psi _{xx}-\sigma \delta (x)|\psi |^{2}\psi =0,
\label{Dlata_NLSE}
\end{equation}%
where $\psi $ is the mean-field wave function in the BEC, or the local
amplitude of the guided electromagnetic field in the context of optics (in
the latter case, time $t$ is replaced by the propagation distance), with $%
\sigma =+1$ and $-1$ corresponding to the repulsive and attractive
nonlinearity, respectively. Obviously, Eq. (\ref{Dlata_NLSE}) amounts to the
simple linear equation valid at $x<0$ and $x>0$, $i\psi _{t}+(1/2)\psi
_{xx}=0$, which is supplemented by the derivative-jump condition at $x=0$,
produced by the integration of Eq. (\ref{Dlata_NLSE}) over an infinitely
small vicinity of $x=0$:%
\begin{equation}
~\psi _{x}\left( x=+0\right) -\psi _{x}\left( x=-0\right) =2\sigma
\left\vert \psi (x=0)\right\vert ^{2}\psi (x=0).  \label{xx}
\end{equation}

Stationary states are looked for in the ordinary form, $\psi (x,t)=e^{-i\mu
t}\phi (x)$, where $\mu $ is the chemical potential in BEC ($-\mu $ is the
propagation constant in optics), and real function $\phi $ obeys equation
\begin{equation}
\mu \phi +(1/2)\phi ^{\prime \prime }-\sigma \delta (x)\phi ^{3}=0.
\label{phi-simple}
\end{equation}%
An exact soliton solution to Eq. (\ref{phi-simple}) with $\sigma =-1$ is
obvious:
\begin{equation}
\phi _{0}(x)=Ae^{-\sqrt{-2\mu }|x|},~A^{2}=\sqrt{-2\mu }.  \label{simple}
\end{equation}%
The norm of this solution,
\begin{equation}
N=\int_{-\infty }^{+\infty }|\psi (x)|^{2}dx,  \label{Norm}
\end{equation}%
is independent of $\mu $, taking a constant value, $N=1$, hence the formal
application of the Vakhitov-Kolokolov (VK) criterion, $dN/d\mu <0$ \cite%
{VK,Berge}, predicts neutral stability. In fact, numerical simulations of
the evolution of pinned solitons (\ref{simple}) with small perturbations
added to them demonstrate a strong instability (not displayed here in
detail): the soliton either decays or suffers the \textit{collapse}
(formation of a singularity), if the exact norm of the perturbed soliton is,
respectively, $N<1$ or $N>1$. These features, including the degeneracy of
the norm and the instability leading to the decay or collapse, resemble
those known for the 2D \textit{Townes solitons} \cite{Berge} in the free 2D
space with the cubic attractive nonlinearity, or in the 1D space with the
quintic nonlinearity \cite{AbdSal}. Moreover, it is easy to find a family of
\emph{exact} analytical solutions to Eq. (\ref{Dlata_NLSE}), with $\sigma
=-1 $, which explicitly describe the approach to the collapse at $%
t\rightarrow 0^{-} $:%
\begin{equation}
\psi \left( x,t\right) =\sqrt{-\frac{x_{0}}{t}}\exp \left\{ i\left[ \frac{%
\left( |x|-ix_{0}\right) ^{2}}{2t}\right] \right\} ,  \label{coll}
\end{equation}%
where $x_{0}$ is an arbitrary real positive constant, the solution being
valid at $t<0$. Decaying solutions, at $t>0$, are described by the same
solution (\ref{coll}), with $-x_{0}$ replaced by $x_{0}>0$, the decay taking
place at $t\rightarrow +\infty $. The norm of solution (\ref{coll}) is
exactly $N=1$ (irrespective of the value of $x_{0}$), i.e., the same as that
of stationary solution (\ref{simple}).

In this work we introduce a natural extension of Eq. (\ref{Dlata_NLSE}),
adding to it the periodic OL\ potential, with the objective to stabilize the
solitons:
\begin{equation}
i\psi _{t}+\frac{1}{2}\psi _{xx}+\varepsilon \cos (2x)\psi -\sigma \delta
(x-\xi )|\psi |^{2}\psi =0.  \label{Single_Dlata_NLSE}
\end{equation}%
Here, the OL potential, whose period is normalized to be $\pi $, is $%
V(x)=-\varepsilon \cos (2x)$. In its first period, $0\leq x<\pi $, the
minimum and maximum of the potential are located, respectively, at $x=0$ and
$x=\pi /2$ (and vice versa for $\varepsilon <0$), while the $\delta $%
-function is set at $x=\xi $, which does not necessarily coincide with the
minimum or maximum of the potential.

A modification of the model, which is considered in section~\ref%
{sec:TwoDeltas}, deals with the nonlinearity represented by a pair of $%
\delta $-functions, placed symmetrically with respect to the minimum or
maximum of the periodic potential:
\begin{equation}
i\psi _{t}+\frac{1}{2}\psi _{xx}+\varepsilon \cos (2x)\psi -\sigma \left[
\delta (x-\xi )+\delta (x+\xi )\right] |\psi |^{2}\psi =0
\label{Two_Dlata_NLSE}
\end{equation}%
(recall the same double-delta nonlinearity, but without the OL potential,
and solely with the attractive sign of the nonlinearity, $\sigma =-1$, was
introduced in Ref. \cite{DeltaSSB}). In this case, the asymmetry measure of
stationary modes is defined as
\begin{equation}
\theta =\frac{\int_{0}^{+\infty }\left\vert \psi (x)\right\vert
^{2}dx-\int_{-\infty }^{0}\left\vert \psi (x)\right\vert ^{2}dx}{%
\int_{-\infty }^{+\infty }\left\vert \psi (x)\right\vert ^{2}dx}\equiv \frac{%
N_{+}-N_{-}}{N}.  \label{Theta}
\end{equation}

Dynamical invariants of Eqs. (\ref{Single_Dlata_NLSE}) and (\ref%
{Two_Dlata_NLSE}) are norm $N$, given by Eq. (\ref{Norm}), and the
Hamiltonian (written here for the latter equation),
\begin{equation}
H=\int_{-\infty }^{+\infty }\left[ \frac{1}{2}\left\vert \psi
_{x}\right\vert ^{2}-\varepsilon \cos \left( 2x\right) \left\vert \psi
(x)\right\vert ^{2}\right] dx+\frac{1}{2}\sum_{+,-}\left\vert \psi \left(
x=\pm \xi \right) \right\vert ^{4}.  \label{Ham}
\end{equation}%
$N$ is proportional to the number of atoms trapped in the BEC, or the total
power of the trapped beam in optics. Control parameters of the models are $%
\varepsilon $ and $\xi $, along with $N$.

As said above, in models with periodic potentials solitons may exist in
bandgaps of the spectrum of the linearized version of the equation -- either
the semi-infinite gap or finite ones, separated by Bloch bands \cite%
{Abdullaev,Kivshar,Morsch}. For the models under consideration, the
spectrum is the same as in the classical Mathieu equation (it is
actually displayed by means of shaded areas in Fig.
\ref{Band_Gap_approximation} below).

\section{Perturbation analysis of the single-delta model}

\label{sec:Approximation} Equation (\ref{Single_Dlata_NLSE}), as well as the
corresponding linear Mathieu equation, can be treated by means of the
perturbation theory if the OL strength, $\varepsilon $, is a small
parameter. Here, we consider the case when the $\delta $-function is placed
symmetrically with respect to the periodic potential, i.e., at $\xi =0$, the
objective being to construct approximate analytical solutions for GSs (gap
solitons) supported by the repulsive ($\sigma =+1$) or attractive ($\sigma
=-1$) point-wise nonlinearity, in the first and second finite bandgaps, as
well as for ordinary solitons in the the semi-infinite gap. The respective
form of the stationary equation is
\begin{equation}
\mu \phi +(1/2)\phi ^{\prime \prime }+\left[ \varepsilon \cos (2x)-\sigma
\delta (x)\phi ^{2}\right] \phi =0,  \label{Single_Dlata_NLSE_symmetric}
\end{equation}%
cf. Eq. (\ref{phi-simple}).

\subsection{The first finite bandgap}

\subsubsection{Solutions of the linear equation}

In the case of small $|\varepsilon |$, the first finite bandgap occupies a
narrow interval of values of $\mu $ around $\mu =1/2$. Accordingly, we set
\begin{equation}
\mu \equiv 1/2+\nu ,~~\mathrm{with}~~|\nu |\ll 1.  \label{nu}
\end{equation}%
Then, approximate solutions corresponding to the GS in the first finite
bandgap can be sought for as%
\begin{equation}
\phi (x)=e^{-\lambda _{1}|x|}\left[ A\cos (x)+B\sin \left( |x|\right) \right]
,  \label{phi}
\end{equation}%
where $\lambda _{1}$ is assumed to be a small positive coefficient
(subscript $1$ indicates that $\lambda _{1}$ pertains to the first finite
bandgap).

The substitution of ansatz (\ref{phi}) into Eq. (\ref%
{Single_Dlata_NLSE_symmetric}), off point $x=0$ (which corresponds to the
linear equation), and the subsequent analysis following the usual
perturbation theory for the Mathieu equation (in other words, the asymptotic
analysis of the linear parametric resonance \cite{Landau}), yield the
following equations for amplitudes $A$ and $B$:
\begin{eqnarray}
\left[ \nu +\frac{1}{2}\left( \lambda _{1}^{2}+\varepsilon \right) \right]
A-\lambda _{1}B &=&0,  \notag \\
\left[ \nu +\frac{1}{2}\left( \lambda _{1}^{2}-\varepsilon \right) \right]
B+\lambda _{1}A &=&0.  \label{AB}
\end{eqnarray}%
The solvability condition for the linear homogeneous system (\ref{AB}) is%
\begin{equation}
\left( \nu +\frac{1}{2}\lambda _{1}^{2}\right) ^{2}-\frac{\varepsilon ^{2}}{4%
}+\lambda _{1}^{2}=0.
\end{equation}%
In the first approximation, $\lambda _{1}^{2}/2$ may be neglected in the
parentheses, which yields%
\begin{equation}
\lambda _{1}\approx\sqrt{\frac{\varepsilon ^{2}}{4}-\nu ^{2}},
\label{lambda}
\end{equation}%
hence the solution exists for
\begin{equation}
|\nu |<|\varepsilon |/2.  \label{width}
\end{equation}%
In fact, Eq. (\ref{width}) is the prediction provided by the perturbation
theory for the width of the first finite bandgap, see Fig.~\ref%
{Band_Gap1_approximation-a}. Further, it follows from Eqs. (\ref{AB}) that,
in the first approximation,%
\begin{equation}
B=\mathrm{sgn}\left( \varepsilon \right) \sqrt{\frac{\varepsilon
/2+\nu }{\varepsilon /2-\nu }}A,  \label{B}
\end{equation}%
where condition (\ref{width}) is taken into regard, to identify the correct
sign.

\subsubsection{The nonlinear part of the solution}

For the stationary solutions, condition (\ref{xx}) takes the form of
\begin{equation}
\Delta \left( \frac{d\phi }{dx}\right) |_{x=0}=2\sigma \phi ^{3}|_{x=0}~,
\label{Delta}
\end{equation}%
where $\Delta $ stands for the jump of the derivative at $x=0$. The
substitution of Eqs. (\ref{phi}), (\ref{lambda}), and (\ref{B}) into Eq. (%
\ref{Delta}) yields
\begin{equation}
-\left( \sqrt{\frac{\varepsilon ^{2}}{4}-\nu ^{2}}-\mathrm{sgn}\left(
\varepsilon \right) \sqrt{\frac{\varepsilon /2+\nu }{\varepsilon /2-\nu }}%
\right) A= \sigma A^{3}.  \label{balance}
\end{equation}%
On the left-hand side of Eq. (\ref{balance}), the term $\sqrt{\varepsilon
^{2}/4-\nu ^{2}}$ is much smaller than the following term [in the general
case, $\sqrt{\varepsilon ^{2}/4-\nu ^{2}}$ is small, while $\sqrt{\left(
\varepsilon /2+\nu \right) /\left( \varepsilon /2-\nu \right) }$ is not; in
special cases, near the upper edge of the bandgap for $\varepsilon >0$, or
the lower one for $\varepsilon <0$, with $\left\vert \varepsilon /2+\nu
\right\vert \ll \left\vert \varepsilon /2-\nu \right\vert $, the term $\sqrt{%
\left( \varepsilon /2+\nu \right) /\left( \varepsilon /2-\nu \right) }$
becomes small, but $\sqrt{\varepsilon ^{2}/4-\nu ^{2}}$ is then still
smaller]. Thus, Eq. (\ref{balance}) yields, for nontrivial solutions ($%
A^{2}\neq 0$):%
\begin{equation}
A^{2}\approx \sigma ~\mathrm{sgn}\left( \varepsilon \right) \sqrt{\frac{%
\varepsilon /2+\nu }{\varepsilon /2-\nu }}.  \label{A}
\end{equation}%
This solution exists provided that $\mathrm{sgn}\left( \varepsilon \right)
=\sigma $, and it does not exist in the opposite case. In other words, it
exists if the \emph{repulsive} $\delta $-function, with $\sigma =+1$, is set
at the local \emph{minimum} of the OL potential, or the \emph{attractive} $%
\delta $-function is placed at the local \emph{maximum}. Note that, although
the result was obtained by means of the perturbation theory, the amplitude
given by Eq. (\ref{A}) is not small (which does not invalidate the
perturbative treatment).

In the first approximation, the calculation of the norm of the weakly
localized soliton solution based on the above formulas yields%
\begin{equation}
N_{1}\approx \left( 2\lambda _{1}\right) ^{-1}\left( A^{2}+B^{2}\right)
\equiv \frac{|\varepsilon |}{2\left( \varepsilon /2-\nu \right) ^{2}}.
\label{N}
\end{equation}%
In the case of the attraction (then, $\varepsilon $ is negative, as shown
above), relation (\ref{N}) yields
\begin{equation}
\frac{dN_{1}}{d\mu }\equiv \frac{dN_{1}}{d\nu }=-\frac{\varepsilon }{%
\left( \varepsilon /2-\nu \right) ^{3}}<0,  \label{VK-wrong}
\end{equation}%
hence, according to the VK criterion, the so predicted family of the GSs
might be stable. Nevertheless, it is actually unstable, as shown below. As
for the repulsive nonlinearity, the VK criterion is irrelevant for it (being
sometimes replaced by an ``anti-VK" condition \cite{HS}).

The calculation of Hamiltonian (\ref{Ham}) in the present approximation
yields%
\begin{equation}
H\approx \frac{|\varepsilon |}{4\left( \varepsilon /2-\nu \right) ^{2}}
\end{equation}%
[in this approximation, $H$ is dominated by the gradient term in Eq. (\ref%
{Ham})]. The fact that this expression is always positive demonstrates that
the GS cannot realize a ground state (it cannot correspond to an absolute
minimum of the energy, which must be either negative or zero). Nevertheless,
this does not mean that these solitons cannot be stable against small
perturbations (actually, they may be \textit{metastable} states, in
comparison with the ground state).

\subsubsection{Comparison with numerical results}

\begin{figure}[tbp]
\subfigure[]{\includegraphics[width=2.7in]{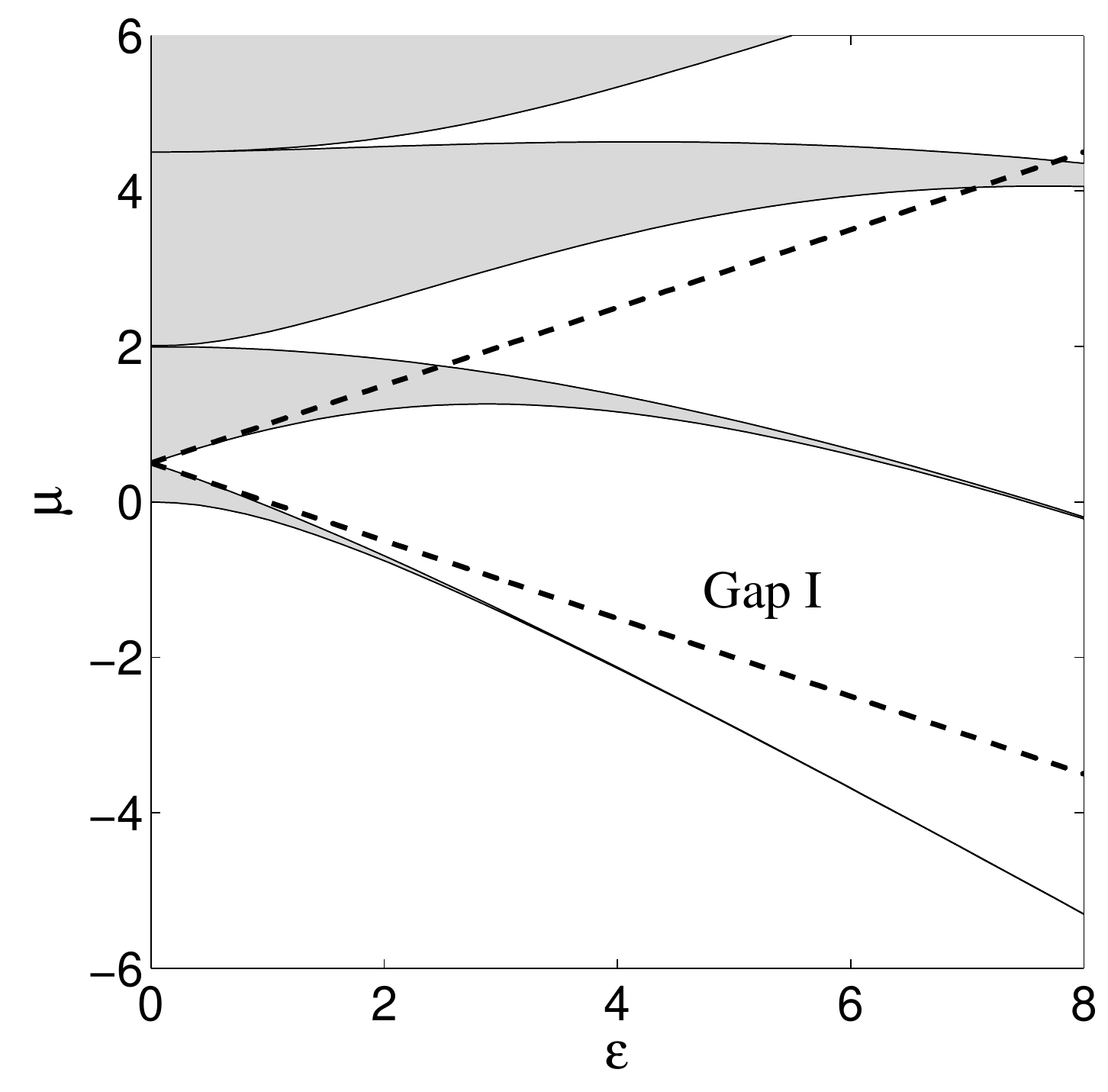}%
\label{Band_Gap1_approximation-a}} \quad \subfigure[]{%
\includegraphics[width=2.7in]{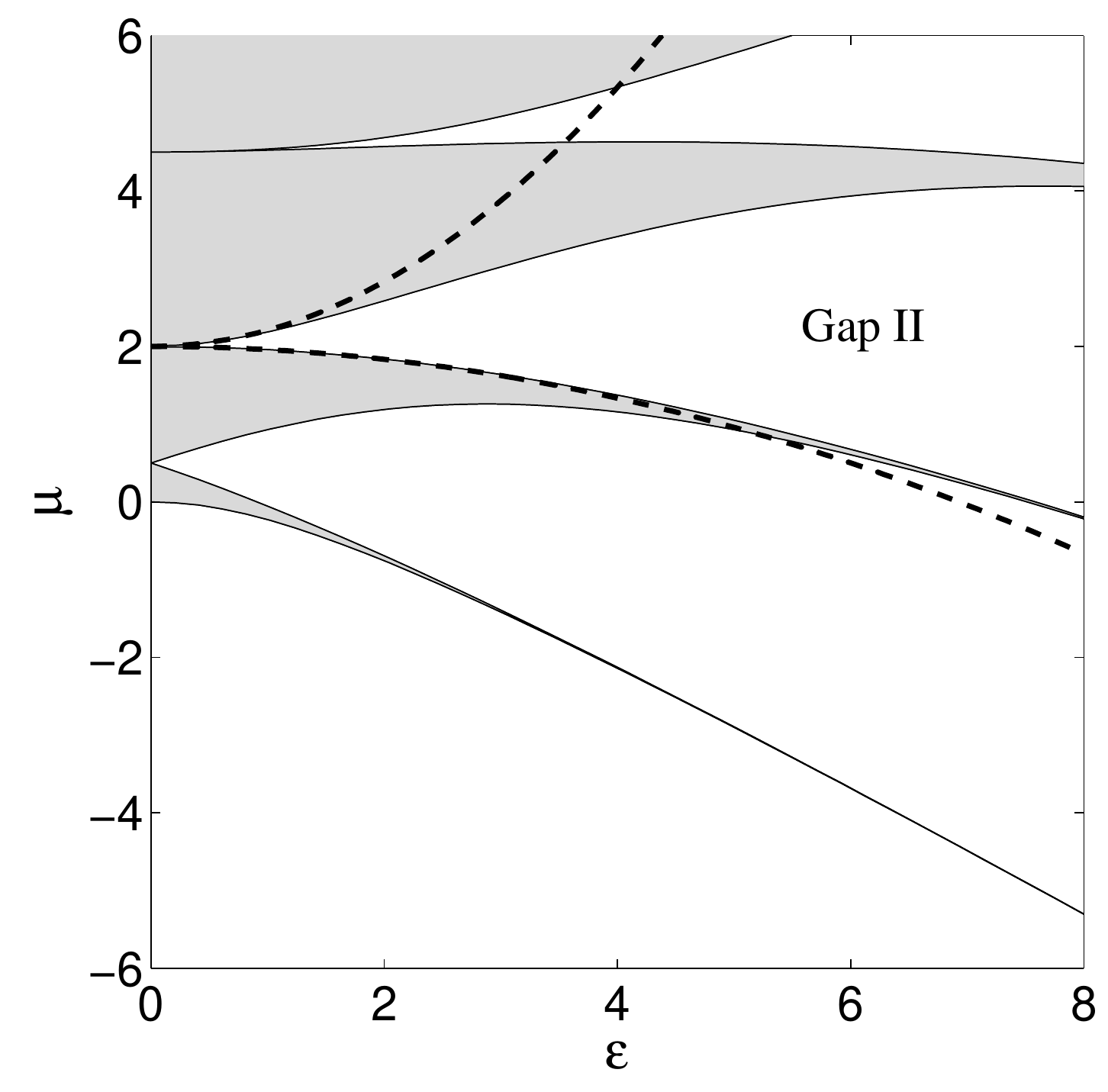}%
\label{Band_Gap2_approximation-b}}
\caption{Borders of the first (a) and second (b) finite bandgaps (labeled as
Gap I and Gap II, respectively), as predicted by the perturbation
approximation, see Eqs. (\protect\ref{width}) and (\protect\ref{SFG}), in
comparison with the numerically constructed bandgap structure, in which
Bloch bands are shaded. The semi-infinite gap is represented by the white
areas at the bottom of the panels.}
\label{Band_Gap_approximation}
\end{figure}

The first finite bandgap, as predicted by the perturbation theory [Eq. (\ref%
{width})], is shown in Fig.~\ref{Band_Gap1_approximation-a}, together with
the numerically found borders of the first bandgap. In accordance with the
above prediction, the soliton in the model with the repulsive or attractive
nonlinearity exists only for $\varepsilon >0$ and $\varepsilon <0$,
respectively (this conclusion is true for all values of $\varepsilon $ and $%
\mu $ in the first bandgap, not only for small $\varepsilon $).

As concerns the stability, the GSs supported by the repulsive $\delta $%
-function placed at the minimum of the OL potential ($\sigma =+1$, $%
\varepsilon >0$) are \emph{stable}, while all the solitons generated in the
finite bandgap by the attractive nonlinearity ($\sigma =-1,~\varepsilon <0$)
are unstable [on the contrary to the formal prediction of the VK criterion,
see Eq. (\ref{VK-wrong})]. The detailed stability analysis is reported in
the next section.

\begin{figure}[tbp]
\subfigure[]{\includegraphics[width=2.2in]{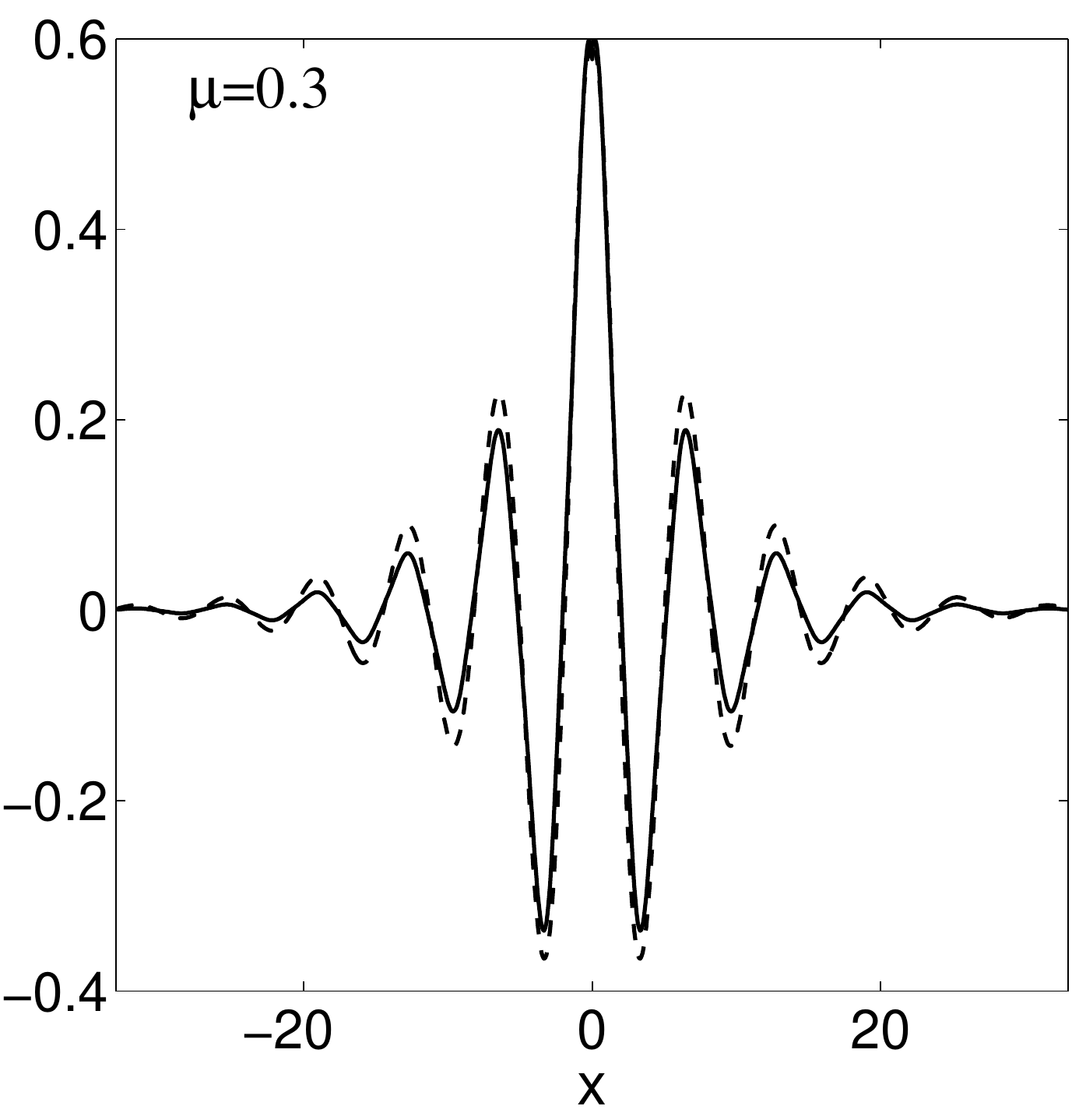}%
\label{Profile_eps05_mu03}} \subfigure[]{%
\includegraphics[width=2.2in]{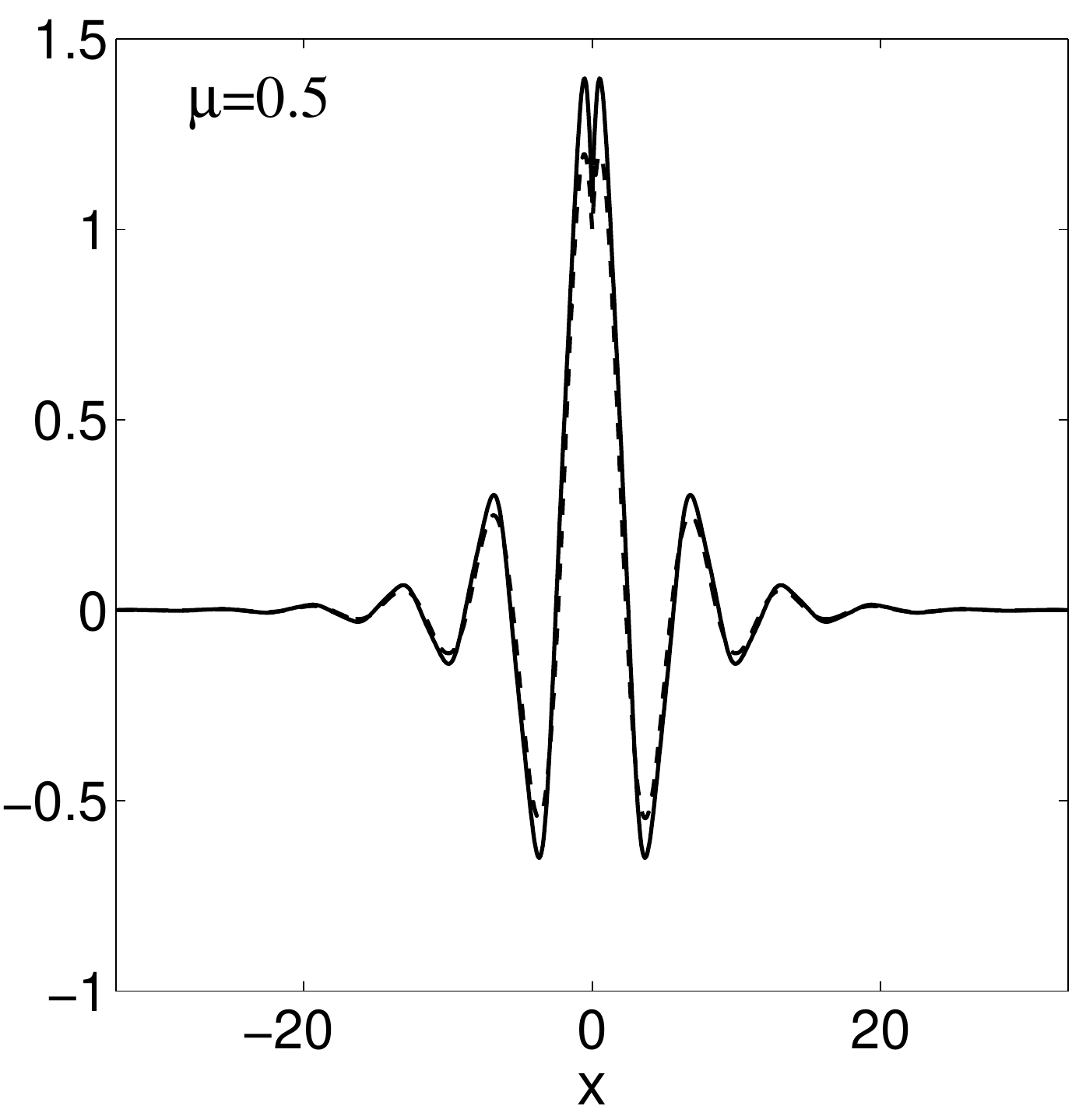}%
\label{Profile_eps05_mu05}} \subfigure[]{%
\includegraphics[width=2.1in]{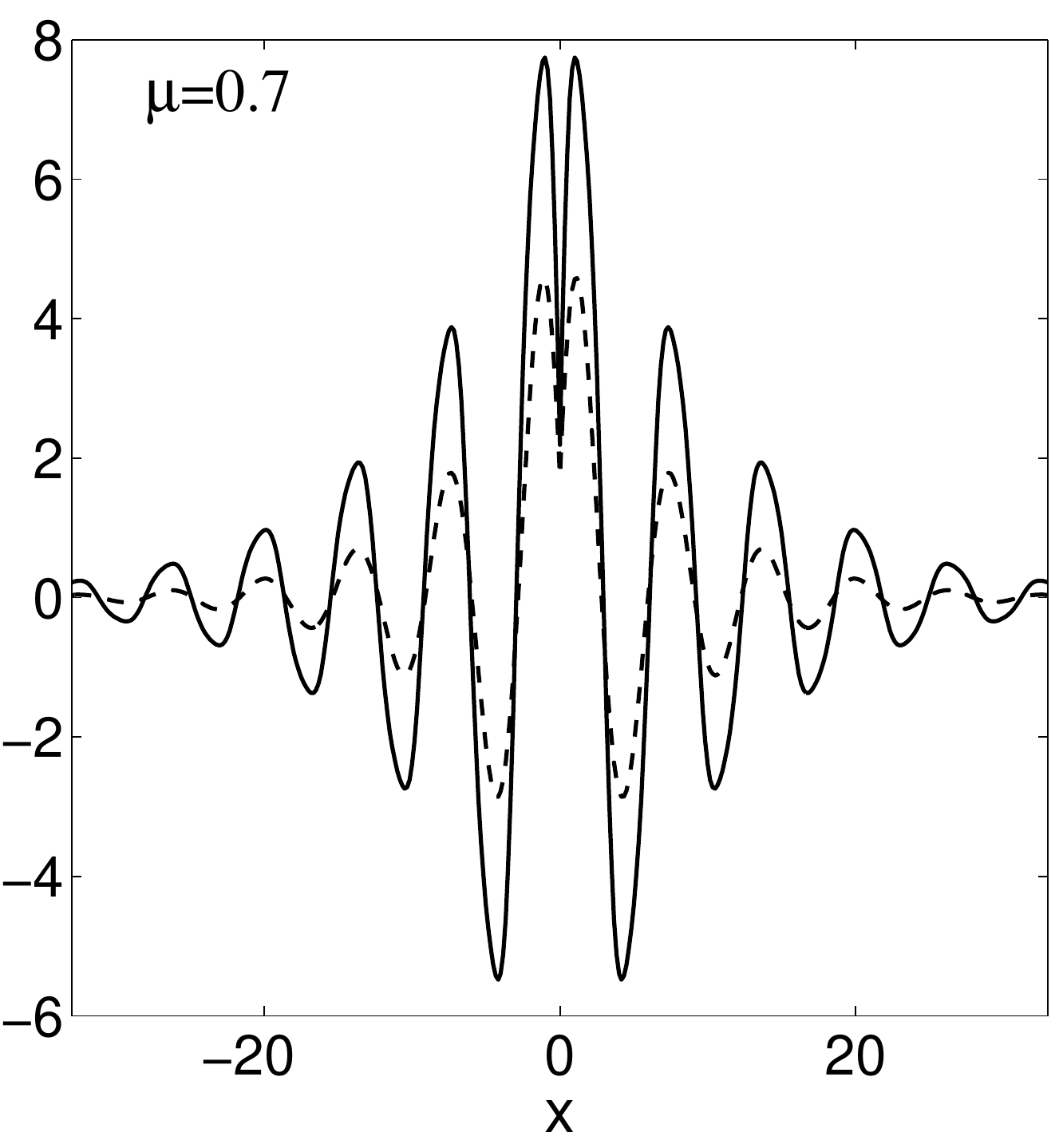}%
\label{Profile_eps05_mu07}} \\ 
\subfigure[]{\includegraphics[width=2.2in]{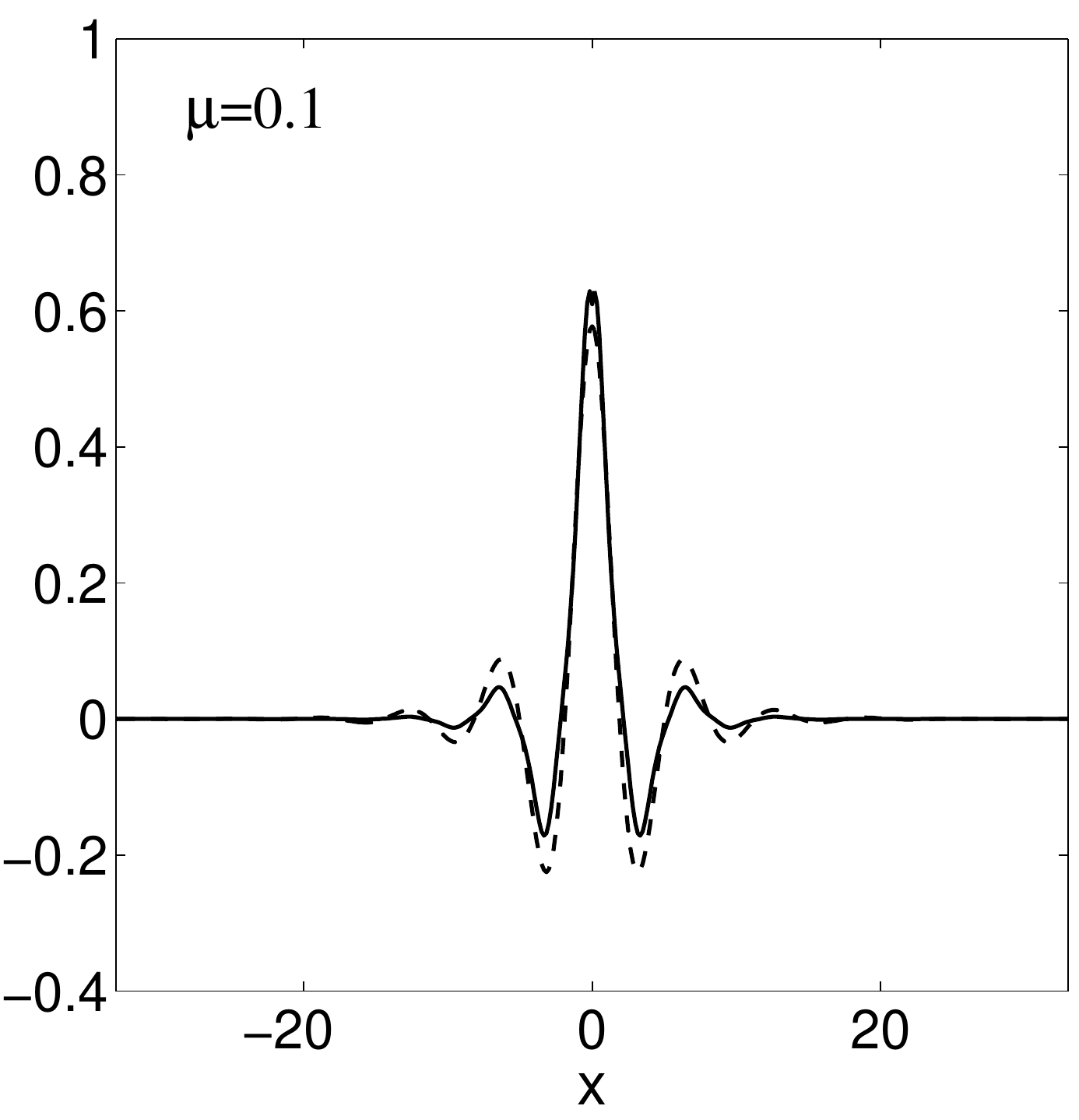}%
\label{Profile_eps1_mu01}} \subfigure[]{%
\includegraphics[width=2.2in]{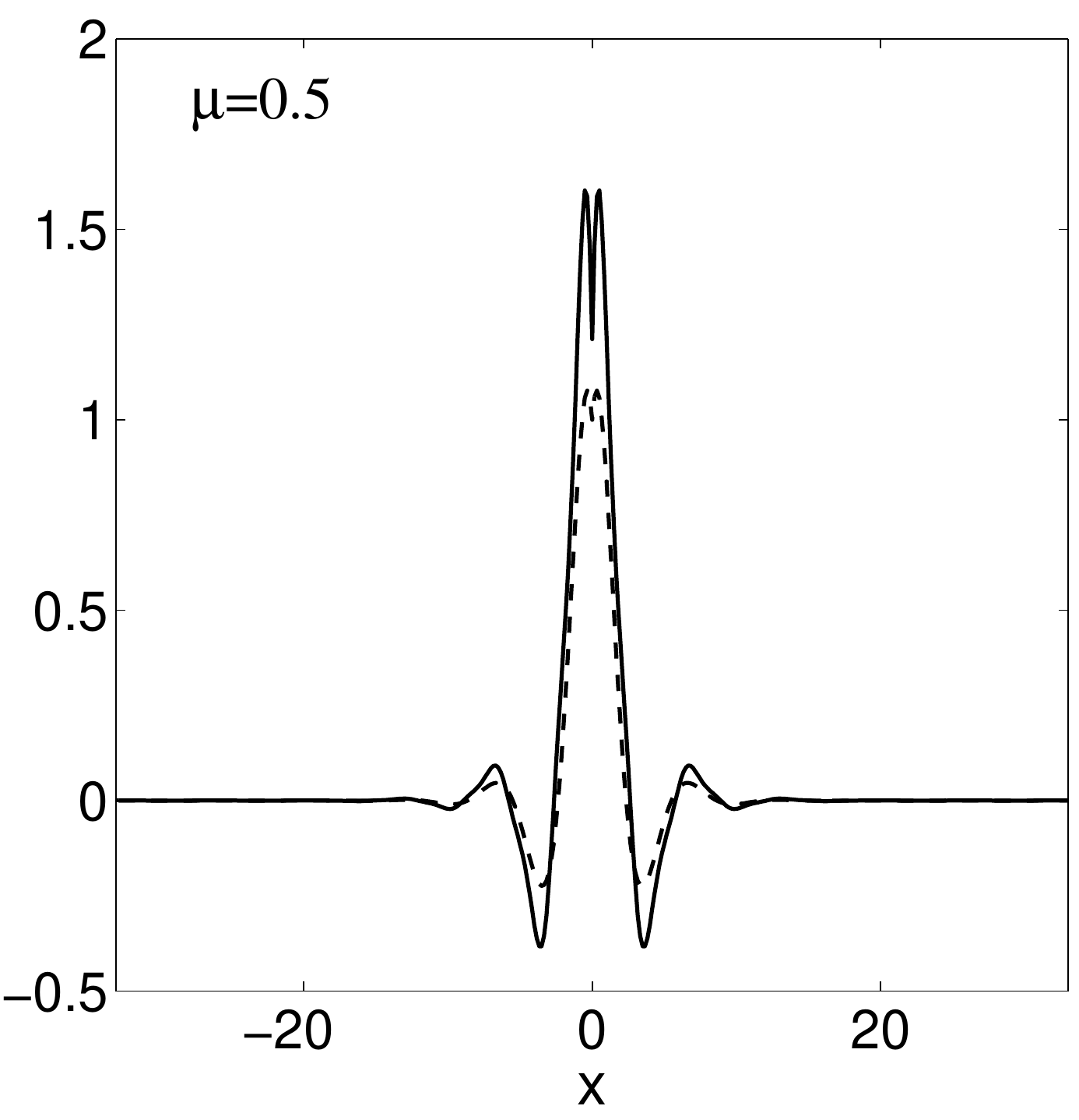}%
\label{Profile_eps1_mu05}} \subfigure[]{%
\includegraphics[width=2.1in]{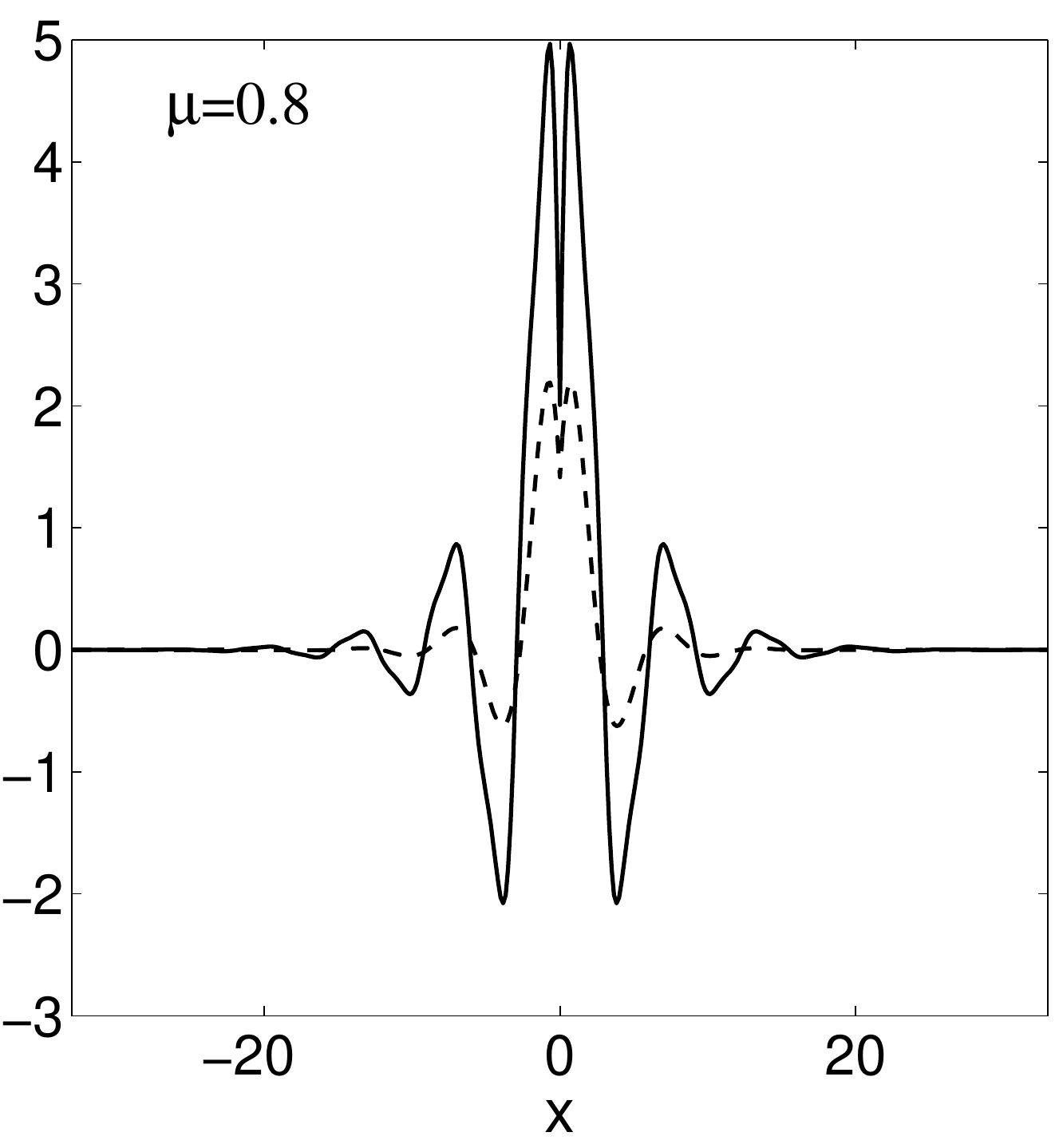}%
\label{Profile_eps1_mu08}}
\caption{Comparison between the analytical approximation (dashed lines) and
the numerically found profiles (solid lines) for solitons in the first
finite bandgap of the repulsive model, for $\protect\varepsilon =0.5$
(a)-(c) and $\protect\varepsilon =1$ (d)-(f). The results are shown for
selected values of $\protect\mu $ (indicated in each panel), close to the
center of the gap or near each of its edges.}
\label{Profiles_FFG_repulsive}
\end{figure}

\begin{figure}[tbp]
\subfigure[]{\includegraphics[width=2.1in]{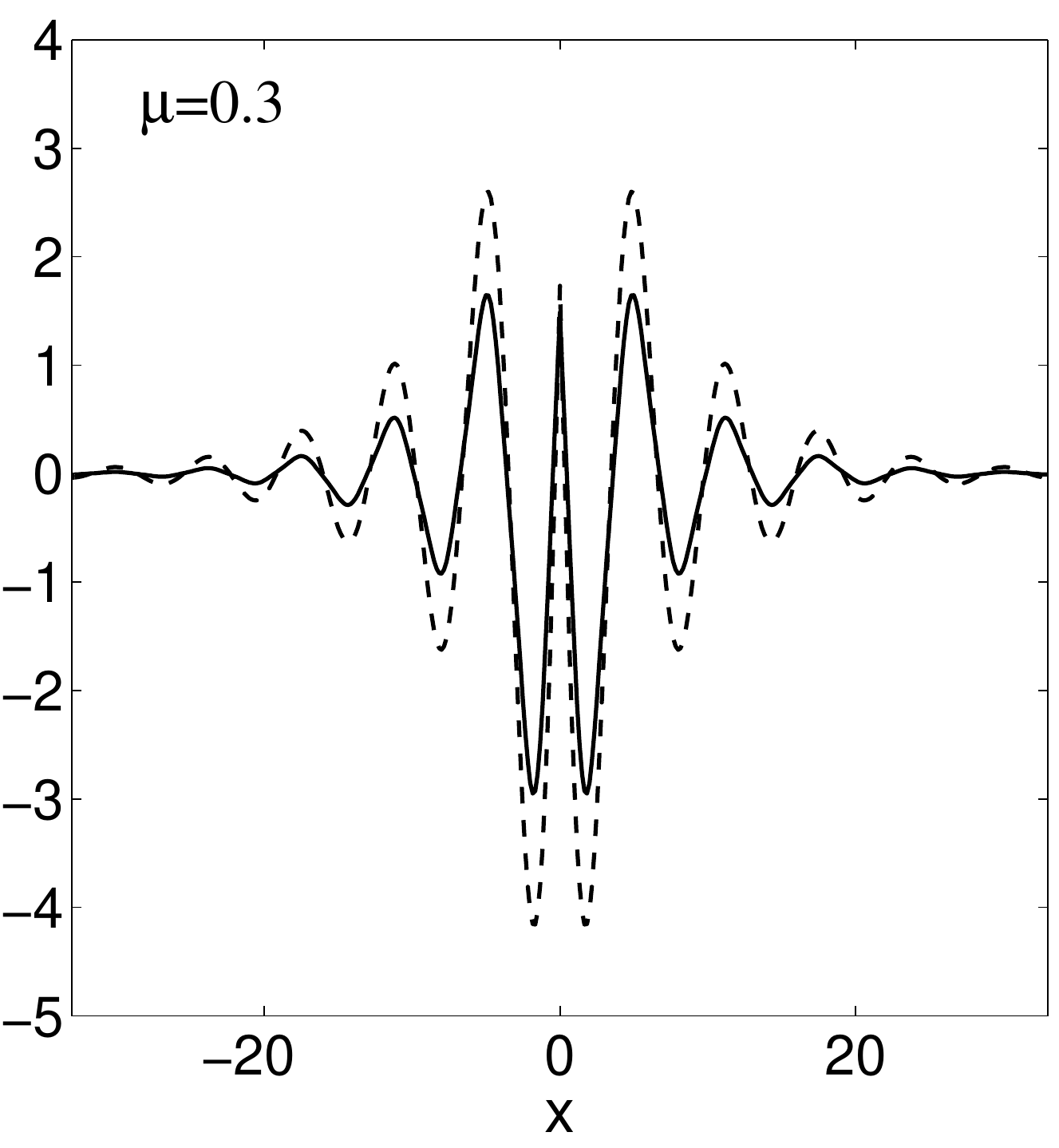}%
\label{Profile_epsm05_mu03}} \subfigure[]{%
\includegraphics[width=2.2in]{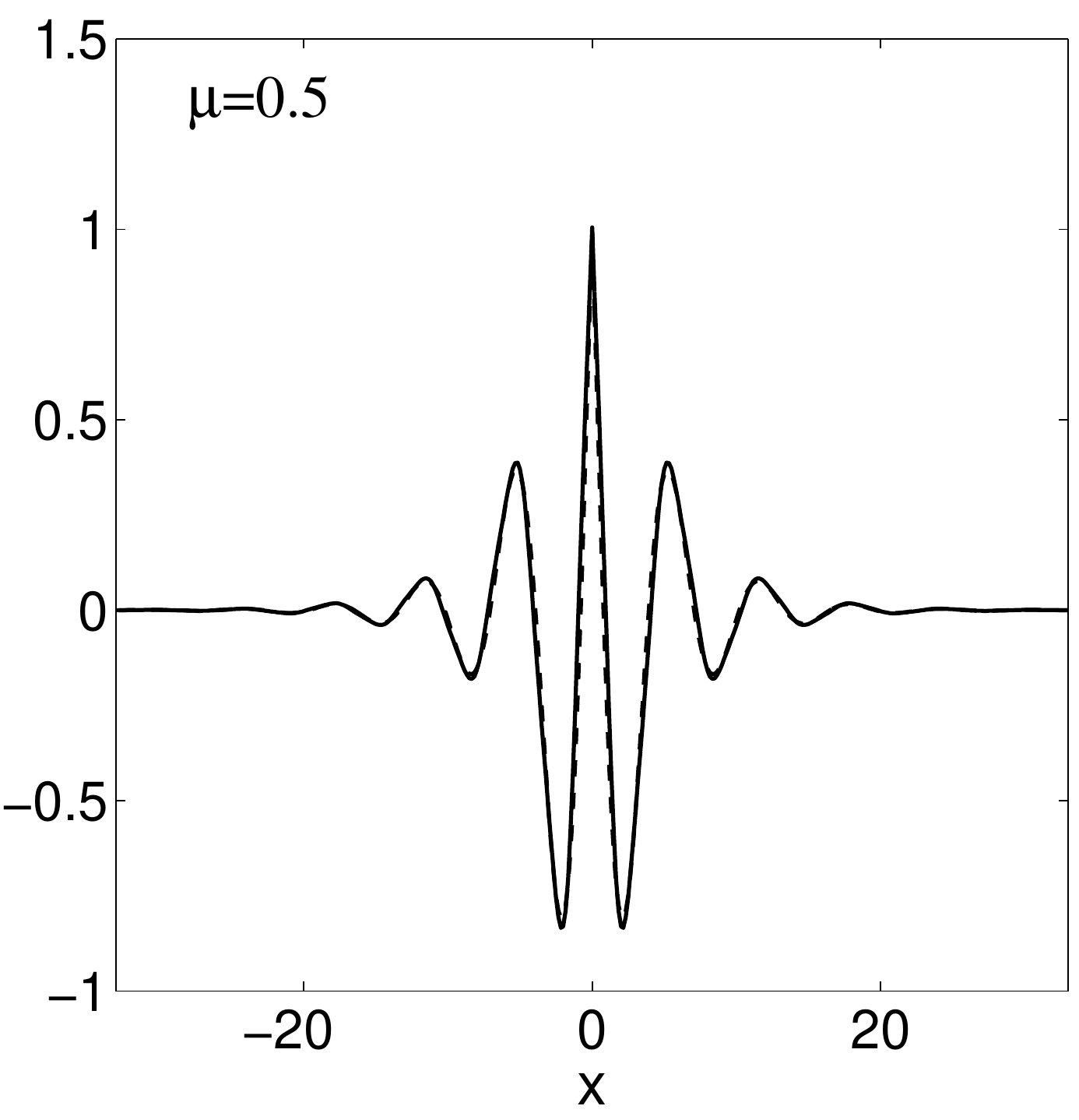}%
\label{Profile_epsm05_mu05}} \subfigure[]{%
\includegraphics[width=2.2in]{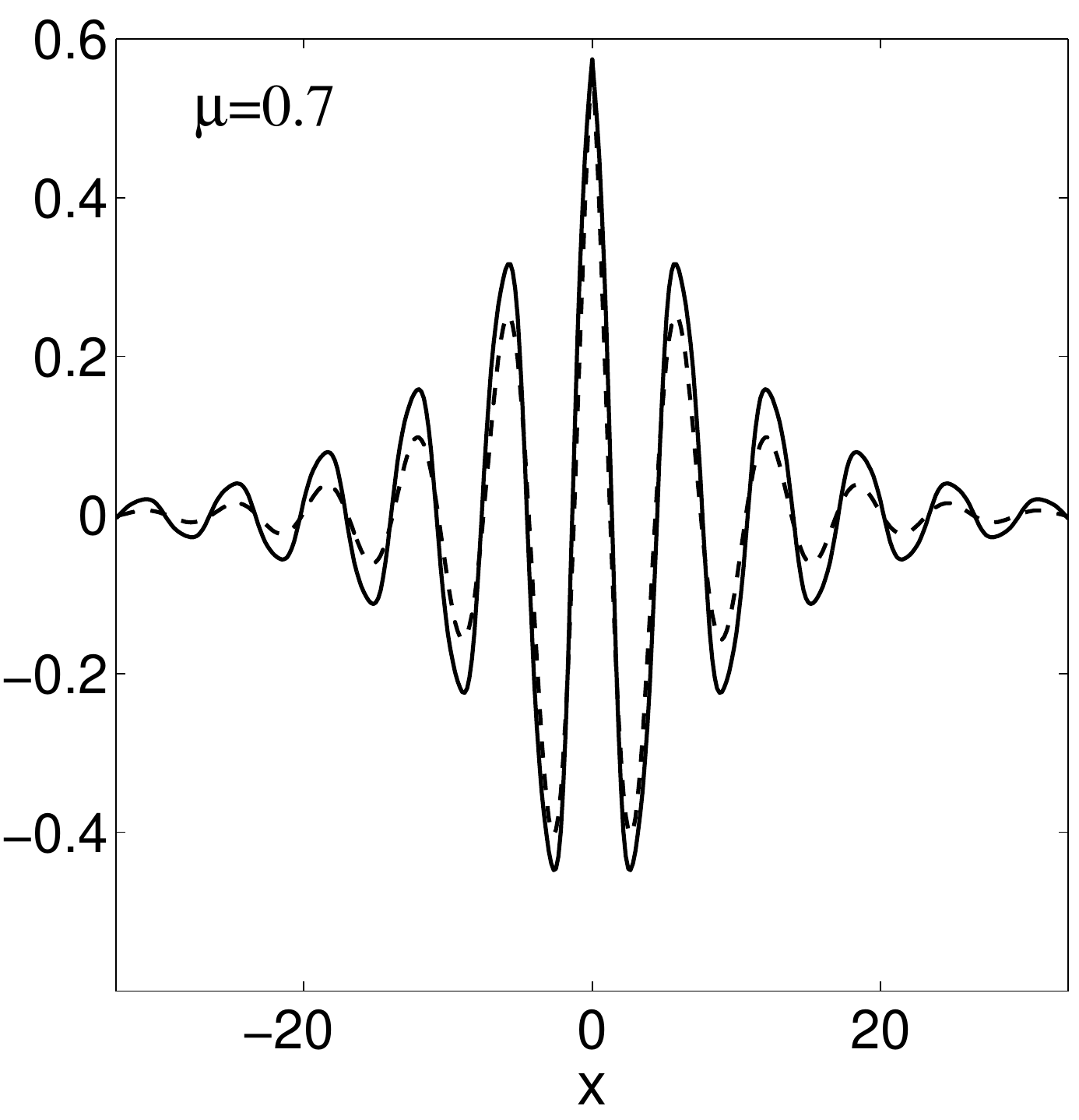}%
\label{Profile_epsm05_mu07}} \\ 
\subfigure[]{\includegraphics[width=2.1in]{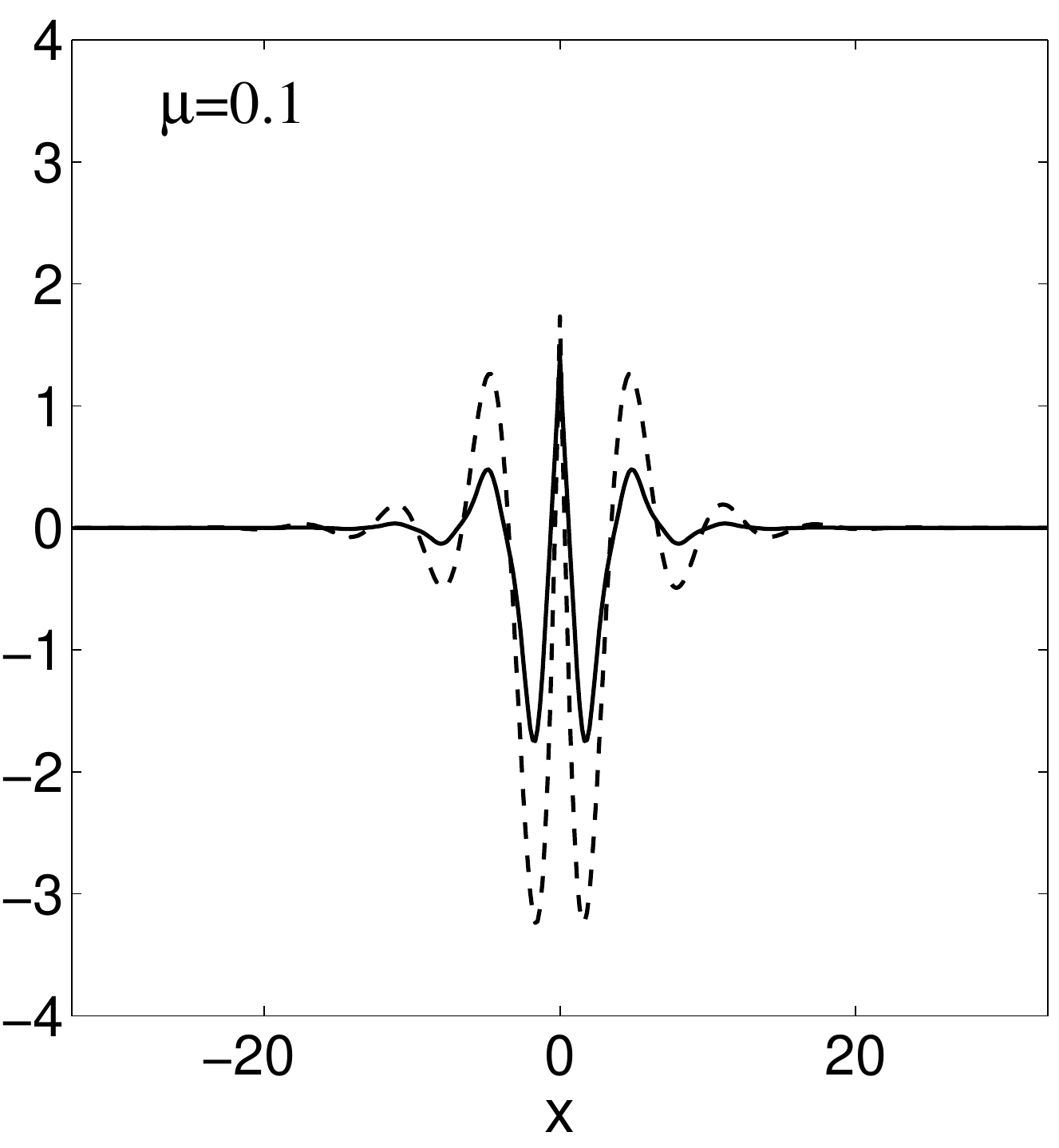}%
\label{Profile_epsm1_mu01}} \subfigure[]{%
\includegraphics[width=2.2in]{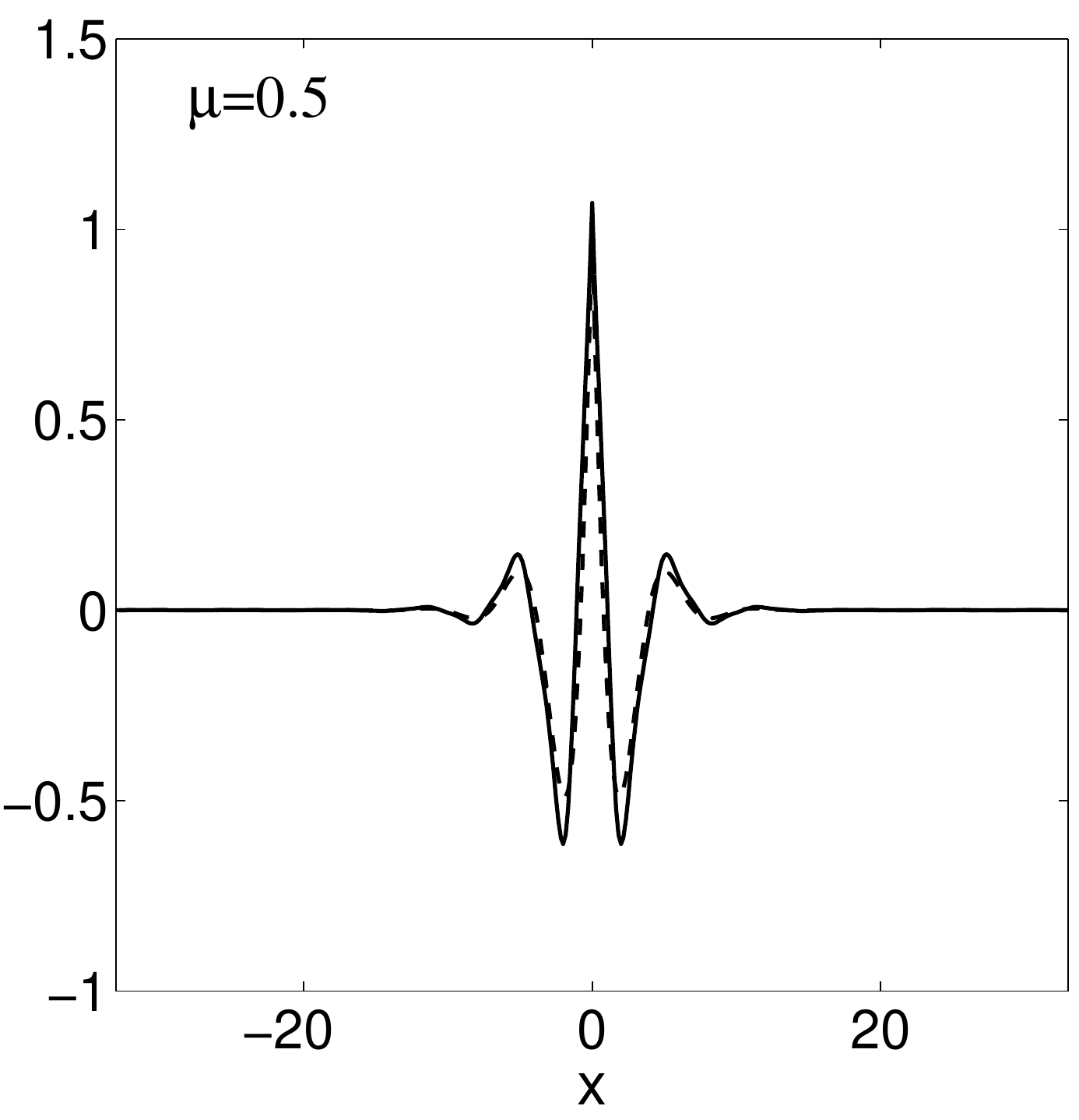}%
\label{Profile_epsm1_mu05}} \subfigure[]{%
\includegraphics[width=2.2in]{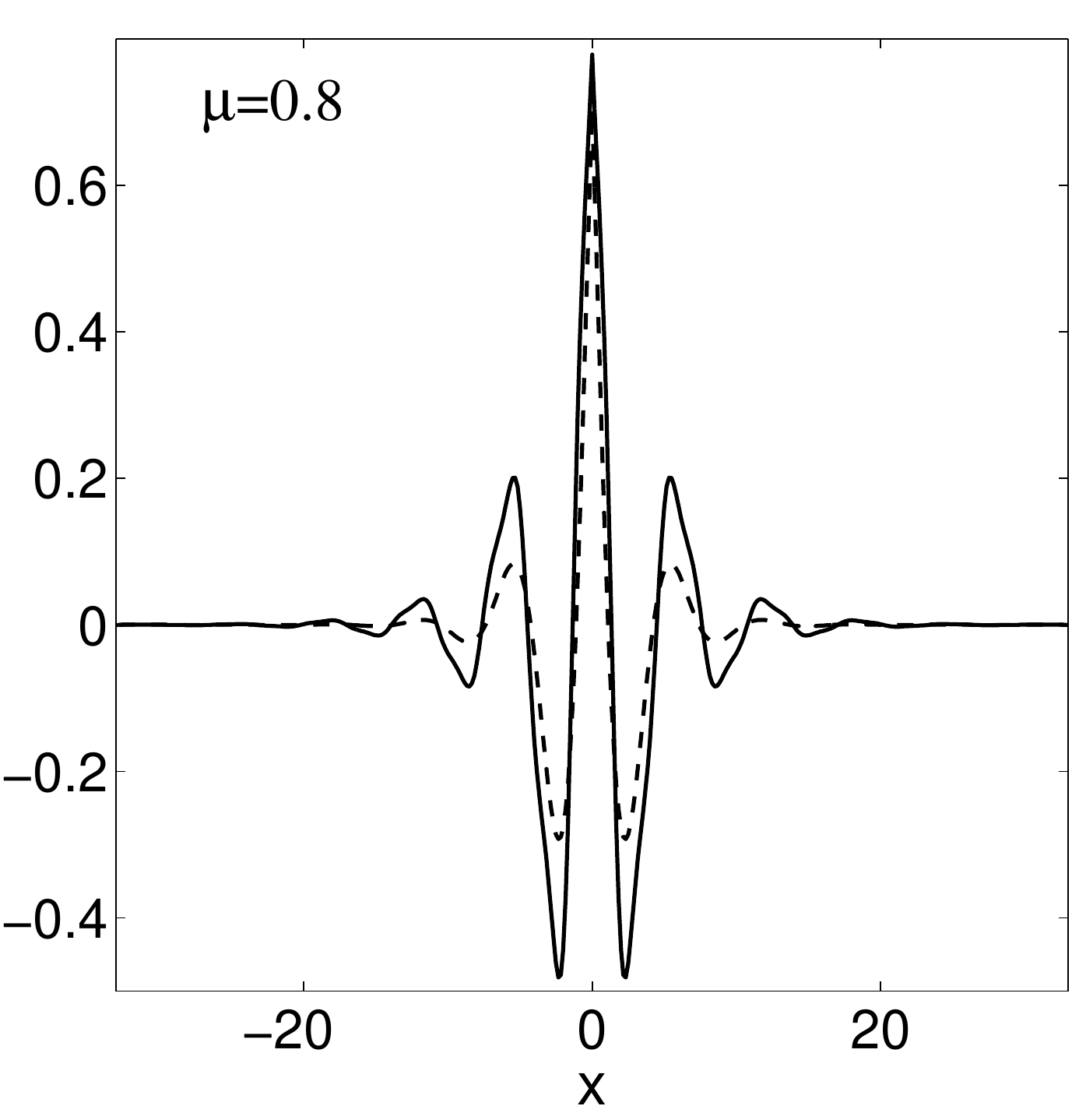}%
\label{Profile_epsm1_mu08}}
\caption{The same as in Fig.~\protect\ref{Profiles_FFG_repulsive}, but for
the attractive nonlinearity. In panels (a)-(c), $\protect\varepsilon =-0.5$,
and in (d)-(f), $\protect\varepsilon =-1$.}
\label{Profiles_FFG_attractive}
\end{figure}

Typical profiles of the soliton solutions are shown in Figs.~\ref%
{Profiles_FFG_repulsive} and~\ref{Profiles_FFG_attractive}. If $\varepsilon $
is not too large, the approximation quite accurately predicts the shape of
the soliton. In particular, the analytical and numerical results agree very
well for $\mu $ taken near the middle of the bandgap. Close to the upper
edge (for $\varepsilon >0$) or lower edge (for $\varepsilon <0$), the
analytical shape of the soliton is less accurate, as seen in Figs. \ref%
{Profile_eps05_mu07},\subref{Profile_eps1_mu08} and~\ref{Profile_epsm05_mu03}%
,\subref{Profile_epsm1_mu01}. In fact, this inaccuracy originates from the
use of approximation (\ref{A}) instead of the more general equation (\ref%
{balance}).

\begin{figure}[tbp]
\subfigure[]{\includegraphics[width=1.7in]{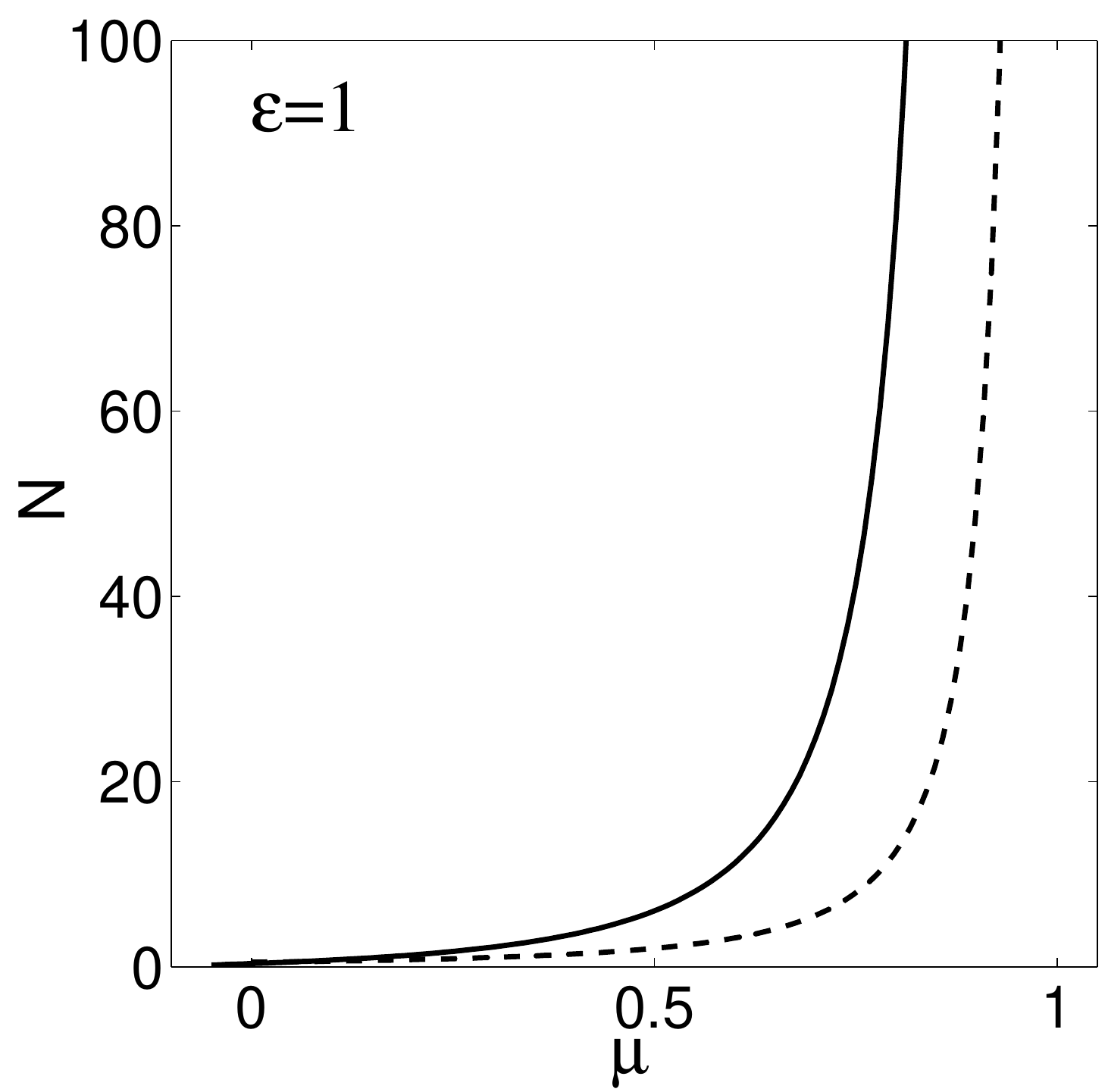}%
\label{NvsMu_approximation_eps1}} \subfigure[]{%
\includegraphics[width=1.7in]{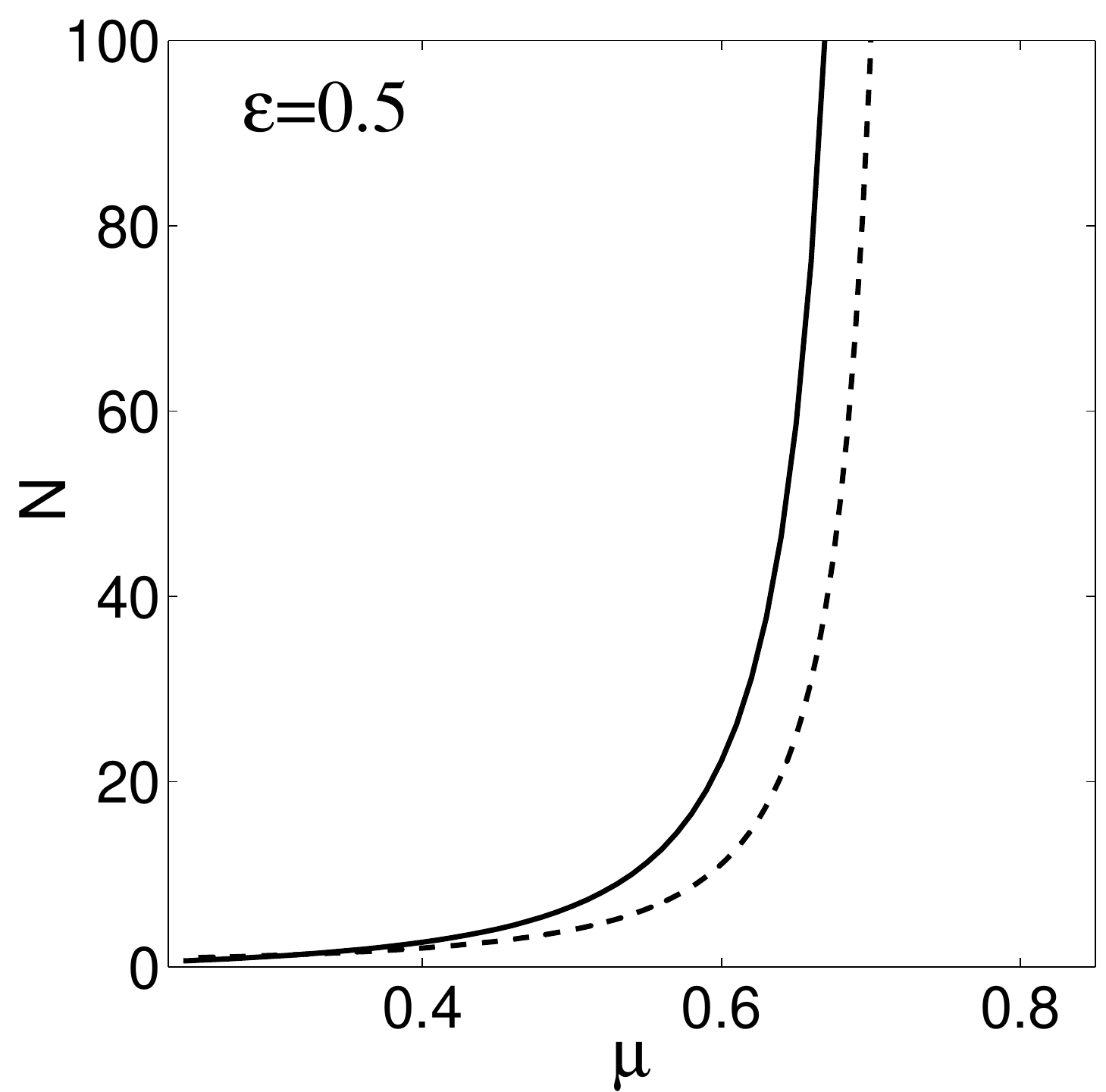}%
\label{NvsMu_approximation_eps05}} \subfigure[]{%
\includegraphics[width=1.7in]{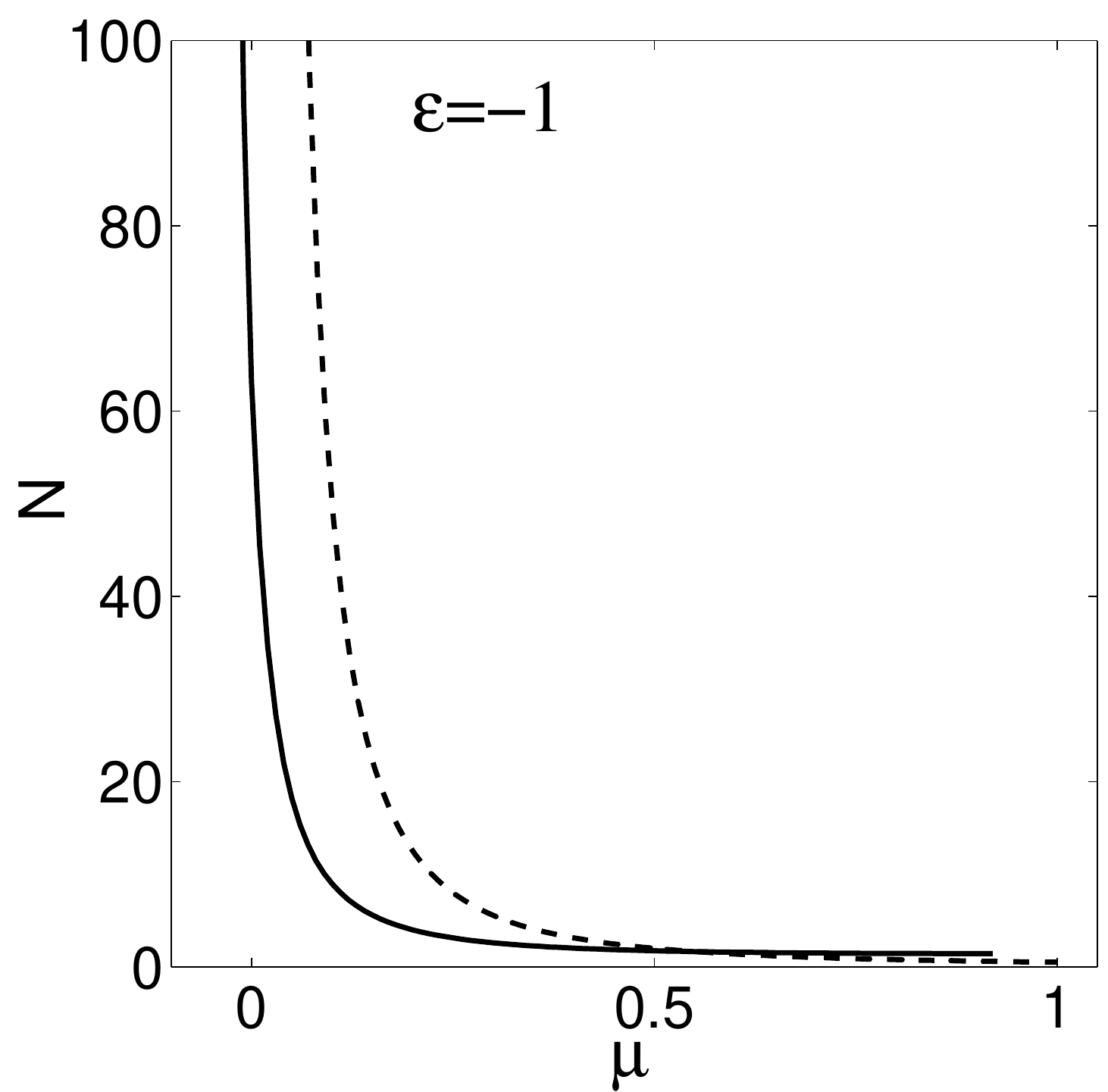}%
\label{NvsMu_approximation_epsm1}} \subfigure[]{%
\includegraphics[width=1.7in]{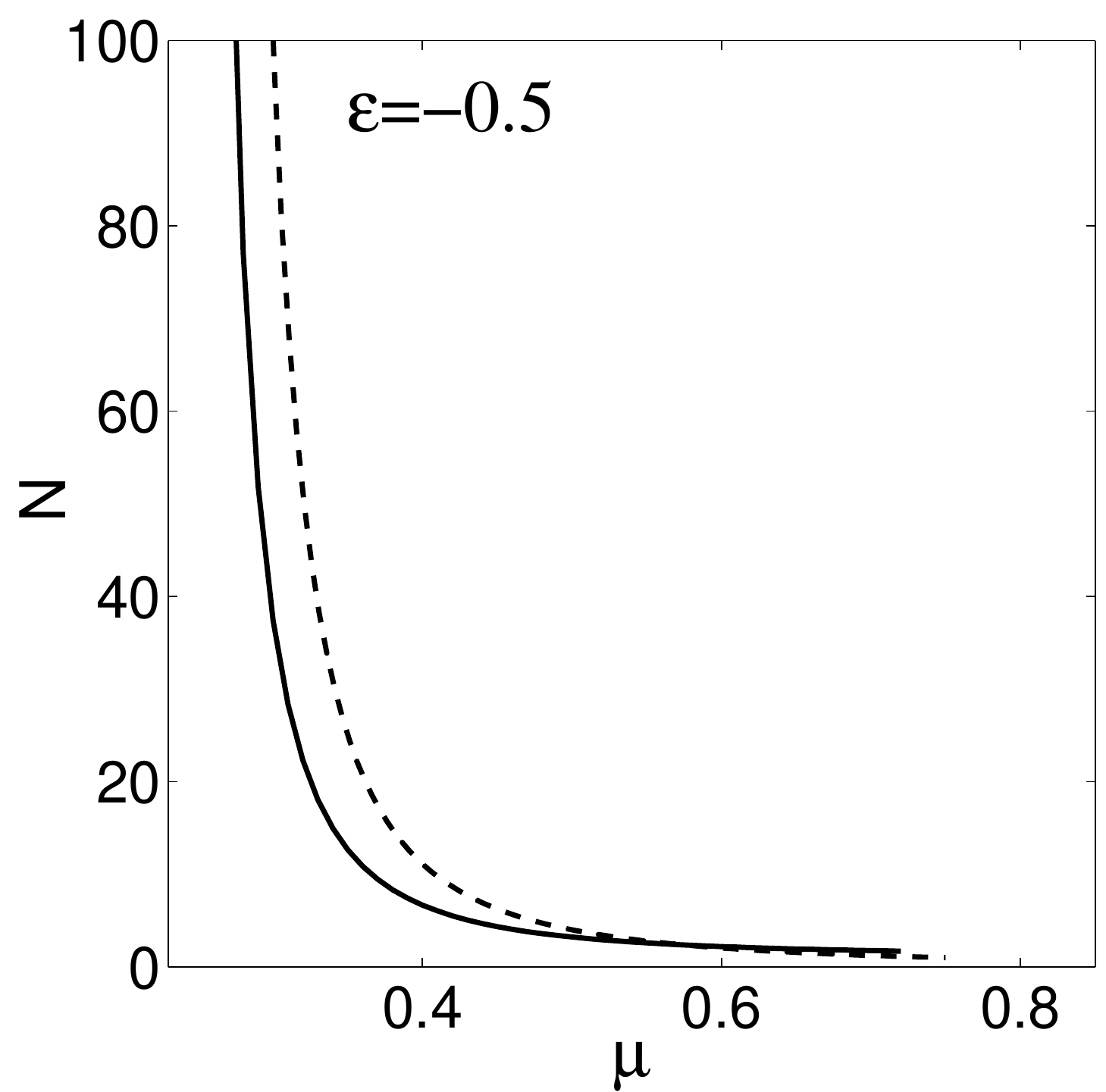}%
\label{NvsMu_approximation_epsm05}}
\caption{The norm of the solitons in the first finite bandgap, as a function
of $\protect\mu $. Examples for the repulsive nonlinearity are given for $%
\protect\varepsilon =1$ (a) and $\protect\varepsilon =0.5$ (b), and the
attractive model is demonstrated for $\protect\varepsilon =-1$ (c) and $%
\protect\varepsilon =-0.5$ (d). Numerical results and the analytical
approximation are depicted by solid and dashed lines, respectively.}
\label{N_vrs_Mu_approximation}
\end{figure}

Figure~\ref{N_vrs_Mu_approximation} shows the norm of the solitons versus $%
\mu $, for both negative and positive $\varepsilon $, which correspond to
the attractive and repulsive nonlinearities, respectively. As expected, the
results become more accurate as $|\varepsilon |$ diminishes, while $\mu $
takes values farther from the upper or lower edges of the bandgap, for $%
\varepsilon >0$ and $\varepsilon <0$, respectively.

\subsection{The second finite bandgap}

The perturbation theory for the Mathieu equation may also uncover the second
finite bandgap, when%
\begin{equation}
\mu =2+\nu ,~~\mathrm{with}~~|\nu |\ll 1\text{,}  \label{nu2}
\end{equation}%
cf. Eq. (\ref{nu}). To this end, an approximate solution to the linear
equation is looked for as
\begin{equation}
\phi (x)=e^{-\lambda _{2}|x|}\left[ A_{0}+A_{2}\cos \left( 4x\right)
+B_{2}\sin \left( 4|x|\right) +A_{1}\cos \left( 2x\right) +B_{1}\sin \left(
2|x|\right) \right] ,  \label{phi2}
\end{equation}%
cf. ansatz (\ref{phi}) adopted in the first finite bandgap. Substitution
ansatz (\ref{phi2}) into Eq. (\ref{Single_Dlata_NLSE_symmetric}), off point $%
x=0$, and neglecting terms $\sim \lambda _{2}$, we find, in the zeroth-order
approximation,%
\begin{equation}
A_{0}=-\frac{\varepsilon }{4}A_{1},~A_{2}=\frac{\varepsilon }{12}%
A_{1},~B_{2}=\frac{\varepsilon }{12}B_{1}~.  \label{0&2}
\end{equation}%
Next, the homogeneous system of equations for $A_{1}$ and $B_{1}$ takes the
following form, cf. Eqs. (\ref{AB}) derived in the first finite bandgap:%
\begin{eqnarray}
\nu A_{1}+\frac{1}{2}\lambda _{2}^{2}A_{1}+\varepsilon \left( A_{0}+\frac{1}{%
2}A_{2}\right) -2\lambda _{2}B_{1} &=&0, \\
\nu B_{1}+\frac{1}{2}\lambda _{2}^{2}B_{1}+\frac{1}{2}\varepsilon
B_{2}+2\lambda _{2}A_{1} &=&0,
\end{eqnarray}%
or, on substituting expressions (\ref{0&2}),%
\begin{eqnarray}
\left( \nu -\frac{5\varepsilon ^{2}}{24}+\frac{\lambda _{2}^{2}}{2}\right)
A_{1}-2\lambda _{2}B_{1} &=&0,  \notag \\
\left( \nu +\frac{\varepsilon ^{2}}{24}+\frac{\lambda _{2}^{2}}{2}\right)
B_{1}+2\lambda _{2}A_{1} &=&0.  \label{AB2}
\end{eqnarray}%
Like in the case of Eqs. (\ref{AB}), terms $\lambda _{2}^{2}/2$ in the
parentheses may be neglected in the lowest approximation, which yields%
\begin{equation}
\lambda _{2}\approx\frac{1}{2}\sqrt{\frac{5}{576}\varepsilon ^{4}+\frac{1}{6}%
\varepsilon ^{2}\nu -\nu ^{2}}.  \label{lambda2}
\end{equation}%
As follows from this expression, the perturbation theory predicts the
following form of the second finite bandgap, see Fig.~\ref%
{Band_Gap2_approximation-b}:%
\begin{equation}
-1/24\leq \tilde{\nu}\equiv \nu /\varepsilon ^{2}\leq 5/24,  \label{SFG}
\end{equation}%
cf. Eq. (\ref{width}) for the first finite bandgap. Then, in the lowest
approximation, the relation between $B_{1}$ and $A_{1}$ is [cf. Eq. (\ref{B}%
)]%
\begin{equation}
B_{1}=-\frac{\left( 5\varepsilon ^{2}/24\right) -\nu }{2\lambda _{2}}%
A_{1}.  \label{B2}
\end{equation}

Further, the jump condition at point $x=0$ keeps the form of Eq. (\ref{Delta}%
), and, in the lowest approximation, the jump of the first derivative is
dominated by term $B_{1}\sin \left( 2|x|\right) $ in ansatz (\ref{phi2}).
With regard to relation (\ref{B2}), this leads to the following prediction
for the solution supported by the single $\delta $-function in the second
finite bandgap:
\begin{equation}
A_{1}^{2}=-\frac{\sigma }{\lambda _{2}}\left( \frac{5}{24}\varepsilon
^{2}-\nu \right) .  \label{A^2-2}
\end{equation}%
As follows from Eqs. (\ref{A^2-2}), inside the second finite bandgap (\ref%
{SFG}) this solution exists for $\sigma =-1$, i.e., solely for the \emph{%
attractive nonlinearity}.

In the first approximation, the norm of the soliton is%
\begin{equation}
N_{2}\approx \frac{2}{\varepsilon ^{2}}\frac{\left( 5/24\right) -\tilde{\nu}%
}{\left( 5/576\right) +\left( \tilde{\nu}/6\right) -\tilde{\nu}^{2}}\left[ 1+%
\frac{\left( \left( 5/24\right) -\tilde{\nu}\right) ^{2}}{\left(
5/576\right) +\left( \tilde{\nu}/6\right) -\tilde{\nu}^{2}}\right] ,
\label{N2}
\end{equation}%
with $\tilde{\nu}$ defined as per Eq. (\ref{SFG}). It follows from the plot
of the norm versus $\mu $, which is shown in Fig.~\ref%
{N_vrs_Mu_approximation_SFB} for $\varepsilon =\pm 1$, that the
corresponding GS family satisfies the VK criterion, $dN/d\mu <0$.
Nevertheless, as well as in the case of the solitons in first finite
bandgap, which was considered above, the GSs in the second bandgap,
supported by the attractive nonlinearity, turn out to be unstable at all
values of $\varepsilon $.

In the general case, when the $\delta $-function is placed asymmetrically
with regard to the OL potential, GSs may exist in the second bandgap in the
case of the repulsive nonlinearity, and may be \emph{stable} in that case. A
detailed analysis of this case is presented in the next section.

\begin{figure}[tbp]
\subfigure[]{\includegraphics[width=2.2in]{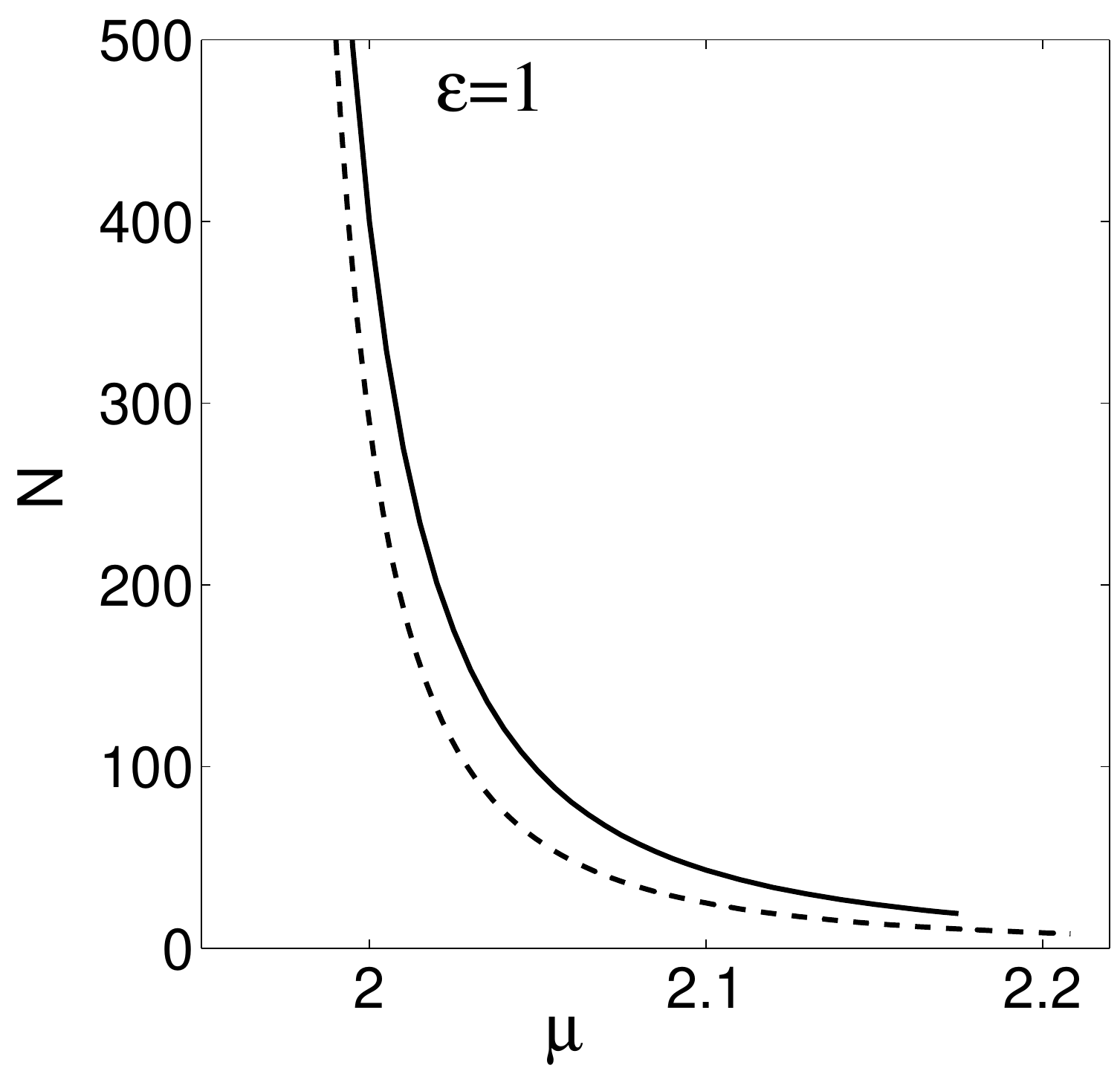}%
\label{NvsMu_approximation_SFG_eps1}} \subfigure[]{%
\includegraphics[width=2.2in]{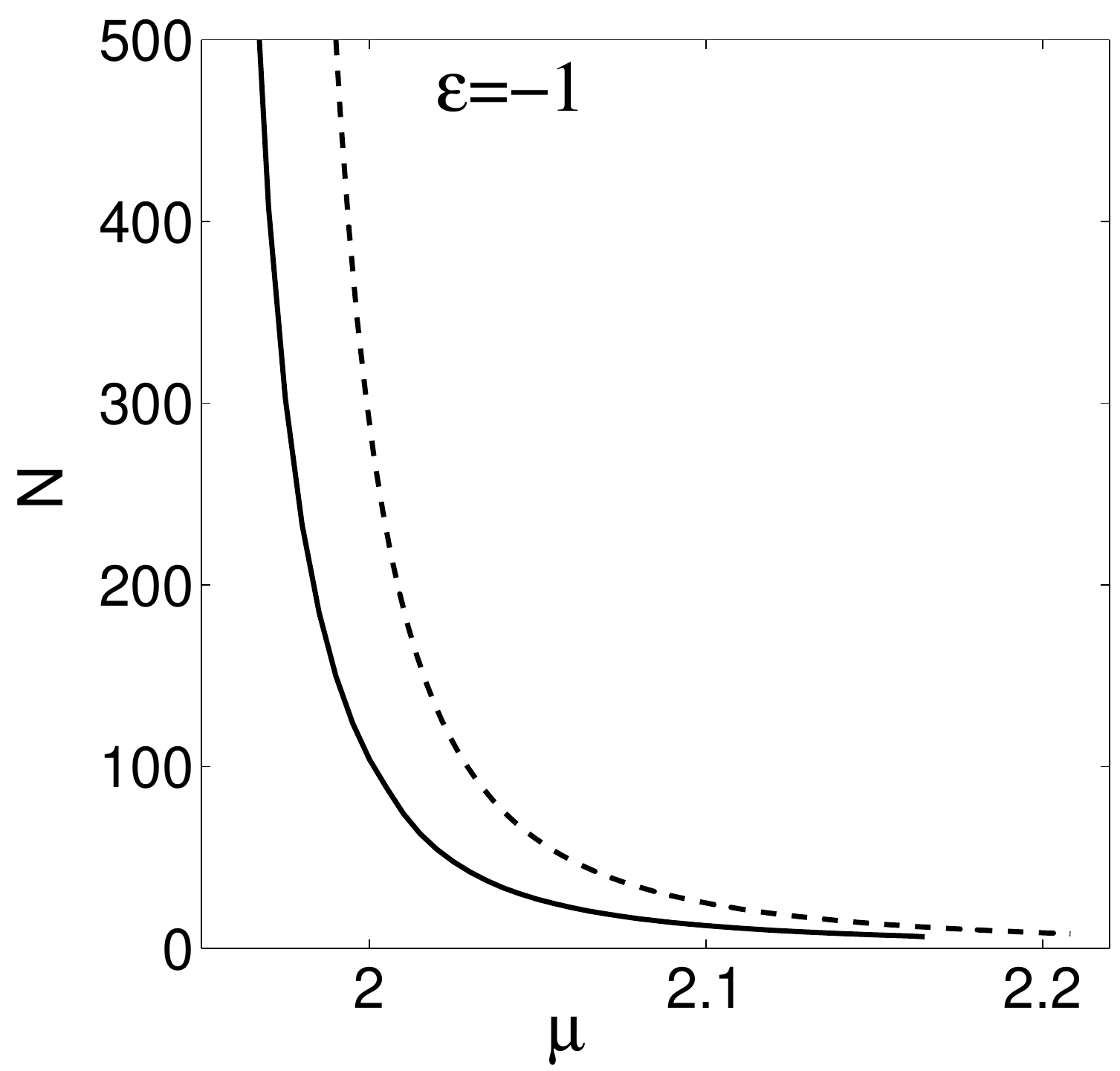}%
\label{NvsMu_approximation_SFG_epsm1}}
\caption{The norm of the analytically predicted (dashed curves) and
numerically found (solid curves) solitons in the second bandgap as a
function of $\protect\mu $, for $\protect\varepsilon =1$ (a) and $\protect%
\varepsilon =-1$ (b), in the case of the attractive nonlinearity.}
\label{N_vrs_Mu_approximation_SFB}
\end{figure}

Examples of numerically found profiles of the solitons in the second finite
bandgap, together with their analytically predicted counterparts [see Eqs.~(%
\ref{B2}) and (\ref{A^2-2})], are displayed in Fig.~\ref%
{Profiles_SFG_attractive}. The results shown in both Figs.~\ref%
{Profiles_SFG_attractive} and \ref{N_vrs_Mu_approximation_SFB} demonstrate
that the perturbative approximation is more accurate for smaller and
positive values of $\varepsilon $, when the attractive $\delta $-function is
set at a local minimum of the OL potential.

\begin{figure}[tbp]
\subfigure[]{\includegraphics[width=2.2in]{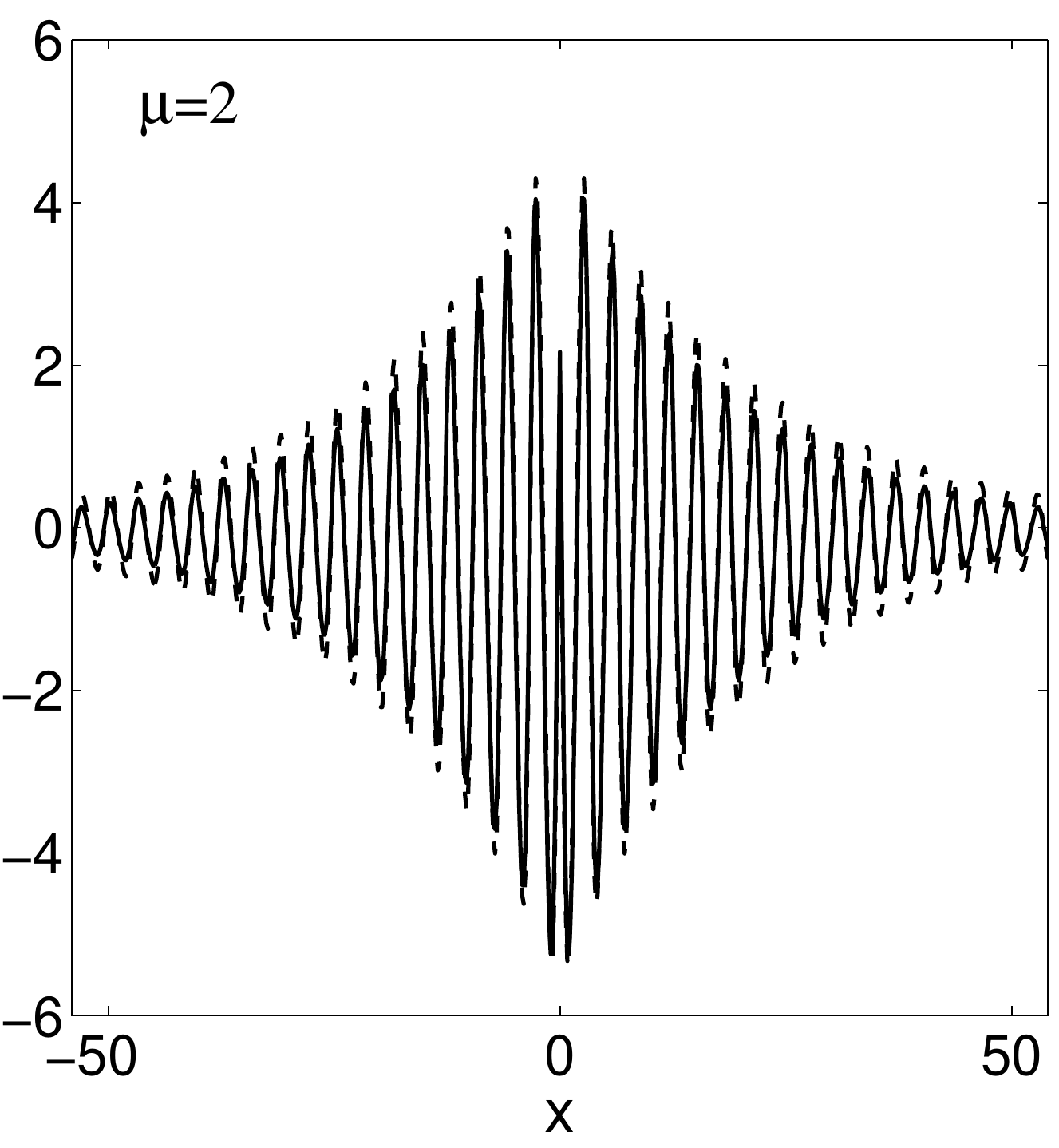}%
\label{Profile_eps1_mu2_SFG}} \subfigure[]{%
\includegraphics[width=2.2in]{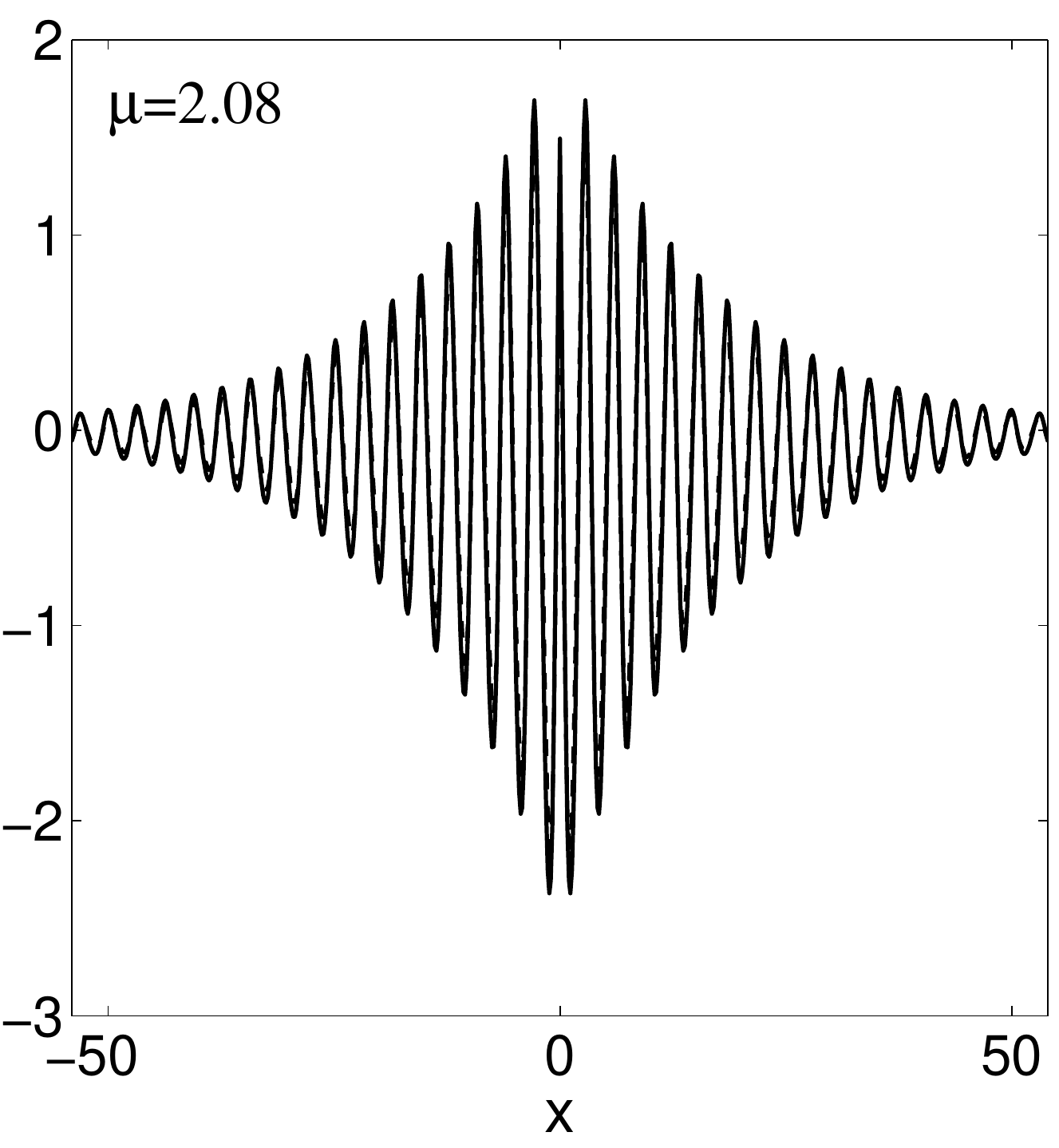}%
\label{Profile_eps1_mu208_SFG}} \subfigure[]{%
\includegraphics[width=2.3in]{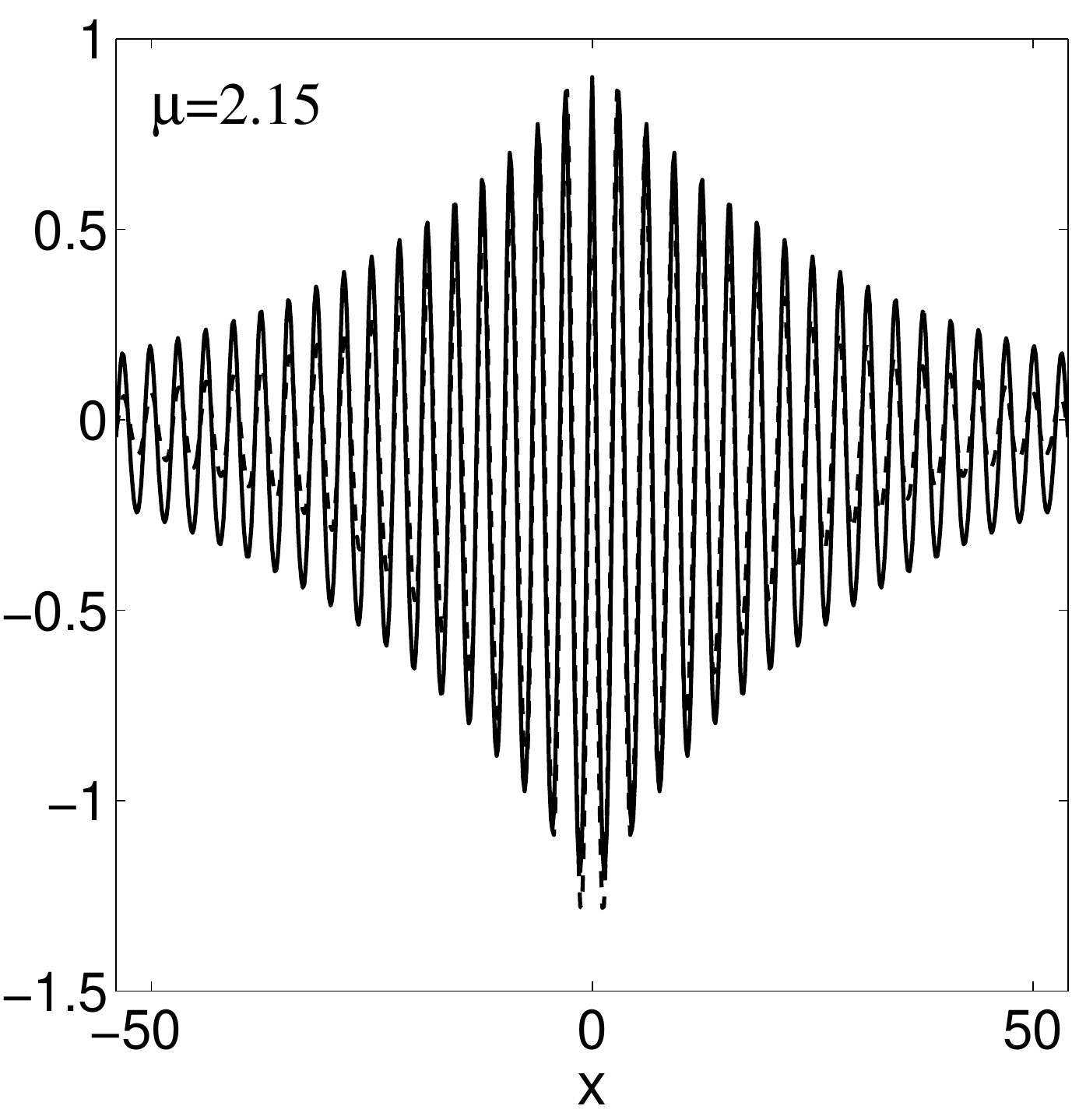}%
\label{Profile_eps1_mu215_SFG}} \\ 
\subfigure[]{\includegraphics[width=2.2in]{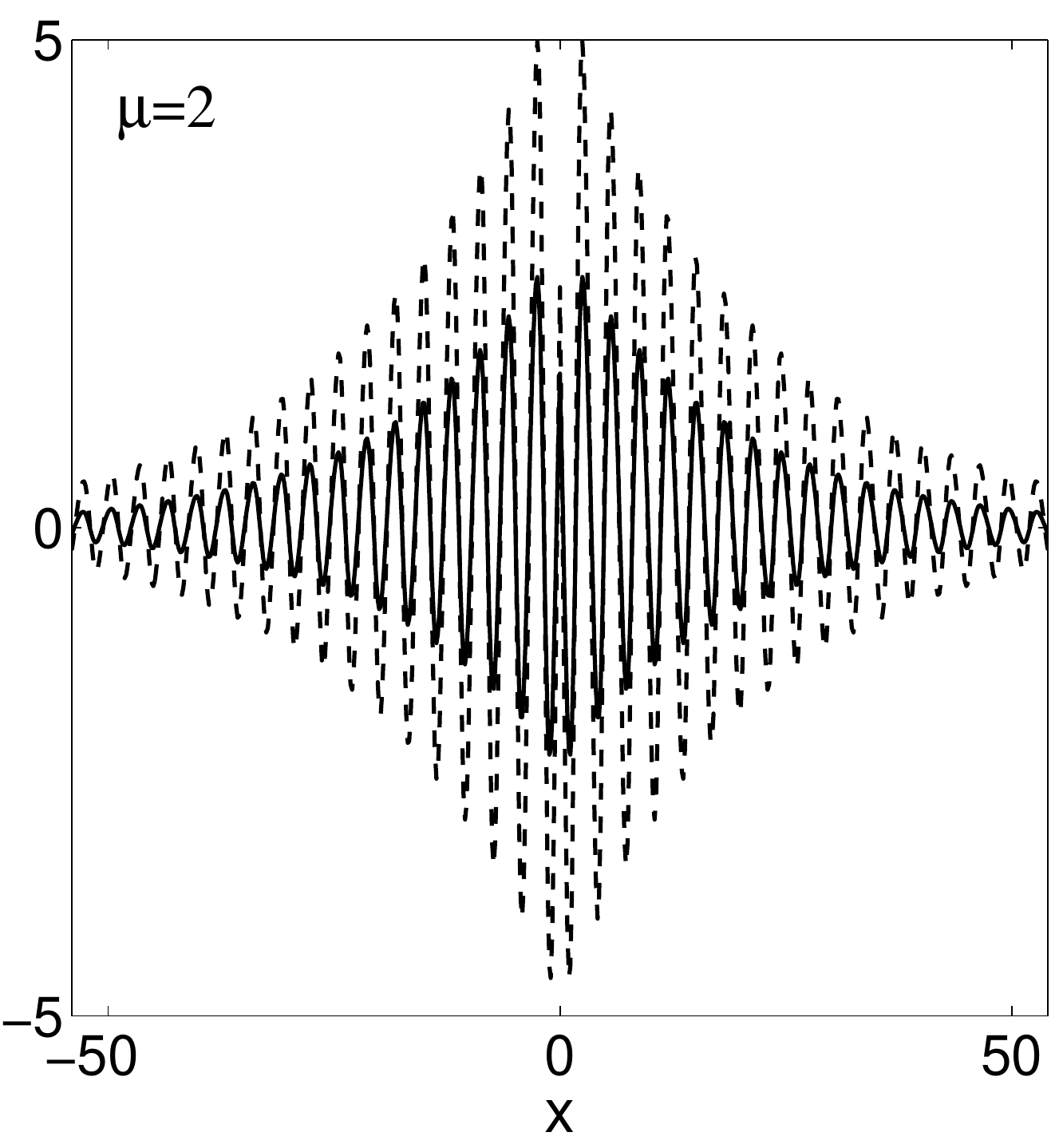}%
\label{Profile_eps1_mu2_SFG}} \subfigure[]{%
\includegraphics[width=2.3in]{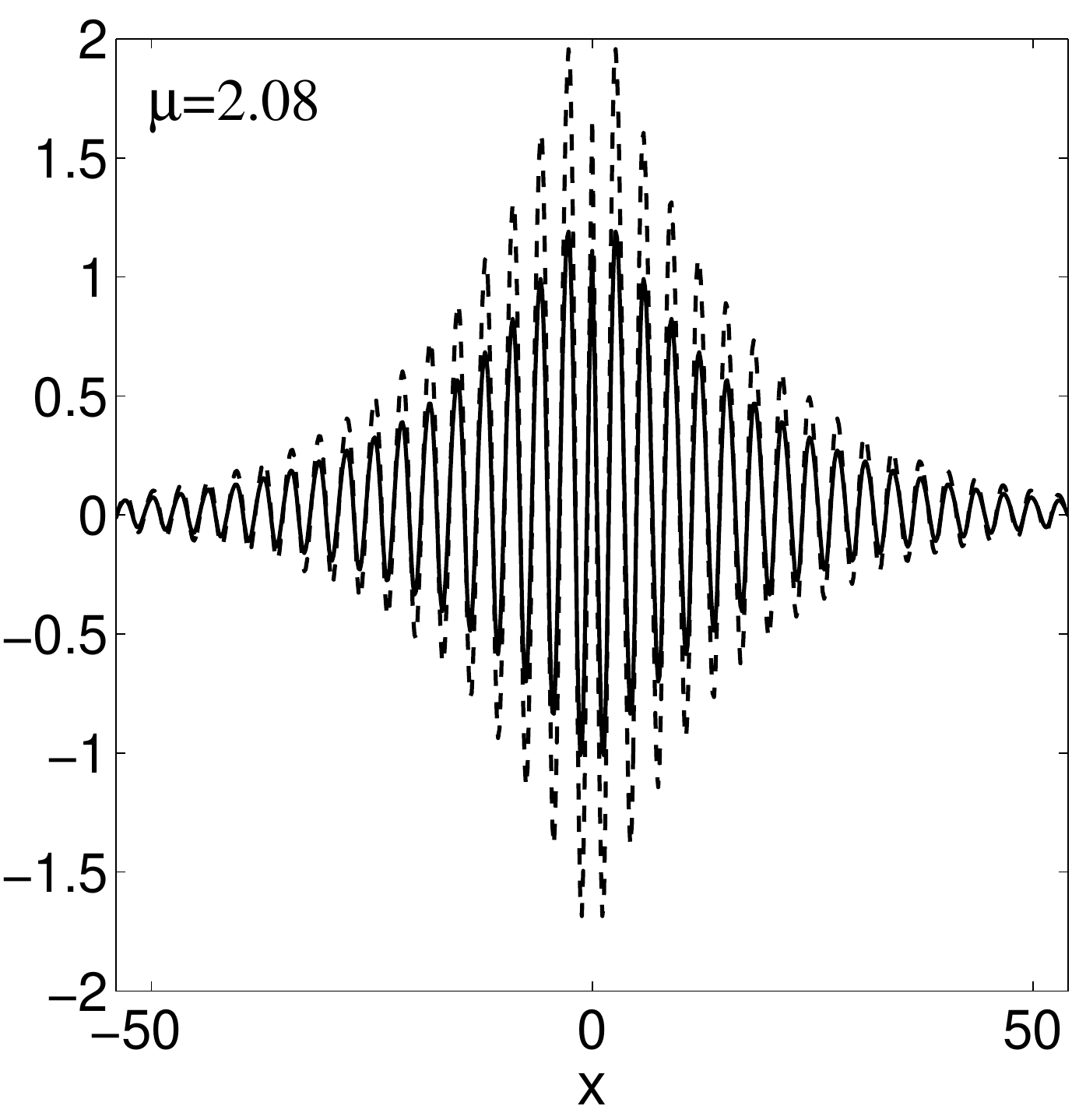}%
\label{Profile_eps1_mu208_SFG}} \subfigure[]{%
\includegraphics[width=2.3in]{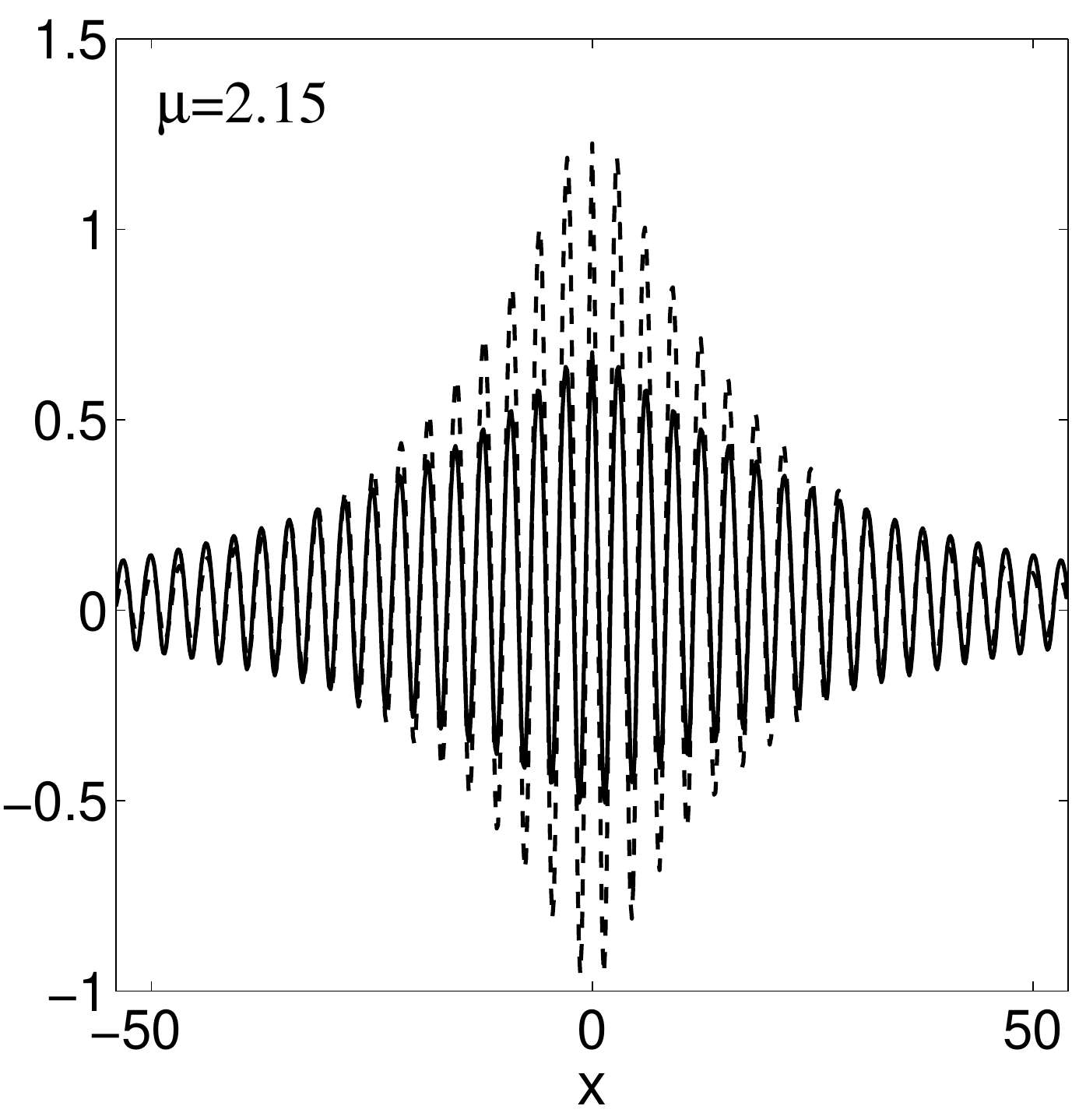}%
\label{Profile_eps1_mu215_SFG}}
\caption{Typical examples of broad profiles of the solitons in the second
finite bandgap, for $\protect\varepsilon =1$ (a)-(c) and $\protect%
\varepsilon =-1$ (d)-(f). The respective values of $\protect\mu $ are
indicated in each panel. Solid curves depict numerically found profiles,
whereas the dashed lines represent the analytical prediction given by Eqs.\ (%
\protect\ref{B2})-(\protect\ref{A^2-2}).}
\label{Profiles_SFG_attractive}
\end{figure}

\subsection{The semi-infinite gap}

For the sake of completeness of the analysis, we also present results of the
application of the perturbation theory (for small $|\varepsilon |$) to
solitons supported by the attractive $\delta $-function ($\sigma =-1$) in
the semi-infinite gap, i.e., at $\mu <0$. In the zeroth approximation ($%
\varepsilon =0$), the soliton is given by solution (\ref{simple}). Then, it
is easy to find the first-order correction to it in the following form:%
\begin{equation}
\phi _{1}(x)=\frac{\varepsilon }{1-2\mu }\phi _{0}(x)\left[ -\frac{5}{6}+%
\frac{1}{2}\cos \left( 2x\right) +\sqrt{-\frac{\mu }{2}}\sin \left(
2|x|\right) \right] ,  \label{phi1}
\end{equation}%
where $\phi _{0}(x)$ is expression (\ref{simple}) [to the first order in $%
\varepsilon $, solution $\phi _{0}(x)+\phi _{1}(x)$ satisfies both the
linear part of Eq. (\ref{Single_Dlata_NLSE_symmetric}) at $x\neq 0$, and the
jump condition (\ref{Delta}) at $x=0$]. To the same order, the norm of this
solution is%
\begin{equation}
N=1-\frac{\varepsilon \left( 5+2\mu \right) }{3\left( 1-2\mu \right)
^{2}},  \label{N1}
\end{equation}%
which demonstrates that the weak OL potential lifts the degeneracy of the
soliton family (\ref{simple}). Note that expression (\ref{N1}) satisfies the
VK stability criterion,
\begin{equation}
\frac{dN}{d\mu }=-\frac{2\varepsilon \left( 11+2\mu \right) }{3\left(
1-2\mu \right) ^{3}}<0,
\end{equation}%
for $\mu >-5.5$ in the case of $\varepsilon >0$, which corresponds to the
attractive $\delta $-function set at a local minimum of the OL potential,
see Eq. (\ref{Single_Dlata_NLSE_symmetric}); the change of the sign of $%
dN/d\mu $ at $\mu <-5.5$ does not really matter, as dependence (\ref{N1}) is
virtually flat in that region.

\begin{figure}[tbp]
\subfigure[]{%
\includegraphics[width=2.3in]{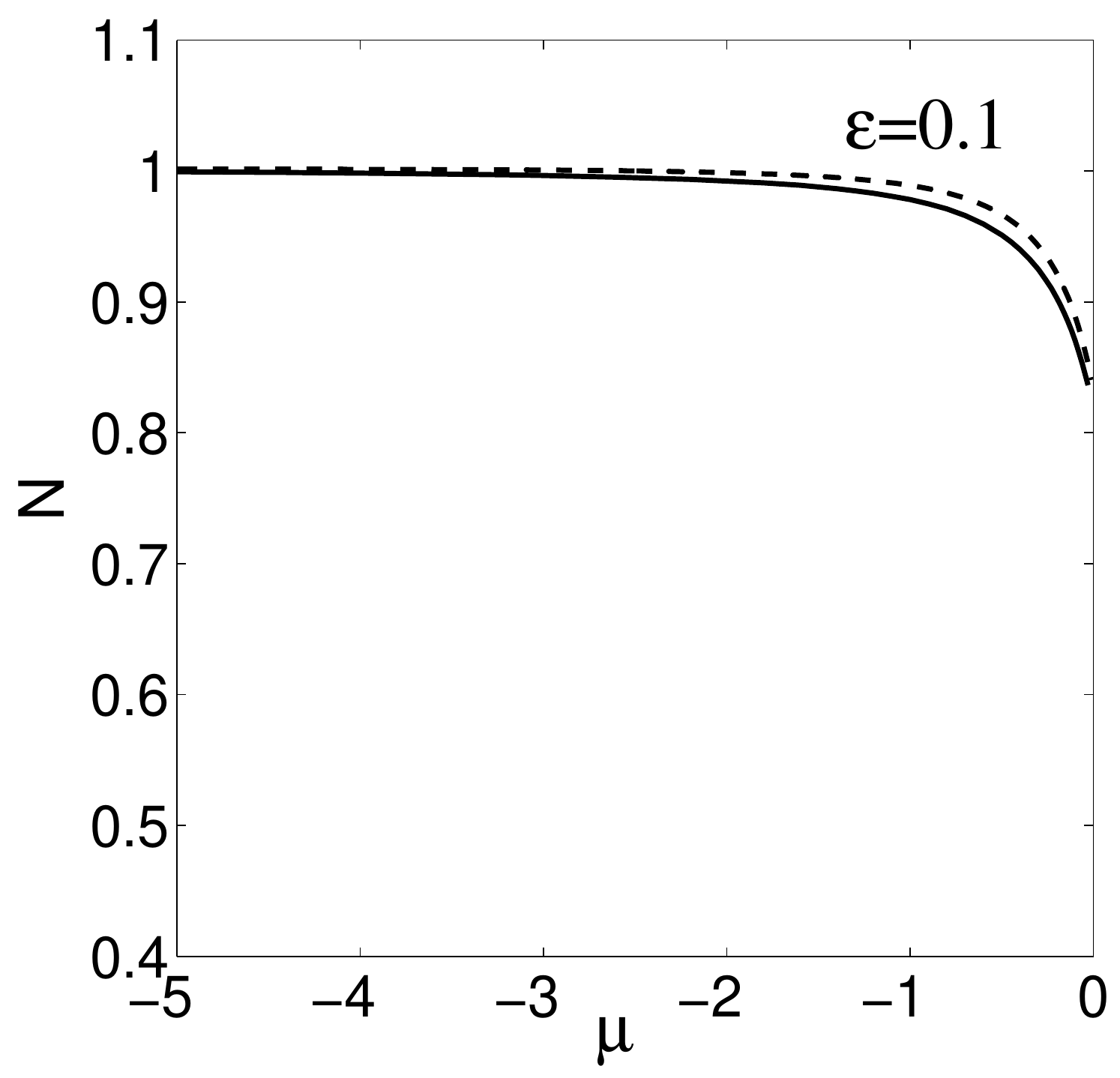}%
\label{NvsMu_approximation_eps01_SIG}} \quad \subfigure[]{%
\includegraphics[width=2.1in]{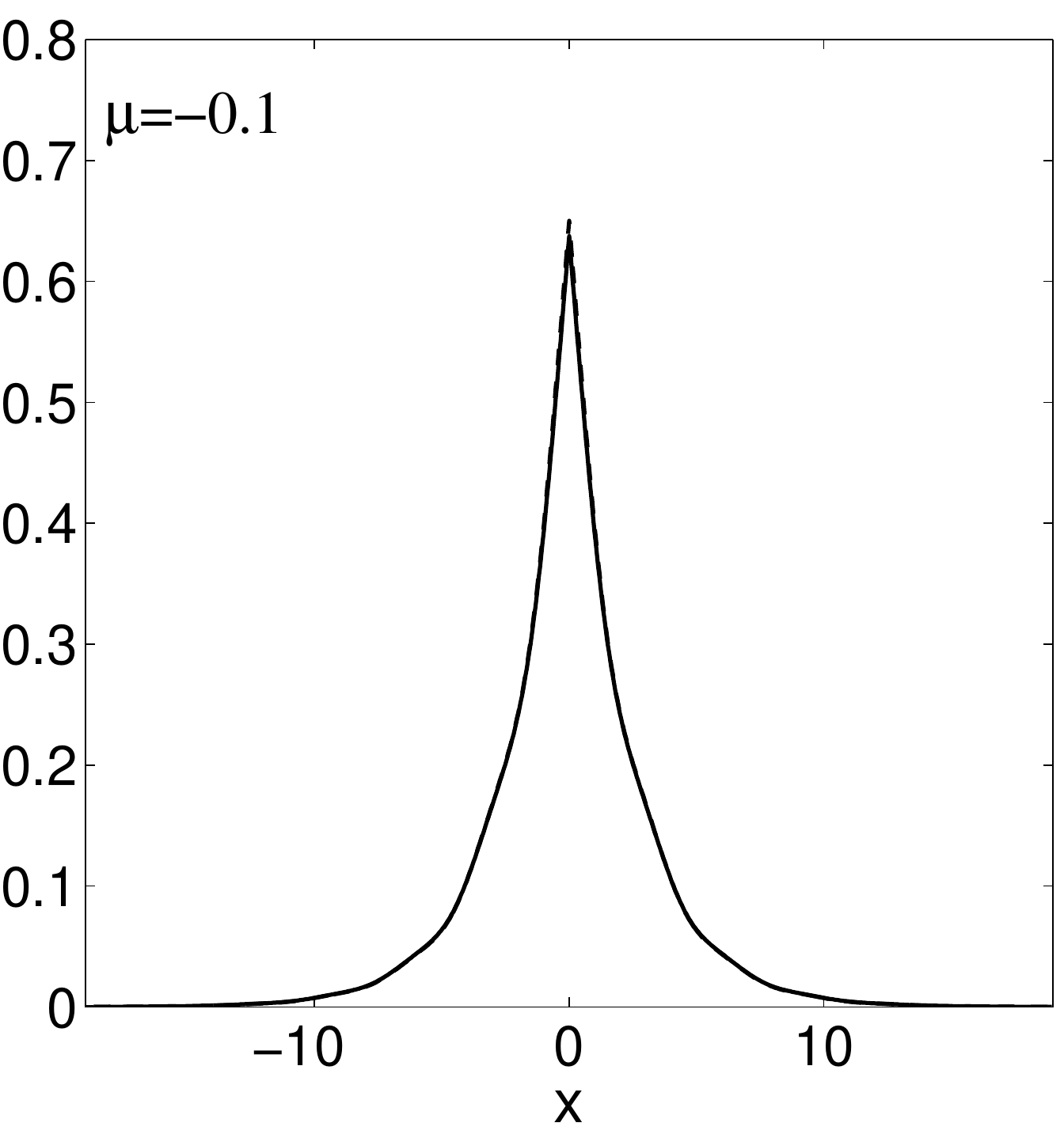}%
\label{Profile_eps01_mum01}} \quad \subfigure[]{%
\includegraphics[width=2.1in]{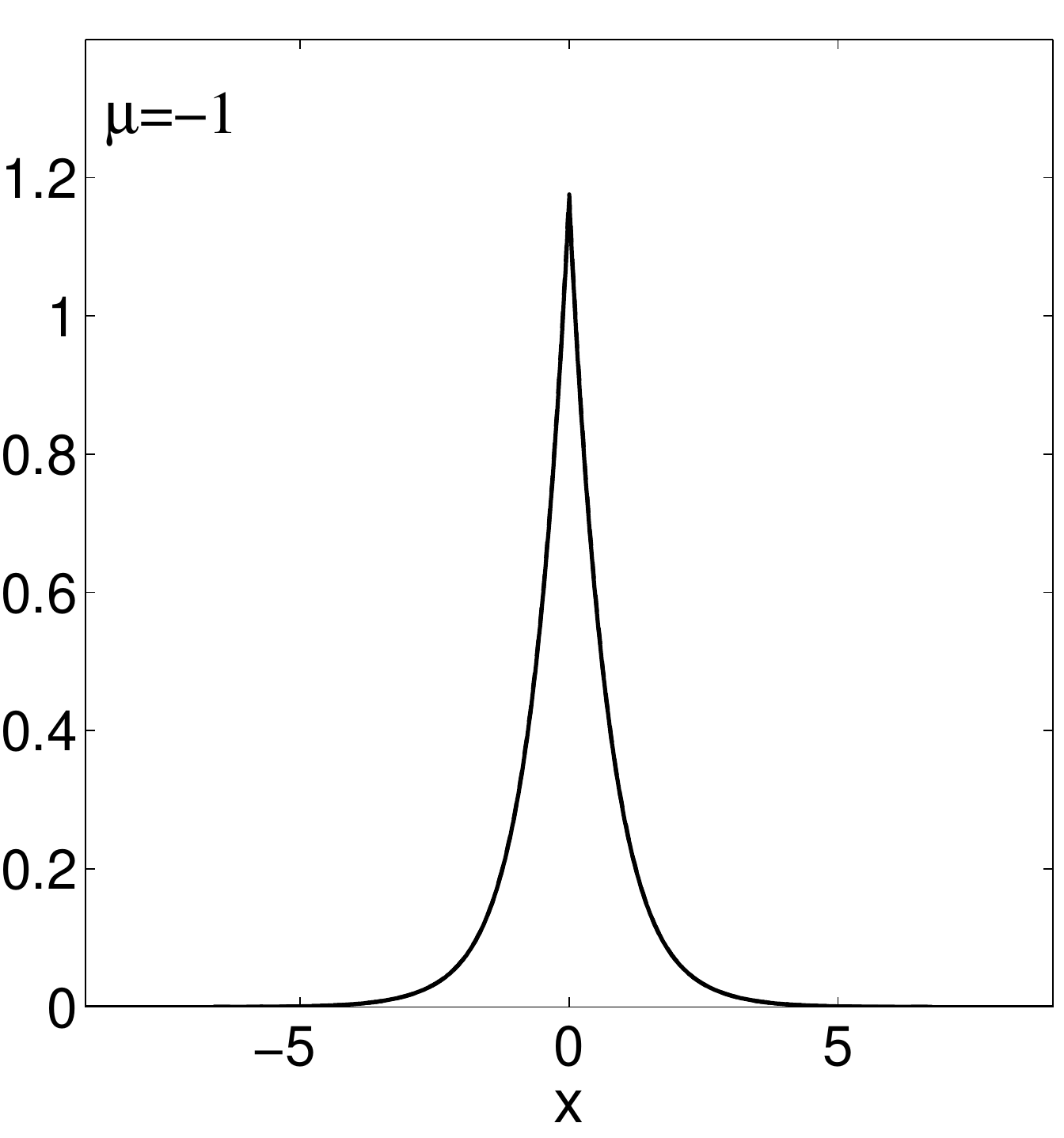}%
\label{Profile_eps01_mum1}} \\ 
\subfigure[]{%
\includegraphics[width=2.3in]{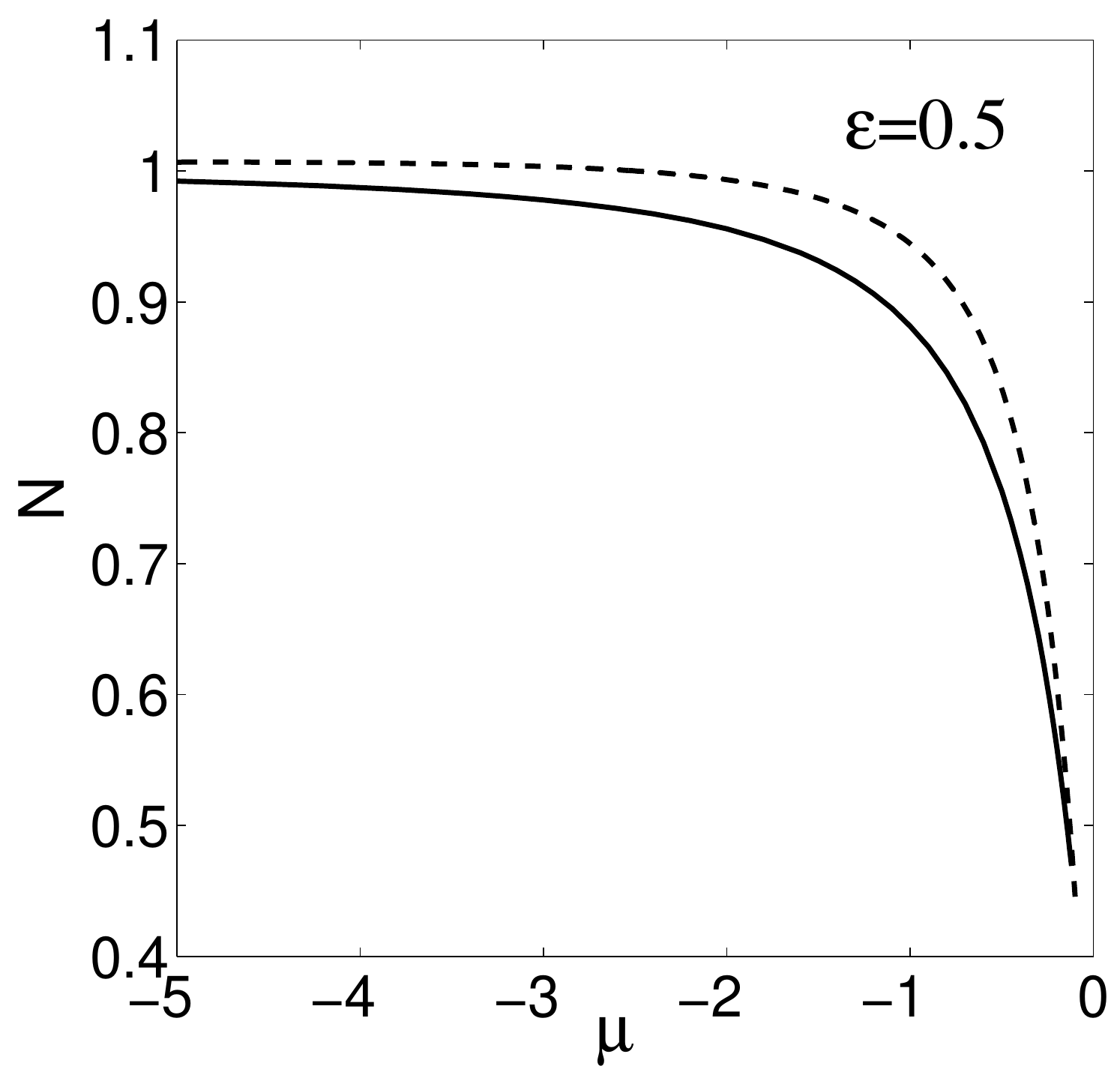}%
\label{NvsMu_approximation_eps05_SIG}} \quad \subfigure[]{%
\includegraphics[width=2.1in]{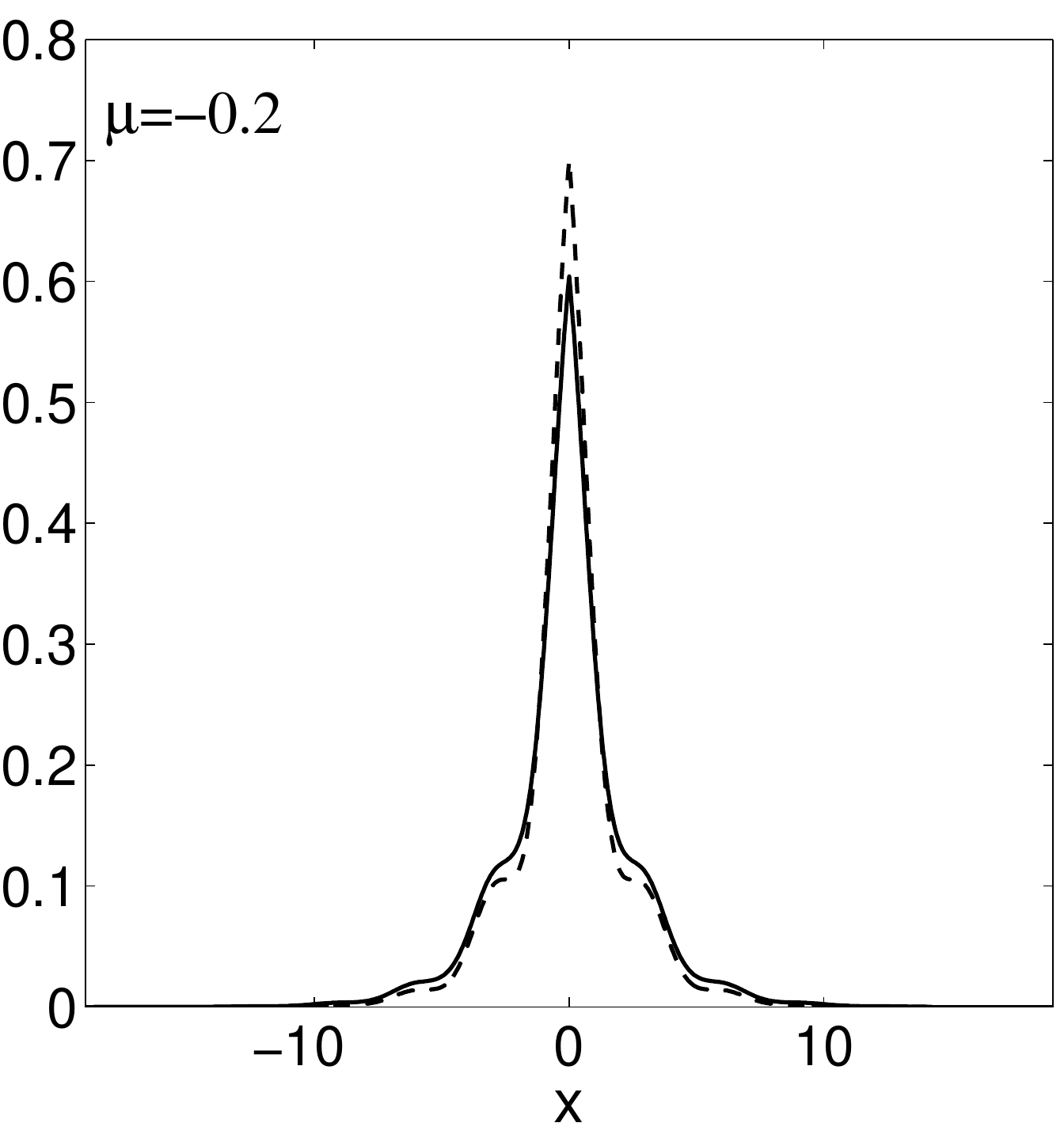}%
\label{Profile_eps05_mum02}} \quad
\subfigure[]{\includegraphics[width=2.1in]{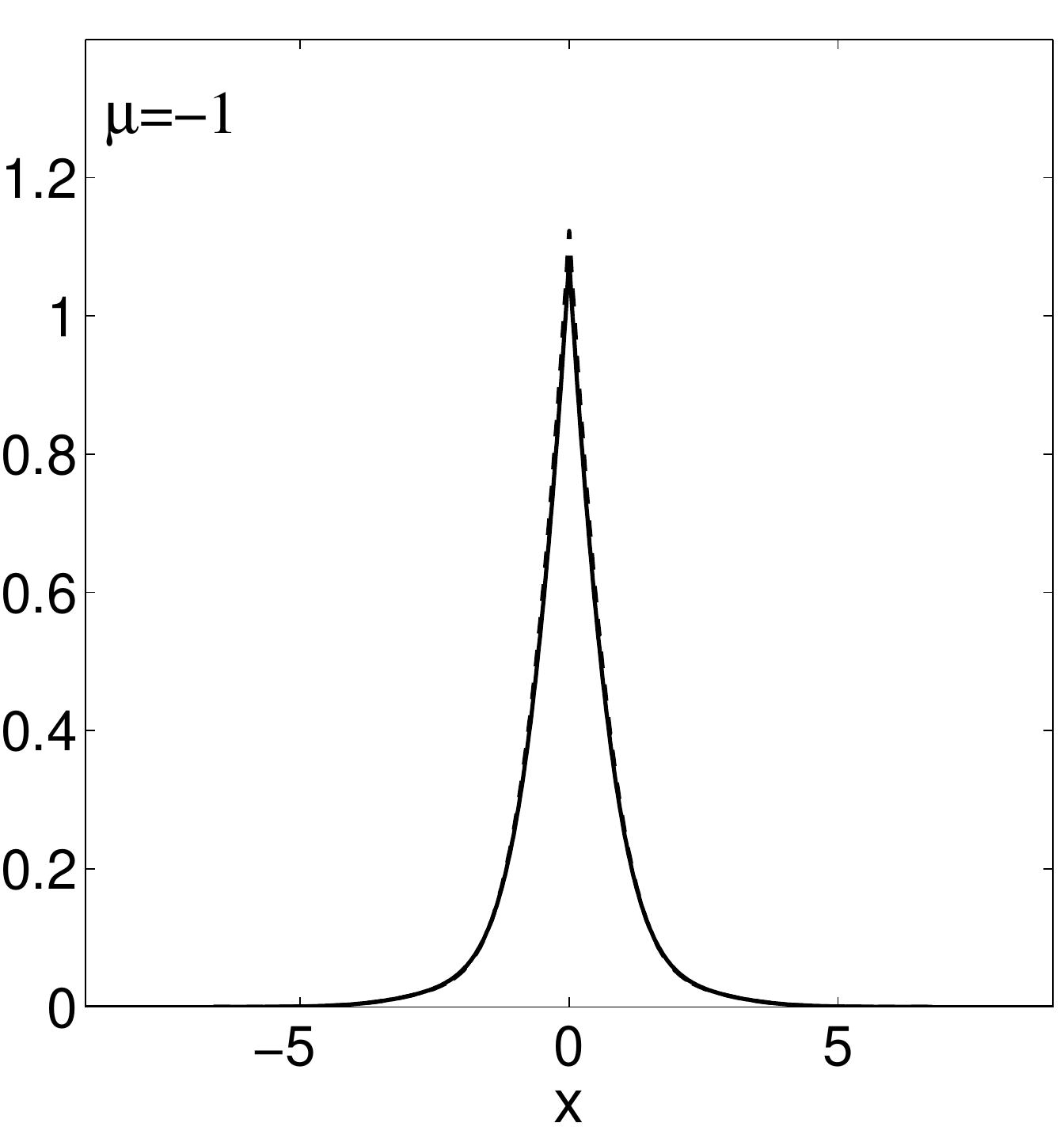}
\label{Profile_eps05_mum1}}
\caption{Panels (a) and (d) display the comparison between the numerical and
analytically estimated norms (the solid and dashed curves, respectively) for
the solitons in the semi-infinite gap, for $\protect\varepsilon =0.1$ and $%
0.5$, respectively. Examples of the soliton profiles, taken close to the
edge of the gap and deeper inside, are displayed in panels (b),(c) and
(e),(f), for the cases corresponding to (a) and (d), respectively. In (b)
and (c), the analytically predicted profiles are very close to their
numerical counterparts, making the dashed and solid lines virtually
indistinguishable.}
\label{Approximation_SIG}
\end{figure}

Numerical results demonstrate that the entire soliton family is stable in
the case of $\varepsilon >0$, obeying the VK criterion everywhere in the
semi-infinite gap (for a further discussion, see the next section). Figure %
\ref{Approximation_SIG} presents the respective comparison between the
analytical approximation and the numerical results. As expected, the
prediction is more accurate for smaller $\varepsilon $, closely
approximating even complex soliton profiles near the edge of the gap, see
panel (e) in Fig. \ref{Approximation_SIG}. On the other hand, the solitons
centered at the local maximum of the OL potential at $\varepsilon <0$ are
definitely unstable (not shown here in detail).

\section{Numerical results for the model with the single $\protect\delta $%
-function}

\label{sec:SingleDelta}

In the case when the local nonlinearity is represented by the single $\delta
$-function in Eq. (\ref{Single_Dlata_NLSE}), we have found soliton modes in
the semi-infinite and two lowest finite gaps, and their stability was
investigated by means of numerical methods. The stationary solutions were
constructed applying the Newton-Raphson method to the respective nonlinear
boundary-value problem. The stability was then examined by considering a
perturbed solution to Eq.~(\ref{Single_Dlata_NLSE}), in the form of
\begin{equation}
\phi (x,t)=\phi _{s}(x)+ge^{-i\lambda t}+f^{\ast }e^{i\lambda ^{\ast }t},
\label{Perturbed_solution}
\end{equation}%
where $\phi _{s}(x)$ is the stationary solution, functions $g$ and $f$ are
eigenmodes of the infinitesimal perturbation, and $\lambda $ is the
corresponding eigenvalue, which may be complex in the general case. When
substituting expression (\ref{Perturbed_solution}) into Eq.~(\ref%
{Single_Dlata_NLSE}) and linearizing, one arrives at the following
eigenvalue problem:
\begin{equation}
\left(
\begin{array}{cc}
\hat{L} & \sigma \delta (x-\xi )\left( \phi _{s}(x)\right) ^{2} \\
-\sigma \delta (x-\xi )\left( \phi _{s}(x)\right) ^{2} & -\hat{L}%
\end{array}%
\right) \left(
\begin{array}{c}
g \\
f%
\end{array}%
\right) =\lambda \left(
\begin{array}{c}
g \\
f%
\end{array}%
\right) ,  \label{Eigenvalue_problem}
\end{equation}%
with $\hat{L}\equiv -\mu -(1/2)d^{2}/dx^{2}-\varepsilon \cos (2x)+2\sigma
\delta (x-\xi )\left( \phi _{s}(x)\right) ^{2}$. This problem can be easily
solved using a simple finite-difference scheme, the solution being stable if
all the eigenvalues are real.

The so predicted stability or instability was verified by means of direct
simulations of the evolution of initially perturbed modes. For this purpose,
the standard pseudospectral split-step method and the Crank-Nicolson
finite-difference algorithm were used. In the course of the analysis, the $%
\delta $-function was set at different positions within half a period of the
OL, $0<\xi <\pi /2$. In the numerical calculations based on the
discretization with stepsize $\Delta x$, the discrete counterpart of the $%
\delta $-function was defined so that it took a nonzero value, $\tilde{\delta%
}=1/(2\Delta x)$, at the single point, $x=\xi $.

\subsection{Solitons in the semi-infinite gap}

\begin{figure}[tbp]
{\includegraphics[width=3.0in]{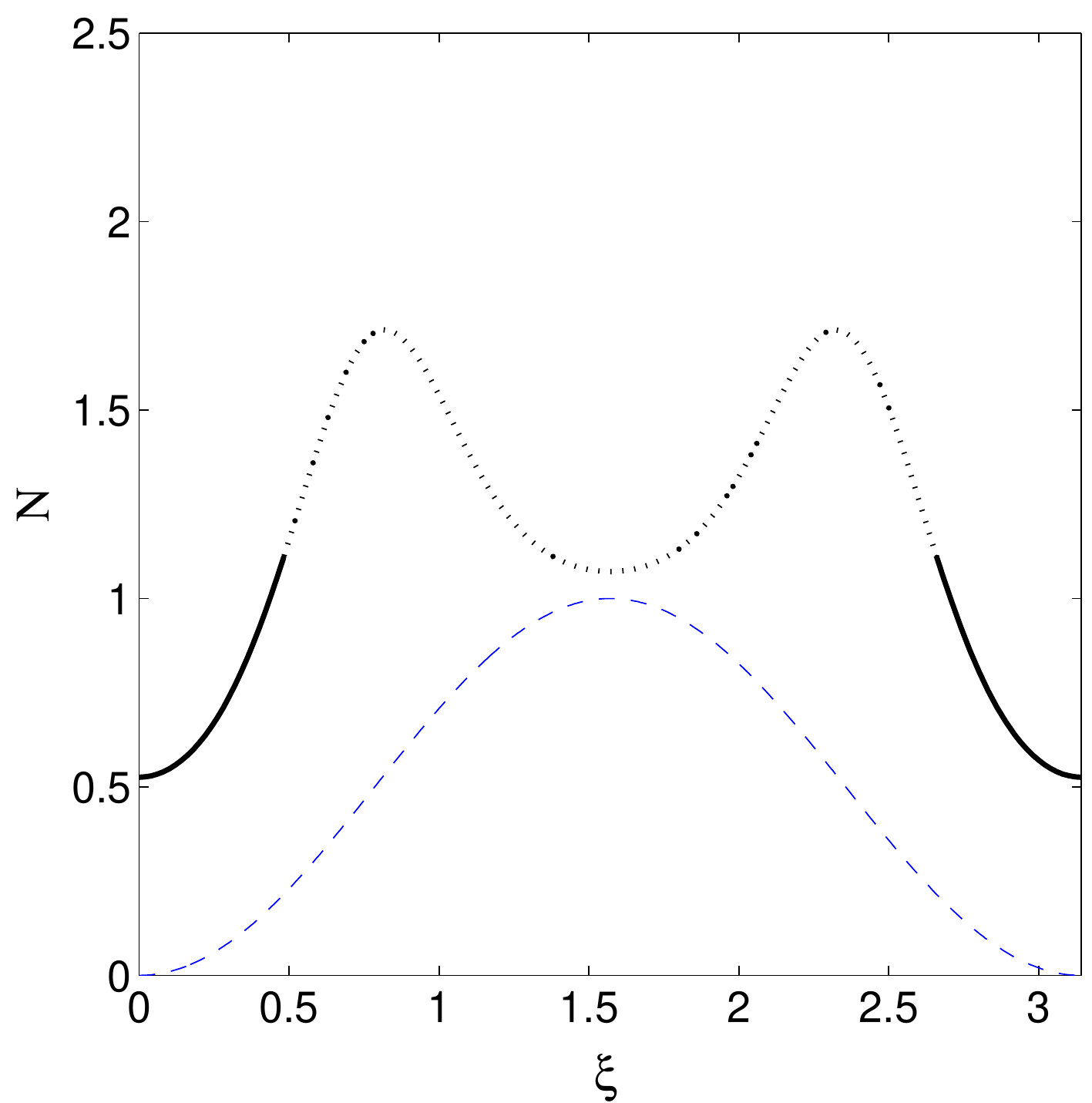}}
\caption{(Color online) The norm versus the position of the attractive $%
\protect\delta $-function, for the soliton family in the semi-infinite gap,
at $\protect\varepsilon =5$ and $\protect\mu =-4$. Here and in similar
figures shown below, the dashed line depicts the underlying periodic
potential (rescaled and shifted upward for the clarity of the picture),
while stable and unstable solitons correspond to continuous and dotted
lines, respectively.}
\label{NvsDeltax_SIG}
\end{figure}

Numerical analysis of the solutions in the semi-infinite gap reveals a
single soliton family, which exists only in the case of the attractive
nonlinearity ($\sigma =-1$). A natural result of the stability analysis is
that these solitons are stable if the attractive $\delta $-function is
located at or near the minimum of the potential. An example for $\varepsilon
=5$ and $\mu =-4$ is displayed in Fig.~\ref{NvsDeltax_SIG}, where shift $\xi
$ of the $\delta $-function from the potential minimum, $x=0$, takes values
within a half of the spatial period of the potential (recall the period is $%
\pi $). In the semi-infinite gap, the soliton's stability complies with the
VK criterion. In particular, the soliton family is completely stable for $%
\xi =0 $. On the other hand, the family is (quite naturally) completely
unstable for $x=\pi /2$, when the $\delta $-function is set at the point of
the potential maximum.

\subsection{Solitons in the first finite bandgap}

\begin{figure}[tbp]
\subfigure[]{\includegraphics[width=2.7in]{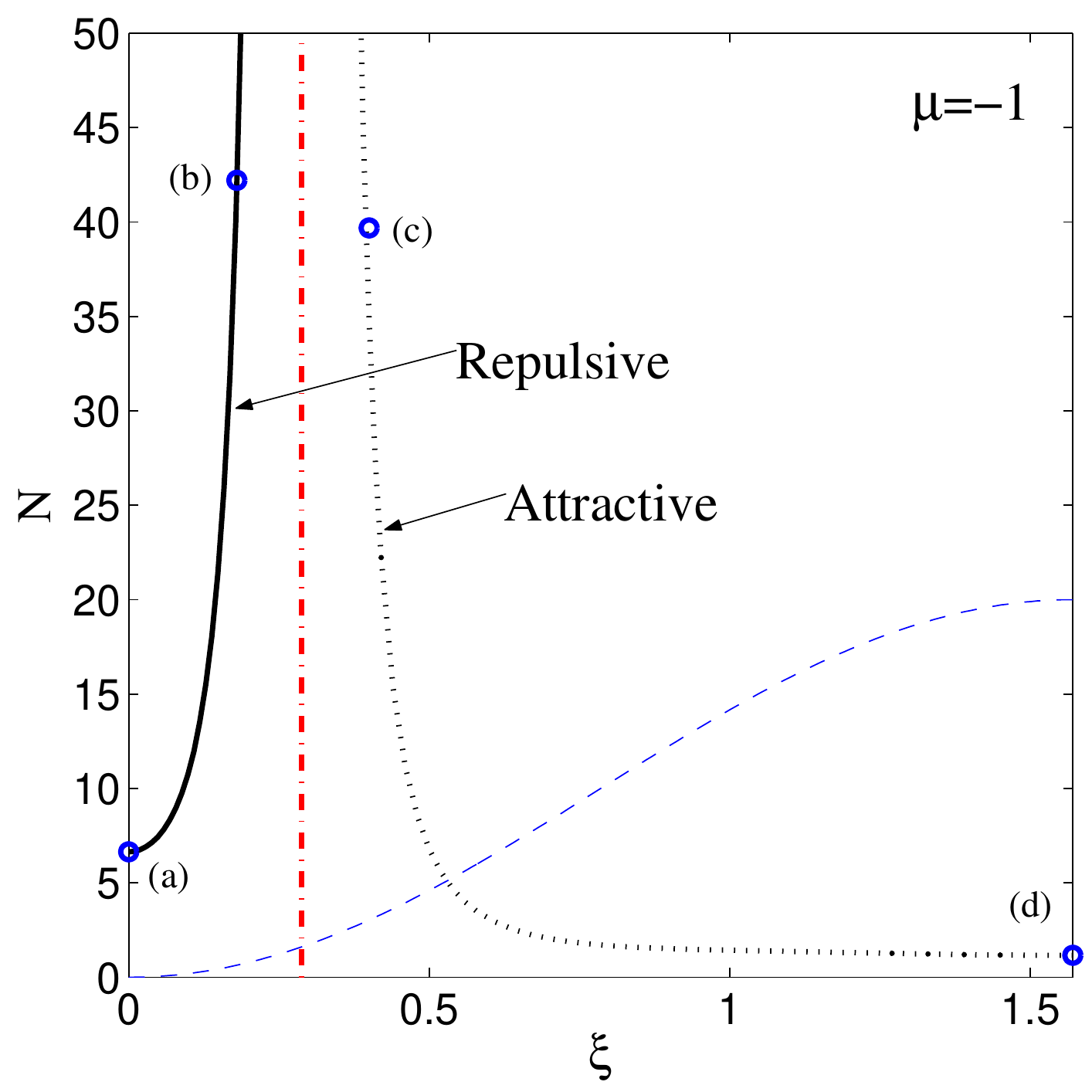}%
\label{NvsDeltax_FFG_Mum1}} \quad \subfigure[]{%
\includegraphics[width=2.7in]{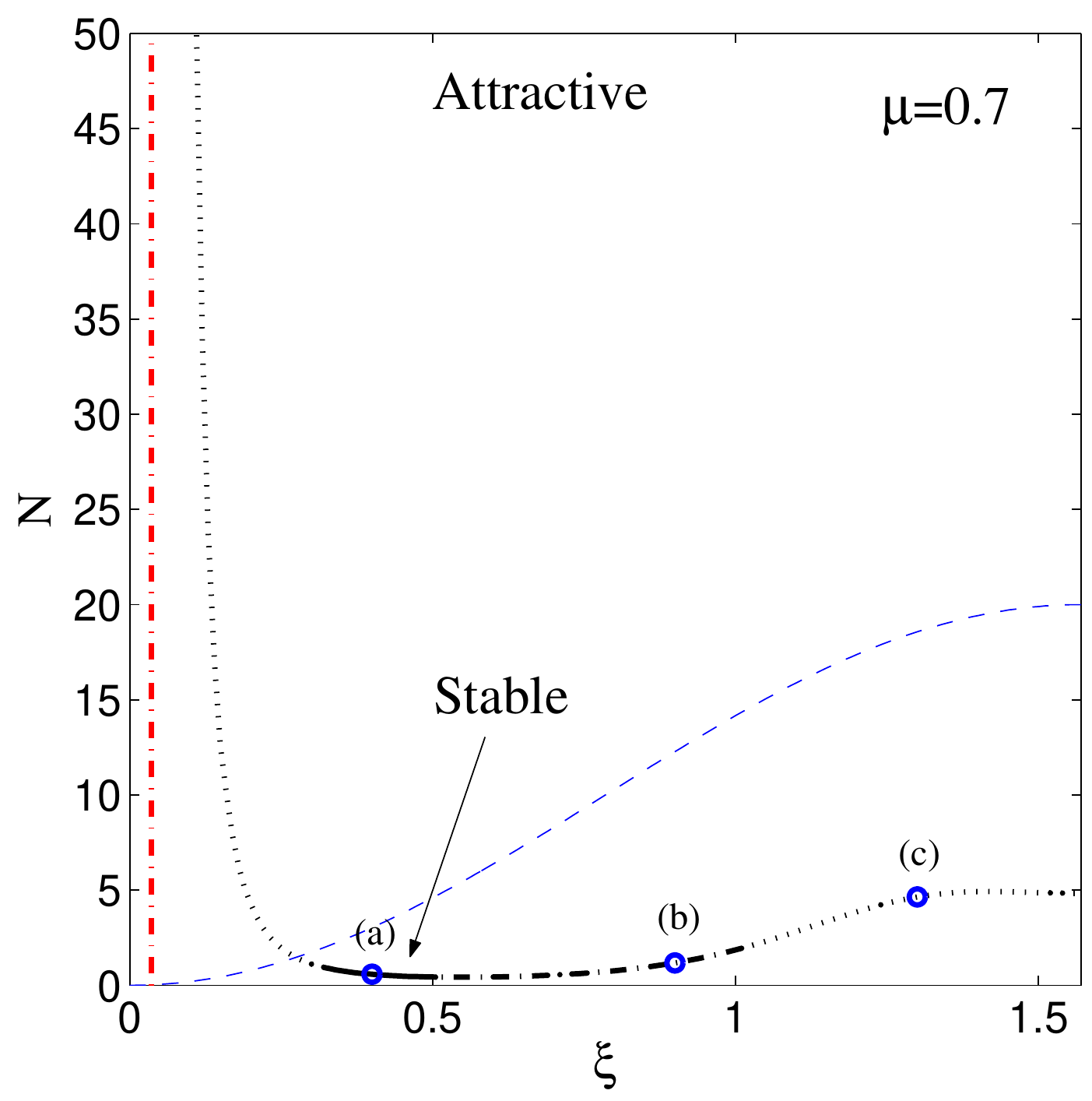}%
\label{NvsDeltax_FFG_Mup07}}
\caption{(Color online) The norm versus the position of the
nonlinearity-modulating $\protect\delta $-function, for solitons in the
first finite bandgap, with $\protect\varepsilon =5$ and $\protect\mu =-1$
(a) or $\protect\mu =0.7$ (b), and both signs of the nonlinearity. The
dashed-dotted vertical lines represent the respective thresholds, $\protect%
\xi _{\mathrm{thr}}=0.287$ (a) and $\protect\xi _{\mathrm{thr}}=0.035$ (b).
As before, continuous lines refer to stable solitons, while dotted and
dashed-dotted ones correspond to the strong localized and weak oscillating
instabilities, respectively. Shapes of the solitons corresponding to marked
points in (a) are displayed in Fig.~\protect\ref{Profile_1Delta_mum1}. The
evolution of the solitons marked in (b) is displayed in Fig.~\protect\ref%
{Evolution_mup07_1Delta_FFG}. In panel (b), only the soliton branch
corresponding to the attractive nonlinearity appears, as the region of $%
\protect\xi $ for the repulsive sign is too small, while the corresponding
solitons have a very large norm.}
\label{NvsDeltax_FFG}
\end{figure}

\begin{figure}[tbp]
\subfigure[]{\includegraphics[width=1.7in]{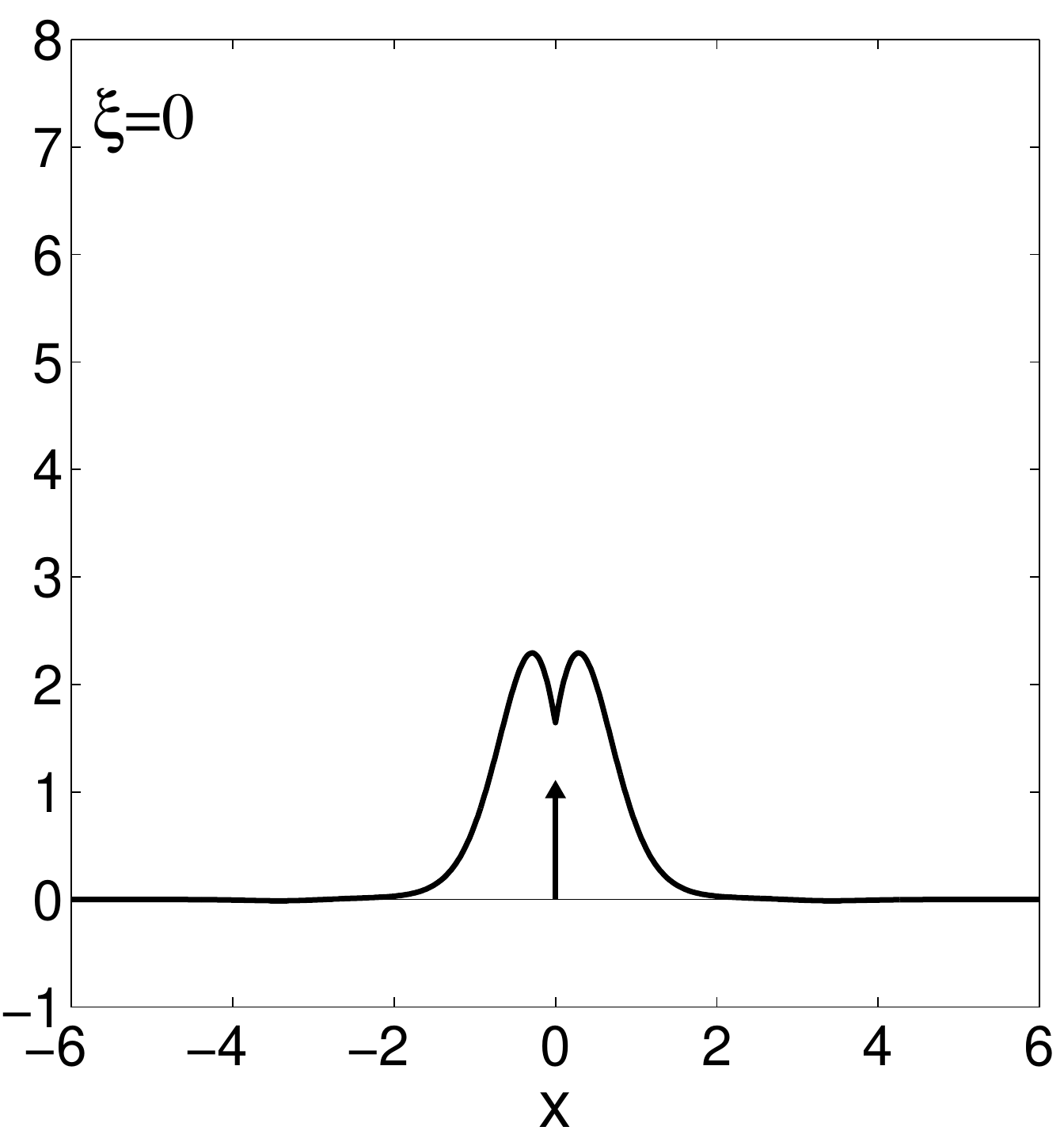}%
\label{Profile_1Delta_deltax0_mum1}} \subfigure[]{%
\includegraphics[width=1.7in]{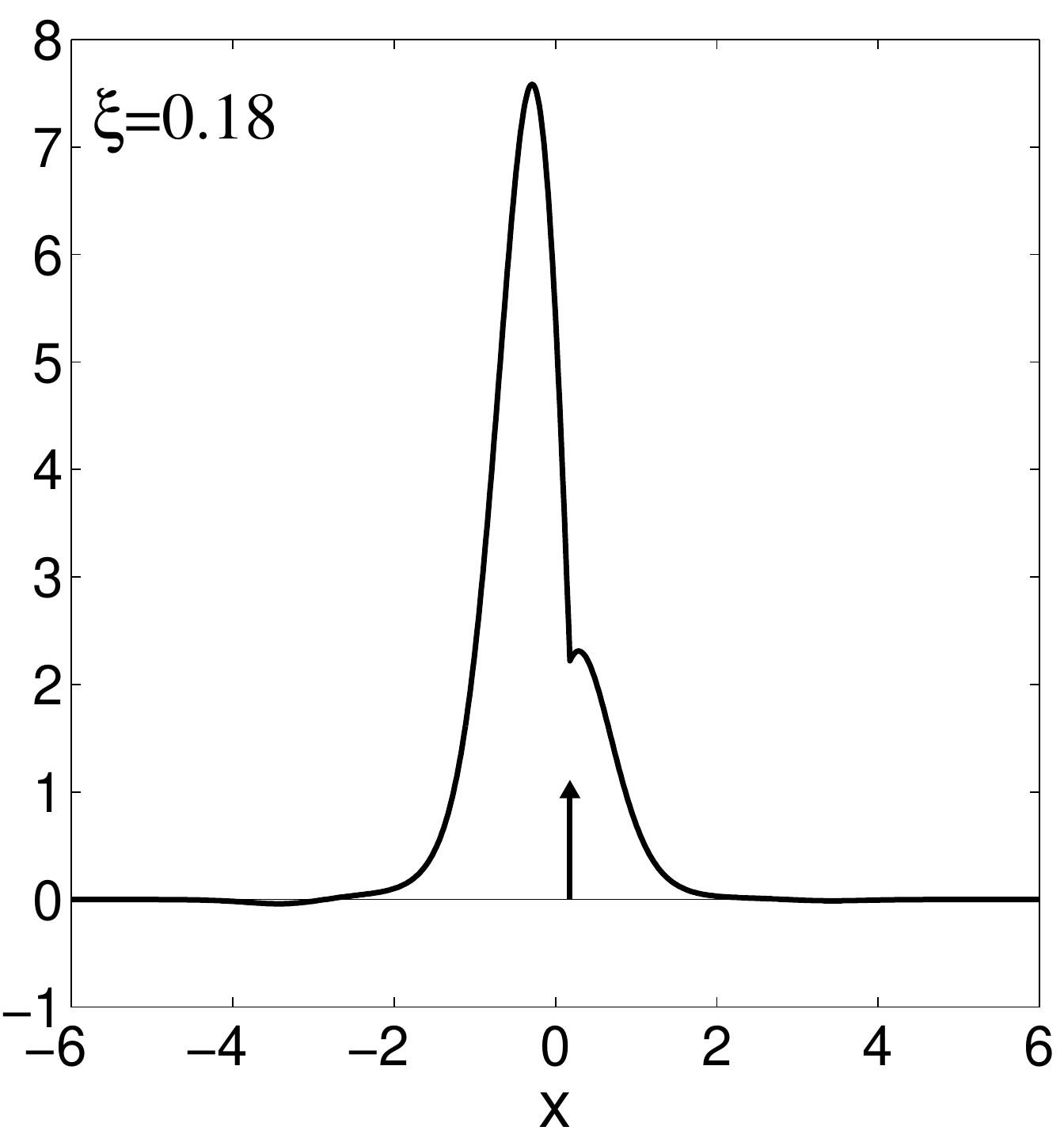}%
\label{Profile_1Delta_deltax018_mum1}} \subfigure[]{%
\includegraphics[width=1.7in]{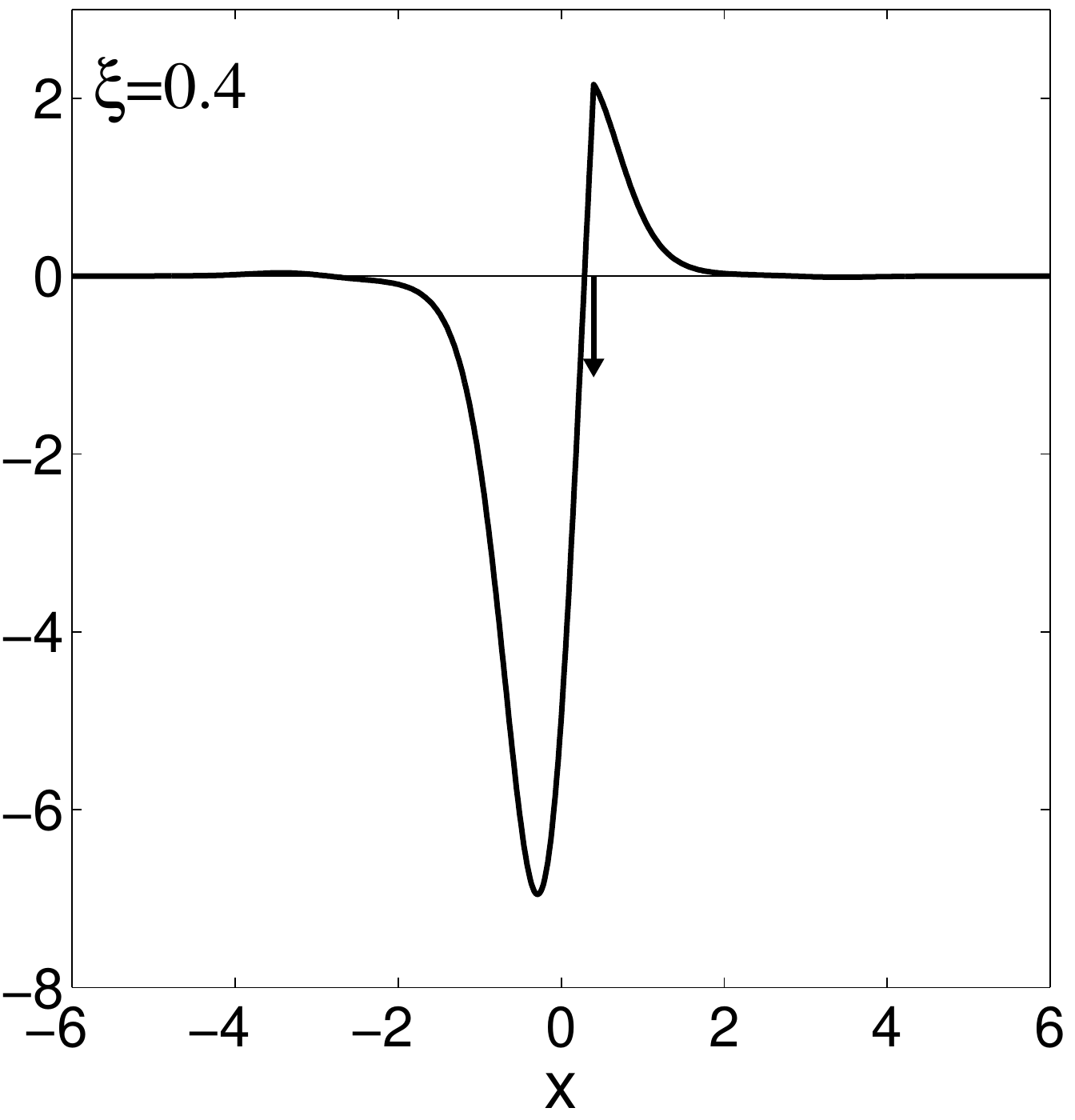}%
\label{Profile_1Delta_deltax04_mum1}} \subfigure[]{%
\includegraphics[width=1.7in]{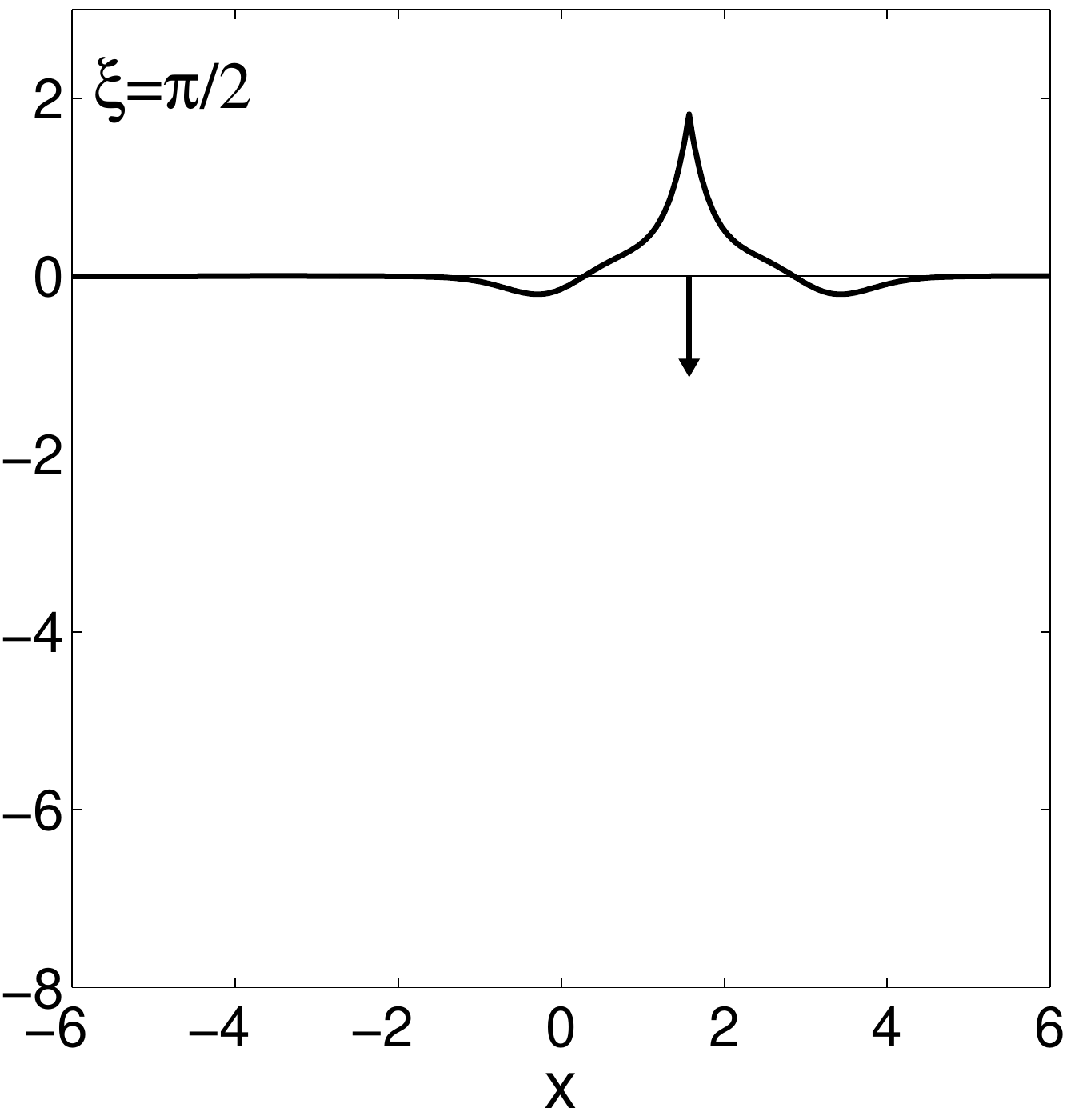}%
\label{Profile_1Delta_deltax157_mum1}} \\ 
\subfigure[]{%
\includegraphics[width=2.2in]{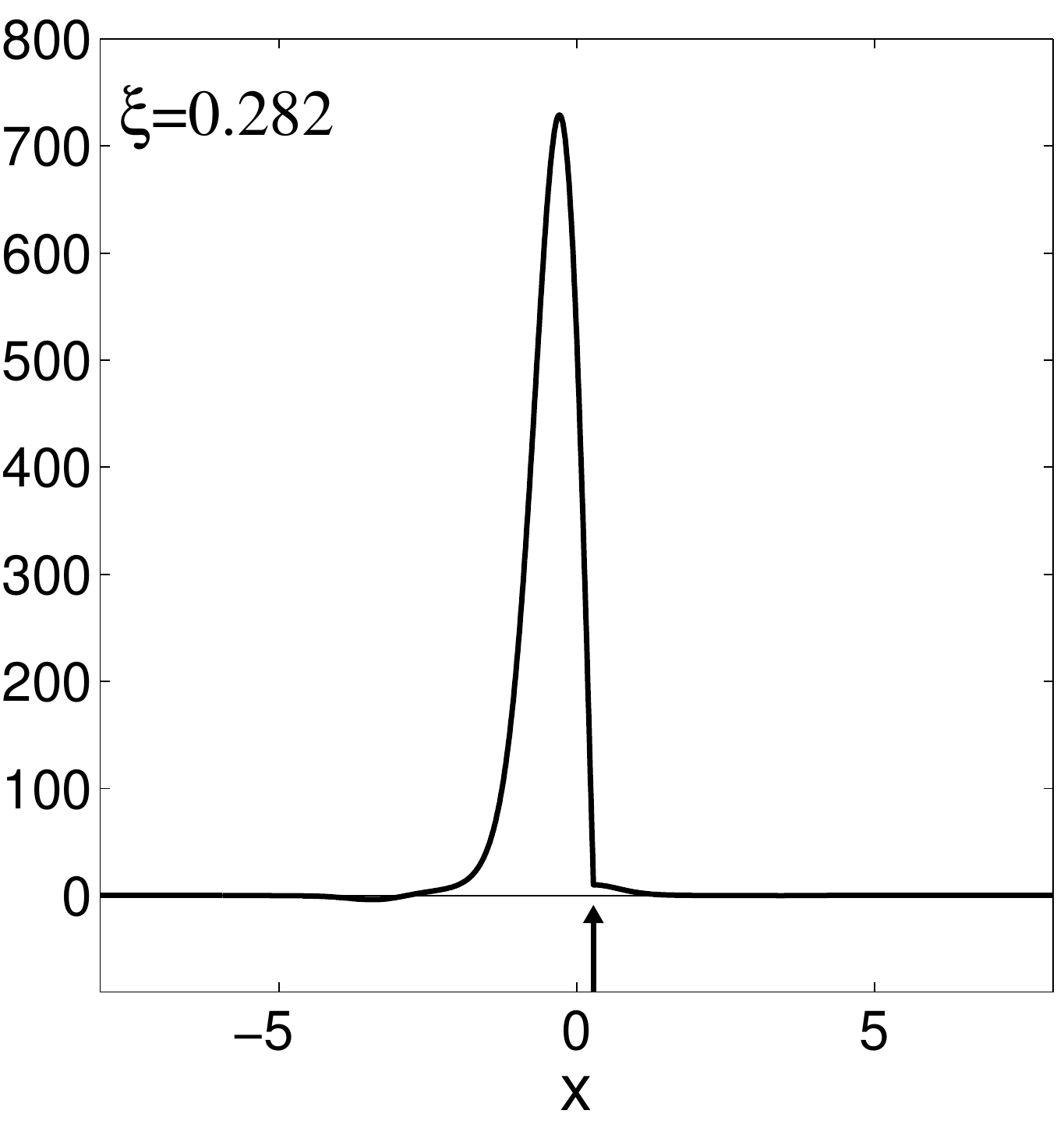}%
\label{Profile_1Delta_deltax028_mum1}} \subfigure[]{%
\includegraphics[width=3.4in]{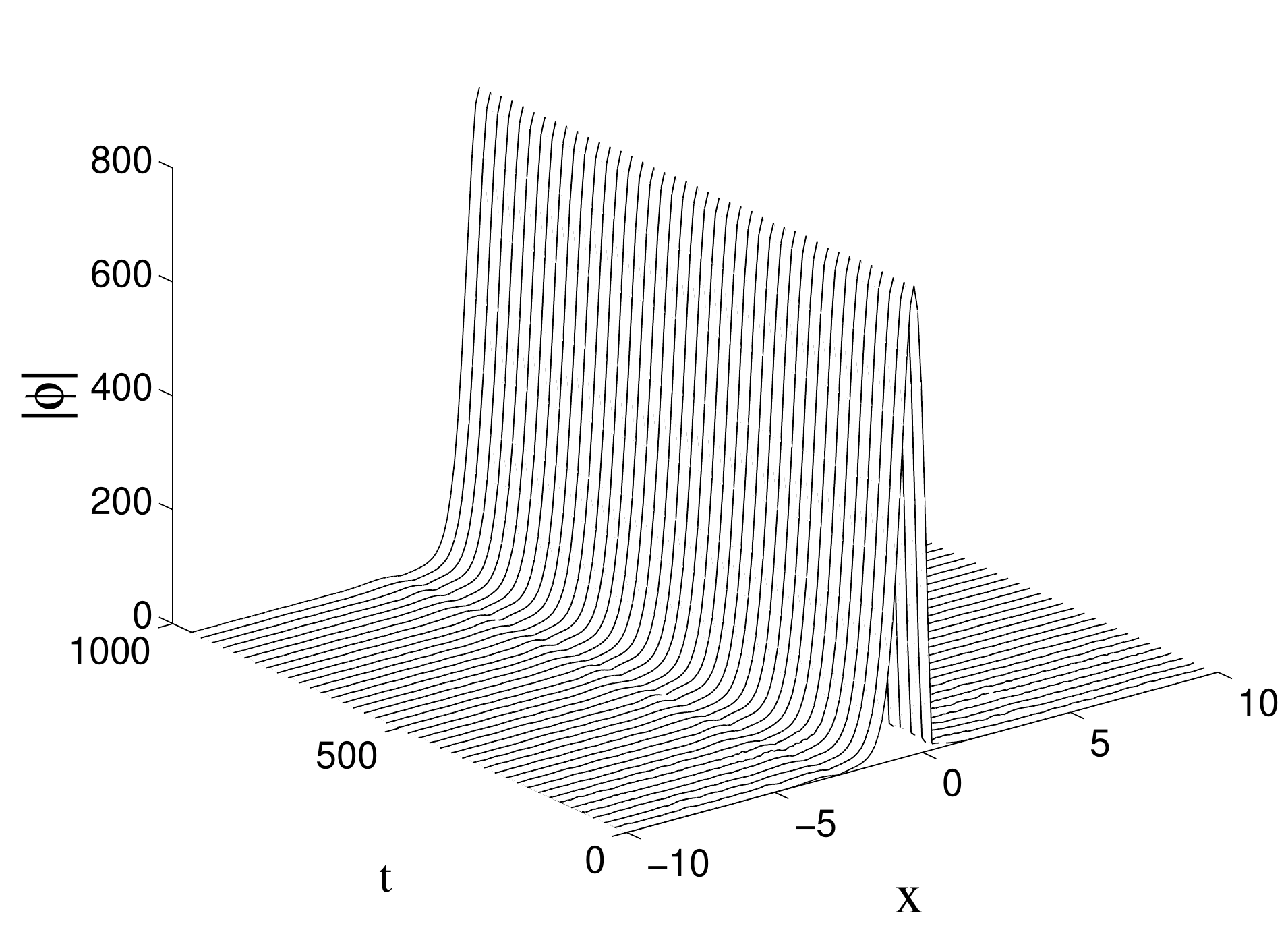}%
\label{Evolution_1Delta_deltax028_mum1}}
\caption{Shapes of the solitons in the first finite bandgap, for the
repulsive (a)-(b) and attractive (c)-(d) single-$\protect\delta $-function
nonlinearities. The solitons correspond to marked points in Fig.~\protect\ref%
{NvsDeltax_FFG_Mum1}, with $\protect\varepsilon =5$, $\protect\mu =-1$ and
(a) $\protect\xi =0$, (b) $\protect\xi =0.18$, (c) $\protect\xi =0.4$ and
(d) $\protect\xi =\protect\pi /2$. The vertical arrow in each panel denotes
the position and sign of the $\protect\delta $-function. A noteworthy
example is the soliton in panel (e), with a particularly large amplitude,
obtained at $\protect\xi =0.282$, which is very close to $\protect\xi _{%
\mathrm{thr}}$. Panel (f) illustrates the stability of this soliton in
direct simulations.}
\label{Profile_1Delta_mum1}
\end{figure}

The numerical analysis of solutions in the first bandgap demonstrates that
GSs cannot exist \emph{simultaneously} for the attractive and repulsive
nonlinearities. If the $\delta $-function is set close to a maximum of the
potential ($\xi =\pi /2$, for $\varepsilon >0$), it can support a soliton
only with the attractive sign of the nonlinearity. Shifting the position of
the attractive $\delta $-function towards an adjacent minimum of the
potential, the soliton ceases to exist at a critical (threshold) value of
the coordinate, $\xi =\xi _{\mathrm{thr}}$. At the same point, a new soliton
appears in the repulsive case, and exists at $\xi <\xi _{\mathrm{thr}}$.
Figure~\ref{NvsDeltax_FFG_Mum1} describes the norm of the GS versus the $%
\delta $-function's position within a half of the OL period, for $%
\varepsilon =5 $ and $\mu =-1$, which is close to the middle of the first
finite bandgap. These results agree with the prediction of the perturbation
theory obtained for $x=0$, according to which the soliton exists only for $%
\sigma =\mathrm{sgn}(\varepsilon )$, see Eq. (\ref{A}) (recall that $%
\varepsilon >0$ and $\xi =\pi /2$ are equivalent to $\varepsilon <0$ and $%
\xi =0$). Close to the existence threshold, the amplitude of the soliton
diverges, while its width remains approximately constant. Representative
examples of such GSs are displayed in Fig.~\ref{Profile_1Delta_mum1}. In
particular, Figs.~\ref{Profile_1Delta_deltax028_mum1}-%
\subref{Evolution_1Delta_deltax028_mum1} demonstrate a stable soliton
(supported by the repulsive nonlinearity) featuring an especially high
amplitude, achieved at $x=0.282$. The threshold value, $\xi _{\mathrm{thr}}$%
, varies with $\varepsilon $ and $\mu $, as shown in Fig.~\ref%
{Deltaxthresh_FFG}. Specifically, for small values of $\varepsilon $ and/or
small values of $\mu $ (close to the lower edge of the first bandgap), the
soliton-existence region expands in the case of the repulsive nonlinearity.

The stability analysis demonstrates that, in the case corresponding to Fig.~%
\ref{NvsDeltax_FFG_Mum1}, the GSs in the first bandgap, supported by the
repulsive $\delta $-function, are \emph{always stable}. On the other hand,
in the case of the attractive nonlinearity, local unstable eigenmodes exist,
with large $(\mathrm{Im}\left\{ \lambda \right\} >1)$ purely imaginary
eigenvalues, making all the solitons unstable. This finding is not
surprising because, as stressed above, in the cases of the repulsion and
attraction the GSs tend to be located, respectively, close to a local
minimum or maximum of the periodic potential.

\begin{figure}[tbp]
\centering
\subfigure[]{\includegraphics[width=2.3in]{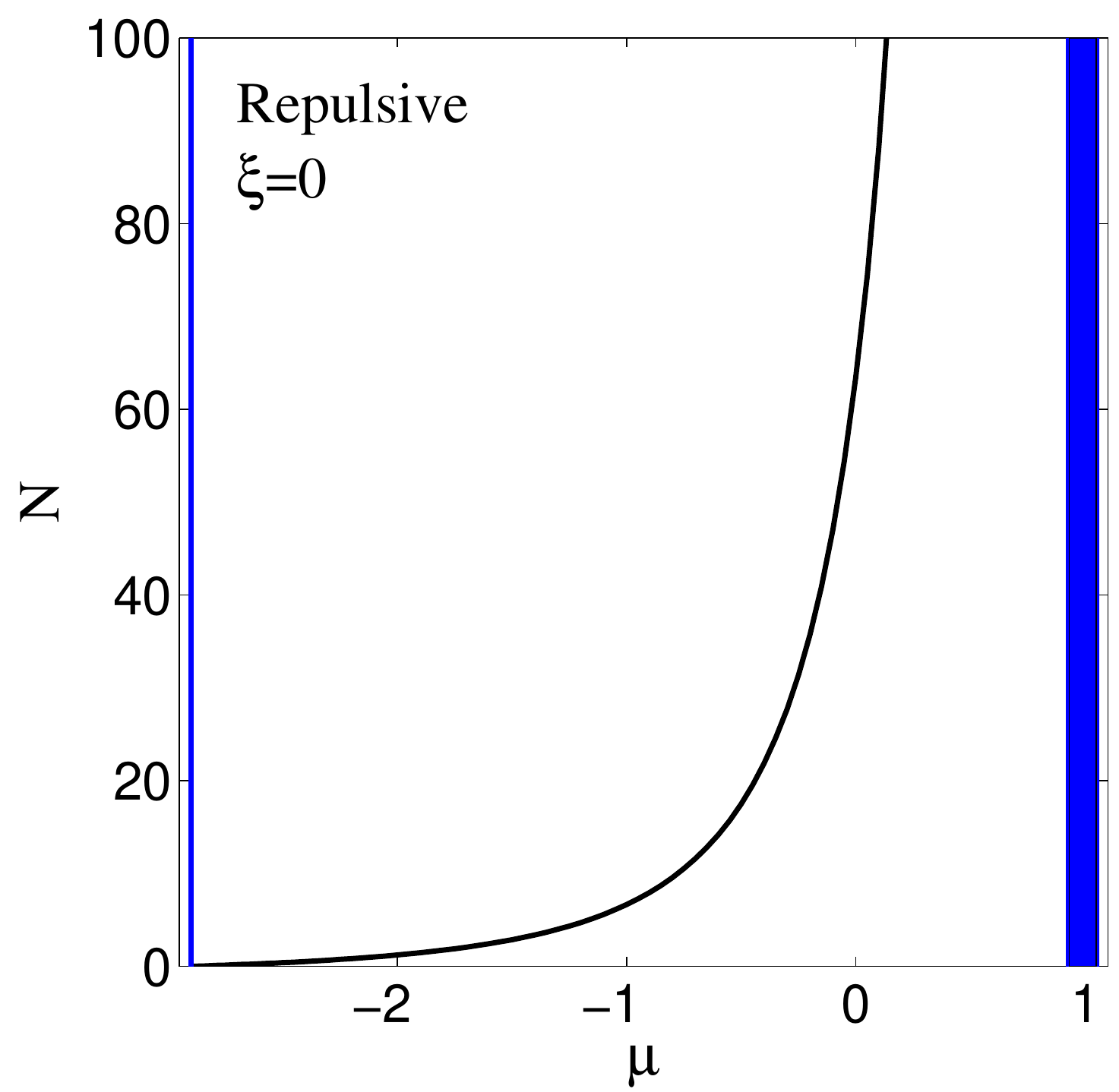}%
\label{NvsMu_FFG_Deltax0}} \subfigure[]{%
\includegraphics[width=2.3in]{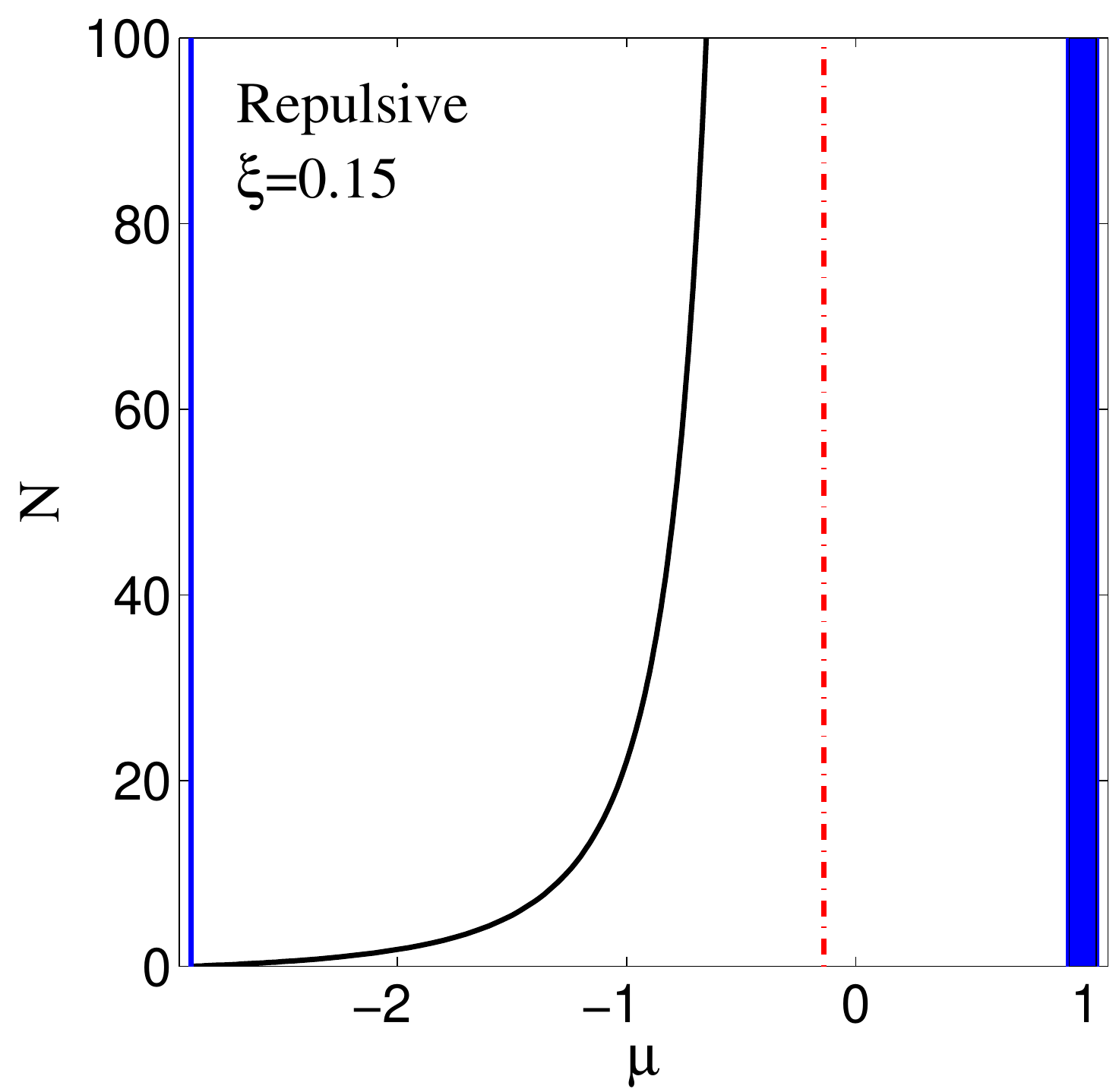}%
\label{NvsMu_FFG_Deltax015}} \subfigure[]{%
\includegraphics[width=2.3in]{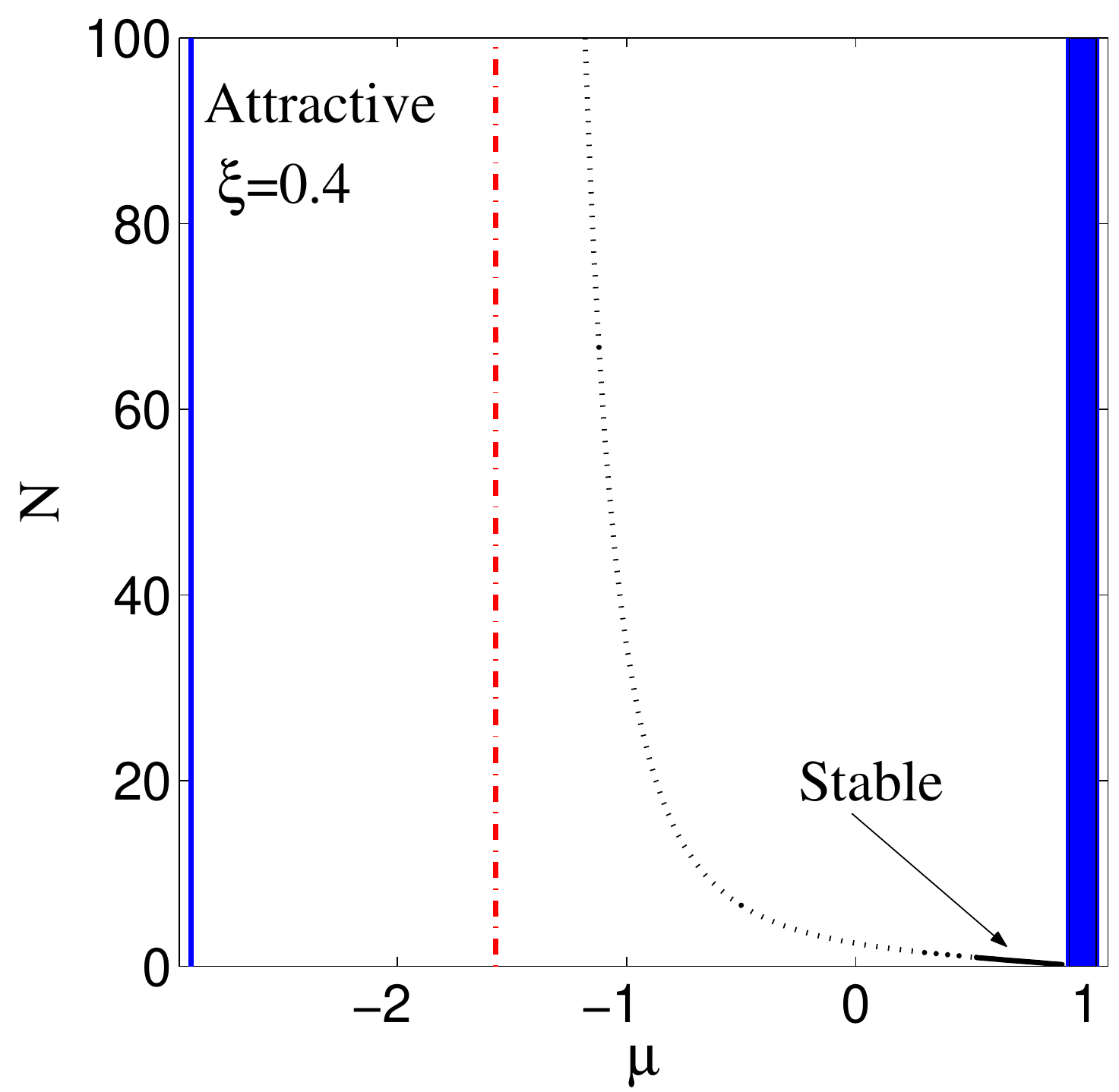}%
\label{NvsMu_FFG_Deltax04}} \\ 
\subfigure[]{\includegraphics[width=2.3in]{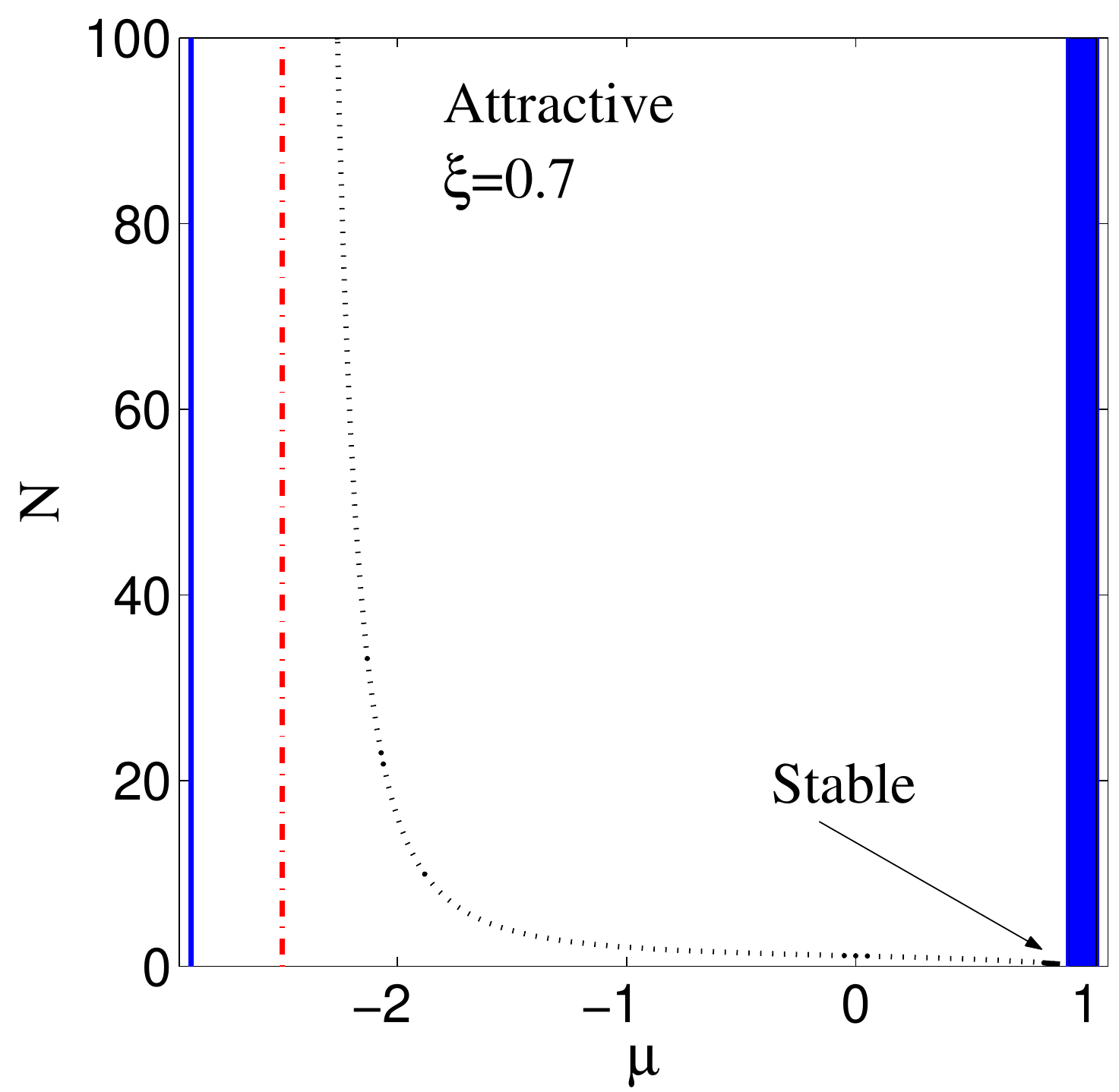}%
\label{NvsMu_FFG_Deltax07}} \subfigure[]{%
\includegraphics[width=2.3in]{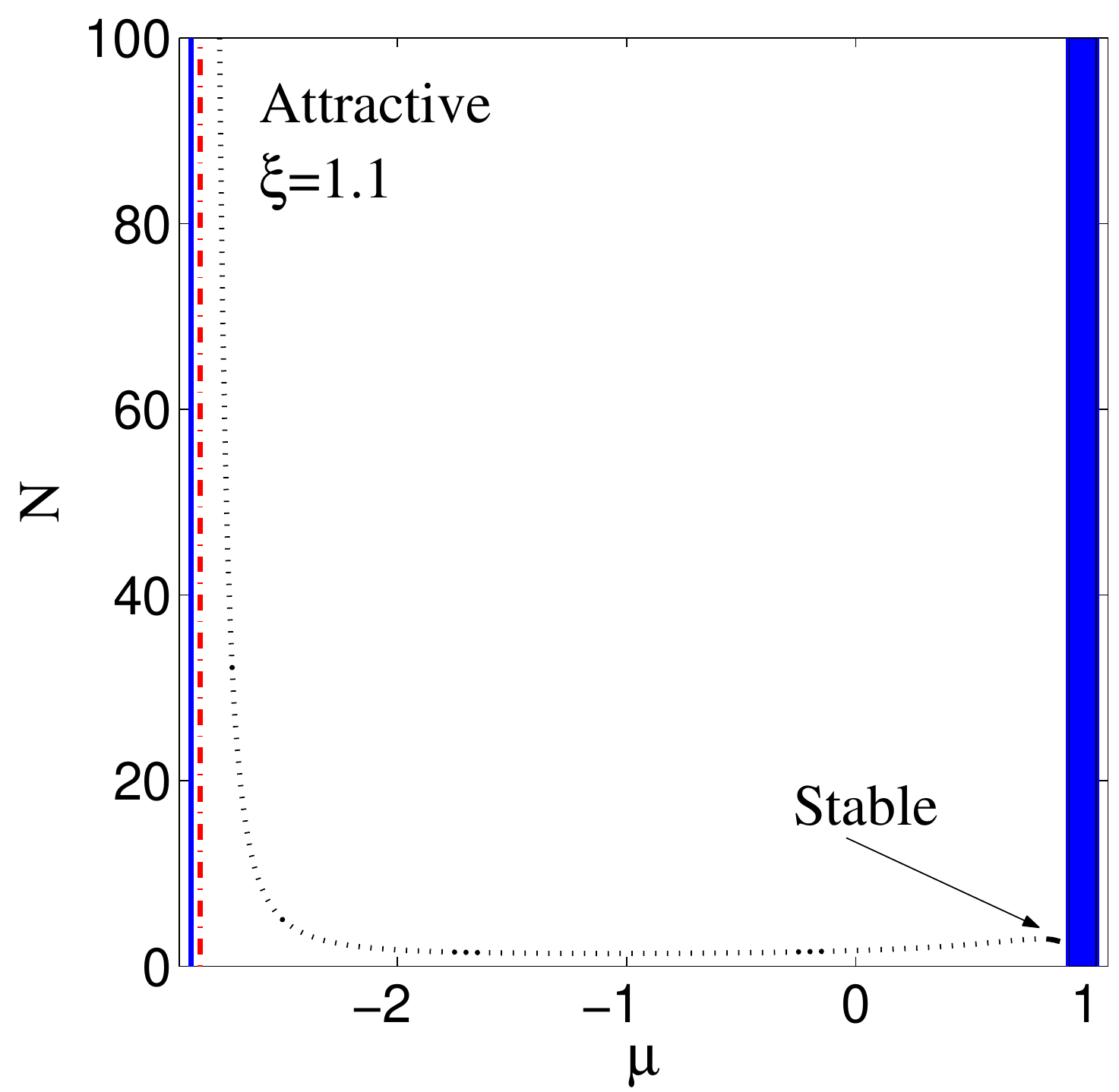}%
\label{NvsMu_FFG_Deltax11}} \subfigure[]{%
\includegraphics[width=2.3in]{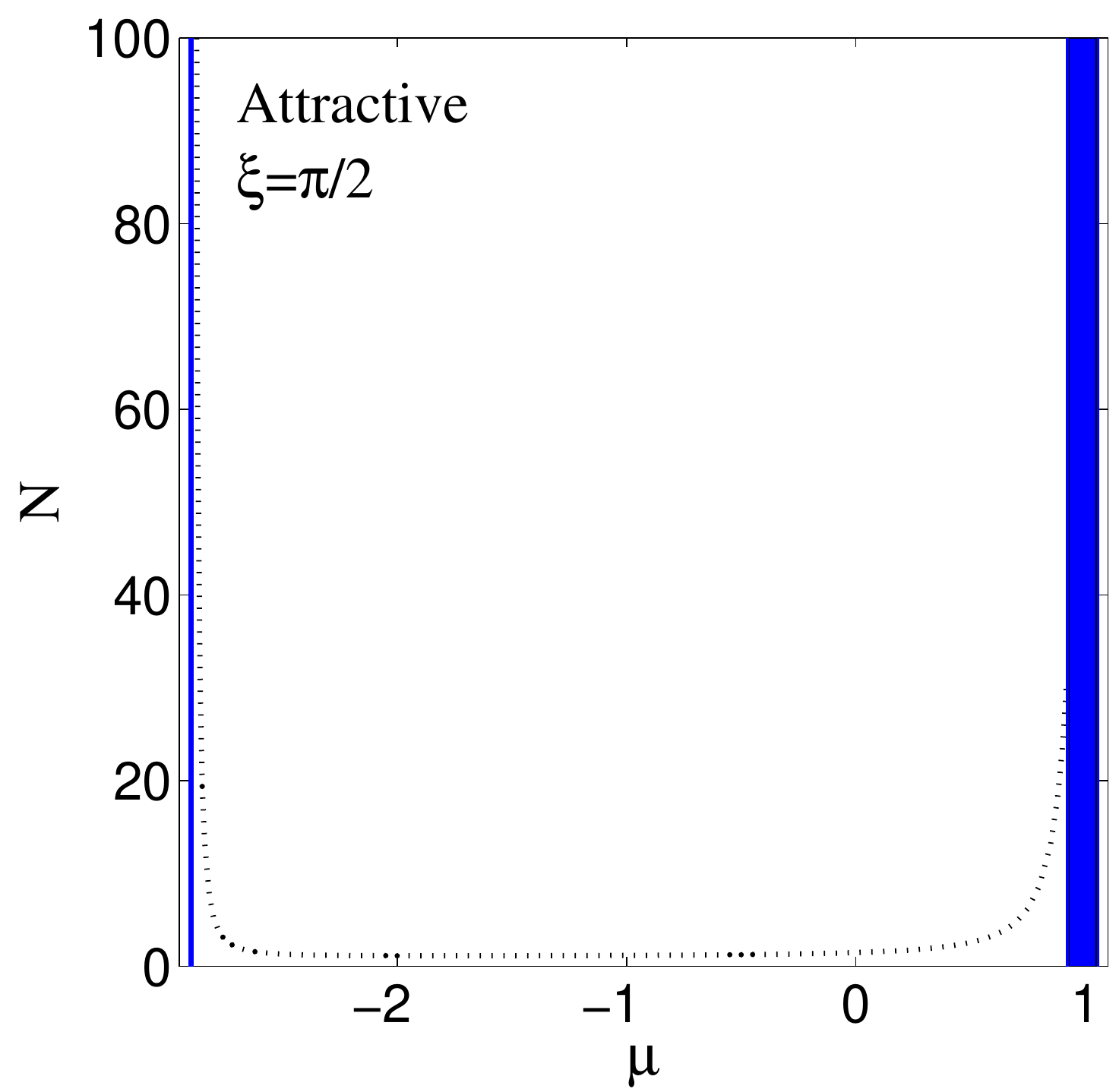}%
\label{NvsMu_FFG_Deltax1p57}}
\caption{(Color online) The norm of the gap solitons versus $\protect\mu $,
in the first finite bandgap, for $\protect\varepsilon =5$, with the
repulsive $\protect\delta $-function placed at $\protect\xi =0$ (a) or $%
\protect\xi =0.15 $ (b), and the attractive $\protect\delta $-function
placed at $\protect\xi =0.4$ (c), $\protect\xi =0.7$ (d), $\protect\xi =1.1$
(e), and $\protect\xi =\protect\pi /2$ (f). As above, the stable and
unstable solitons correspond to continuous and dotted lines, respectively. A
soliton subject to an oscillating instability can be found in the case (e),
in an extremely narrow region (barely visible on the scale of this figure)
between the stable and locally unstable sections. Blue vertical stripes on
both sides indicate Bloch bands between which the first bandgap is
sandwiched. The dashed-dotted vertical lines represent thresholds at which
the solitons' norm diverges: $\protect\mu _{\mathrm{thr}%
}=-0.141,-1.572,-2.5,-2.858$ for (b), (c), (d) and (e), respectively.}
\label{NvsMu_FFG}
\end{figure}

\begin{figure}[tbp]
\subfigure[]{%
\includegraphics[width=2.3in]{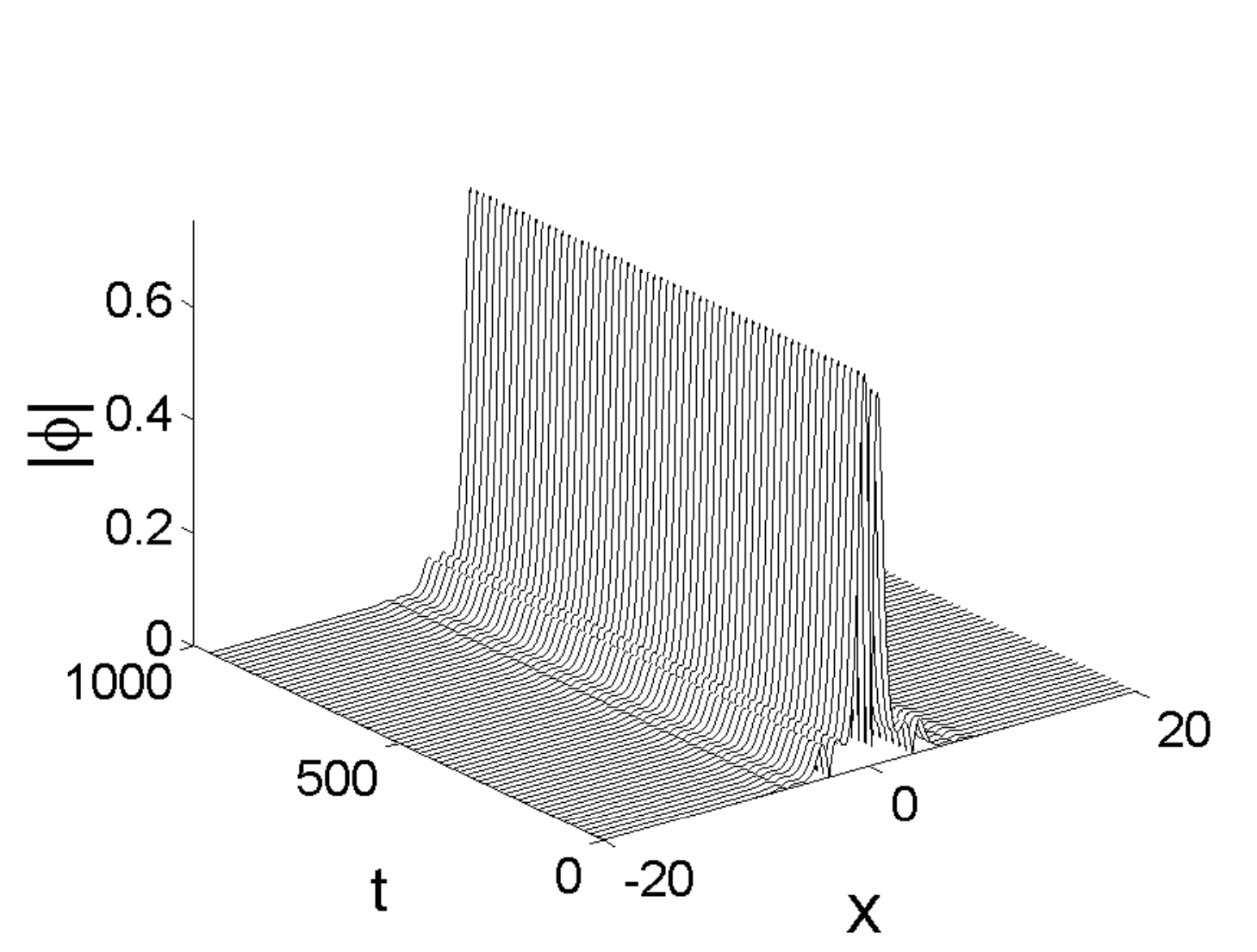}%
\label{LCNS_1000_mup07_eps5_deltax04}} \subfigure[]{%
\includegraphics[width=2.3in]{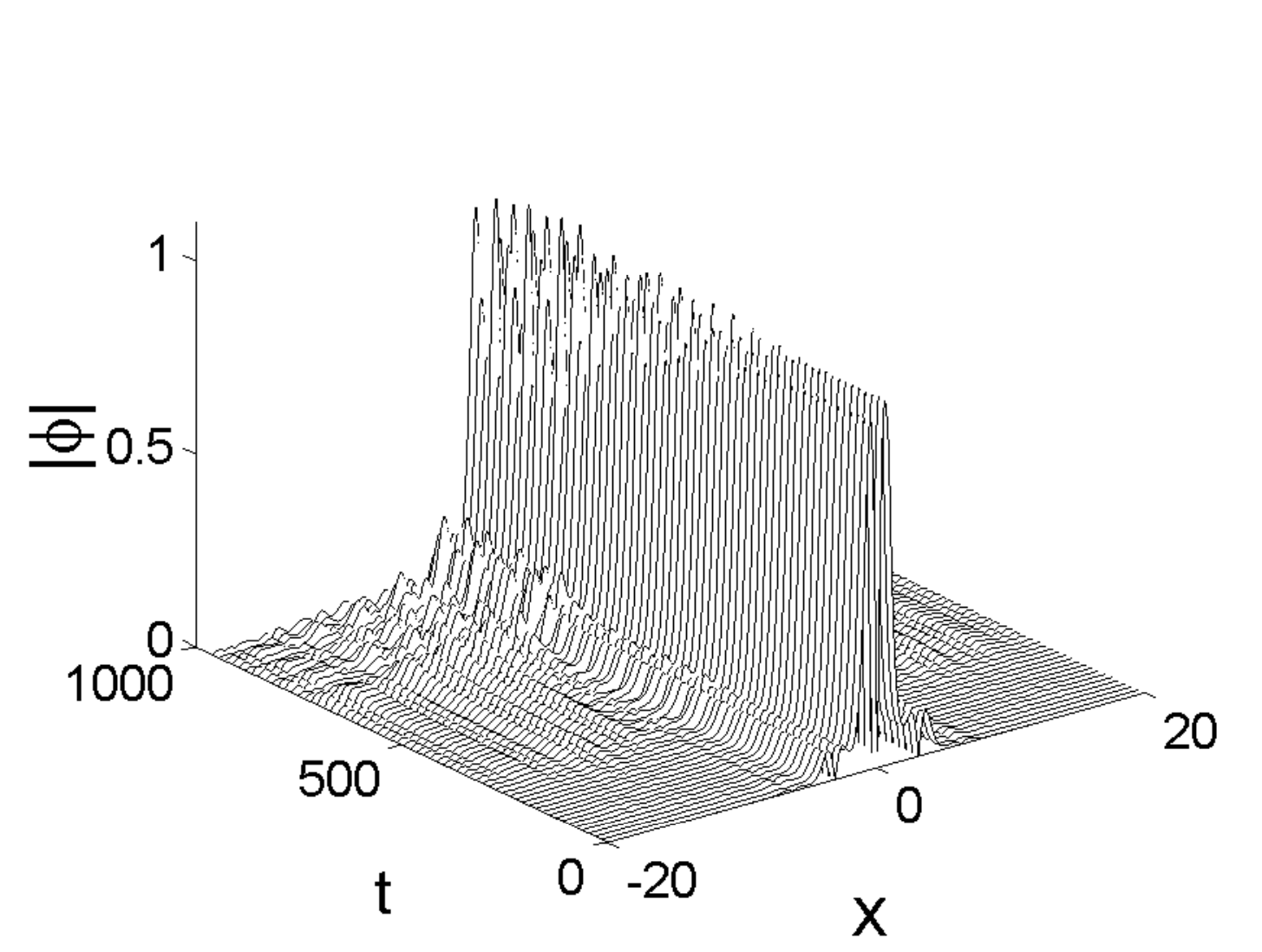}%
\label{LCNS_1000_mup07_eps5_deltax09}} \subfigure[]{%
\includegraphics[width=2.3in]{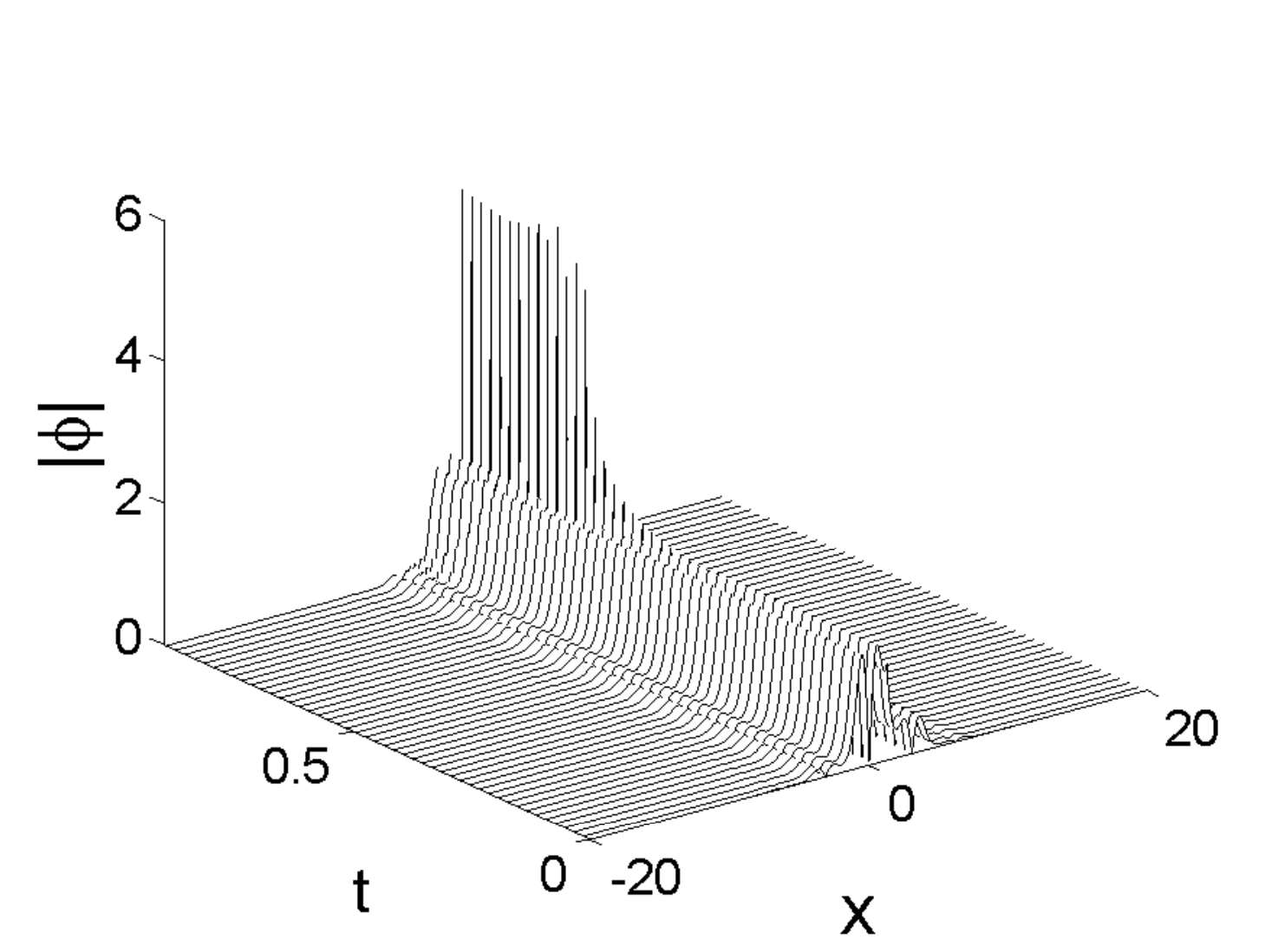}%
\label{LCNS_1000_mup07_eps5_deltax13}}
\caption{The evolution of solitons supported by the single attractive $%
\protect\delta $-function, for the case shown in Fig.~\protect\ref%
{NvsDeltax_FFG_Mup07}. Panel (a) displays an example of a stable soliton,
obtained for $\protect\xi =0.4$. Unstable solitons, subject to the
oscillatory instability or the strong localized instability, are exhibited
for $\protect\xi =0.9$ (b) and $\protect\xi =1.3$ (c), respectively.}
\label{Evolution_mup07_1Delta_FFG}
\end{figure}

The stability analysis was also carried out for other values of $\varepsilon
$ and $\mu $. The solitons are always stable under the repulsive
nonlinearity, while the stability may change in the case of the attraction.
Examples are presented in Fig.~\ref{NvsMu_FFG_Deltax015}-%
\subref{NvsMu_FFG_Deltax11}, for $\varepsilon =5$ and $\xi =0.15$
(repulsion) and $\xi =0.4,~0.7,~1.1$ (attraction). In the latter case, it is
seen that, for $\mu $ close to the upper edge of the first bandgap, there
are small stability areas for the solitons (conspicuous for $\xi =0.4$, and
extremely small for larger $\xi $). For fixed $\mu =0.7$ and varying
position $\xi $ of the attractive $\delta $-function, a small stability
region is found too, as shown in Fig.~\ref{NvsDeltax_FFG_Mup07}. Although it
is small, the existence of the stability area for the GSs in the case of the
attraction is a remarkable fact, as it is often assumed that all solitons are
unstable in finite bandgaps, if the nonlinearity is attractive (see, however,
Ref.~\cite{Kivshar}, where stable GSs were found in the case of attraction).
Apart from the common ``localized" instability discussed above (indicated in
the stability diagrams by dotted lines), there is additional weaker
instability which is often referred to as ``oscillatory" one (see Refs.~\cite%
{Abdullaev,Kivshar} for more details). This instability is characterized by
complex eigenvalues, $\lambda $, and delocalized oscillatory eigenfunctions.
Techniques which can be used for the detection of such instabilities are
described in Ref.~\cite{Kivshar}. Regions corresponding to unstable solitons
of this type can be clearly seen in Fig.~\ref{NvsDeltax_FFG_Mup07},
distinguished in this and other figures by a dashed-dotted marking. The
oscillatory instability can also be found in the case presented in Fig.~\ref%
{NvsMu_FFG_Deltax11}, within a tiny area between the non-oscillating
unstable solitons and the region of stable solitons, which is very small by
itself. Figure~\ref{Evolution_mup07_1Delta_FFG} illustrates the stability
and the simulated development of the solitons' instabilities, for the
representative cases marked in Fig.~\ref{NvsDeltax_FFG_Mup07}.

It is relevant to stress that the stability investigation was carried out
repeatedly, varying the size of the spatial domain and the number of grid
points. By doing so, we have checked that the stable solitons exist indeed,
not being simply a case of a weak instability.

Looking at Fig.~\ref{NvsMu_FFG}, it is easy to conclude that, unlike the
semi-infinite gap, the stability of the GSs in the first finite bandgap does
not obey the VK criterion (cf. Ref. \cite{HS}, where the same conclusion was
made for models with combined linear and nonlinear periodic potentials).
Another noteworthy fact is that, when the $\delta $-function is placed
anywhere except maxima or minima of the potential ($x=0,\pi /2$), the GS
family does not completely fill the first bandgap. For instance, for $\xi
=0.7$, the lower cutoff (existence border) for the soliton in the model with
attraction is $\mu _{\mathrm{thr}}\approx -2.5$, while the lower border of
the bandgap is at $\mu =-2.894$.
\begin{figure}[tbp]
\subfigure[]{\includegraphics[width=2.7in]{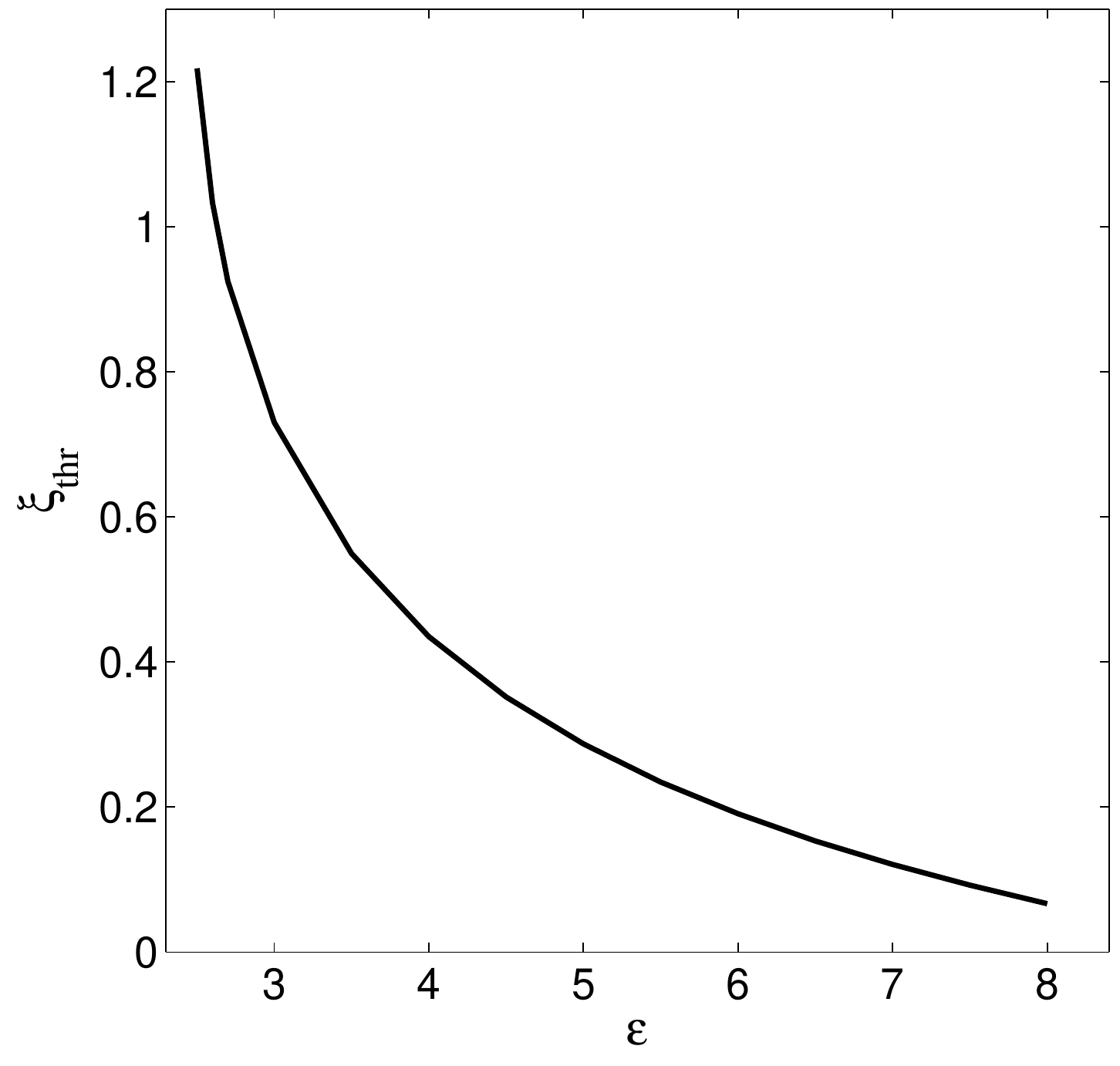}%
\label{Deltaxthresh_vs_eps}} \quad \subfigure[]{%
\includegraphics[width=2.7in]{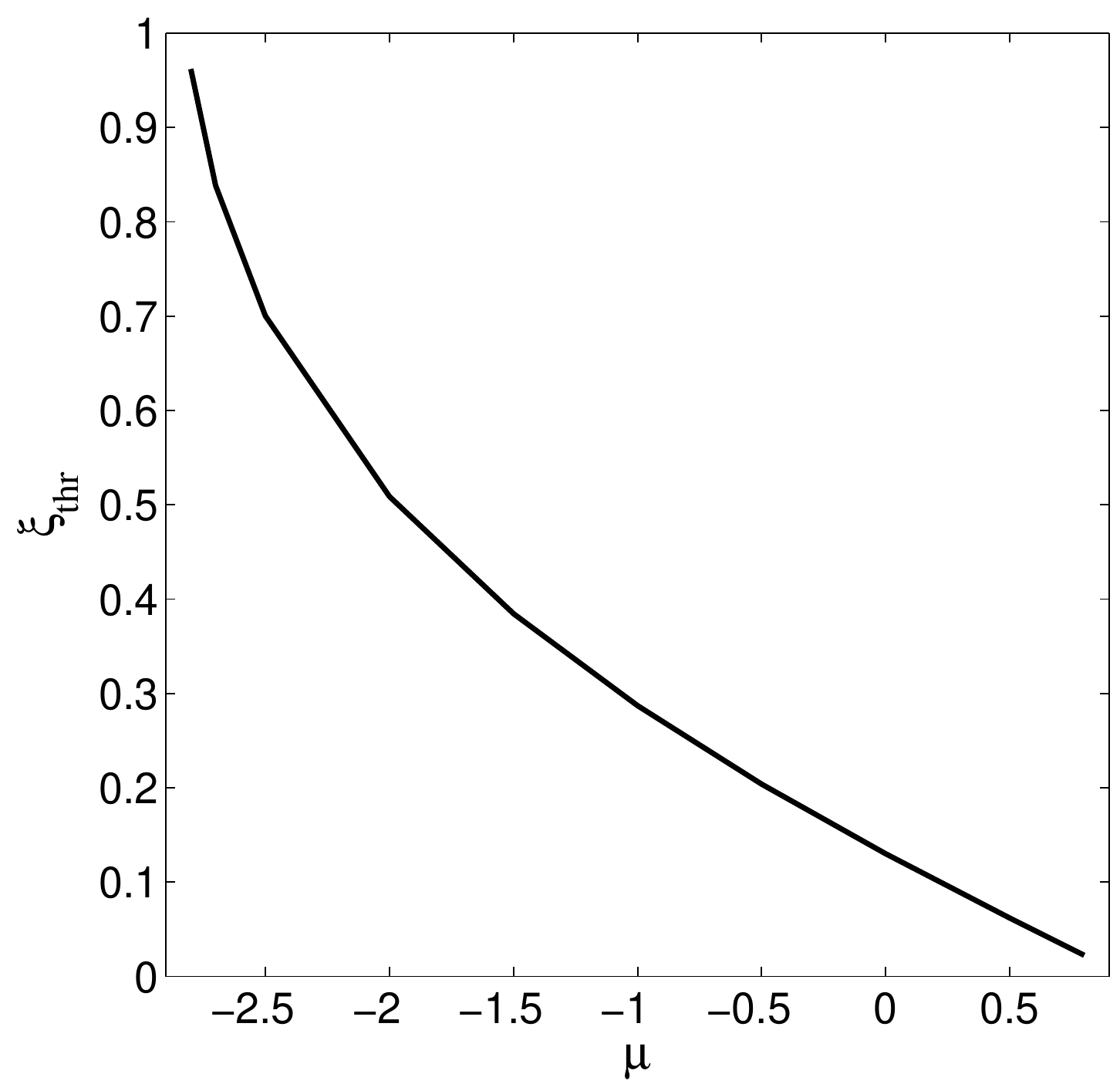}%
\label{Deltaxthresh_vs_mu}}
\caption{(a) The existence threshold, $\protect\xi _{\mathrm{thr}}$, for the
solitons in the first finite bandgap, as a function of $\protect\varepsilon $%
, for $\protect\mu =-1$. (b) The same, but as a function of $\protect\mu $
for fixed $\protect\varepsilon =5$.}
\label{Deltaxthresh_FFG}
\end{figure}

\subsection{Soliton solutions in the second finite bandgap}

The numerical investigation was also performed for GSs in the second finite
bandgap. For the repulsive nonlinearity, a single branch of solitons exists
in one half of the potential period in the $(N,\xi )$ plane. On the other
hand, in the case of the attractive nonlinearity, there are \emph{two}
different branches, one located around the minimum of the potential and the
other one -- around its maximum. A typical example is shown in Fig.~\ref%
{NvsDeltax_SFG_Mup2}, for $\varepsilon =5$ and $\mu =2$, close to the middle
of the second bandgap. In this bandgap, there are two thresholds in the
region of $0<\xi <\pi /2$. Similar to what was seen in the first bandgap,
the soliton's amplitude diverges at the threshold, while the soliton's width
remains finite.

\begin{figure}[tbp]
\subfigure[]{\includegraphics[width=2.7in]{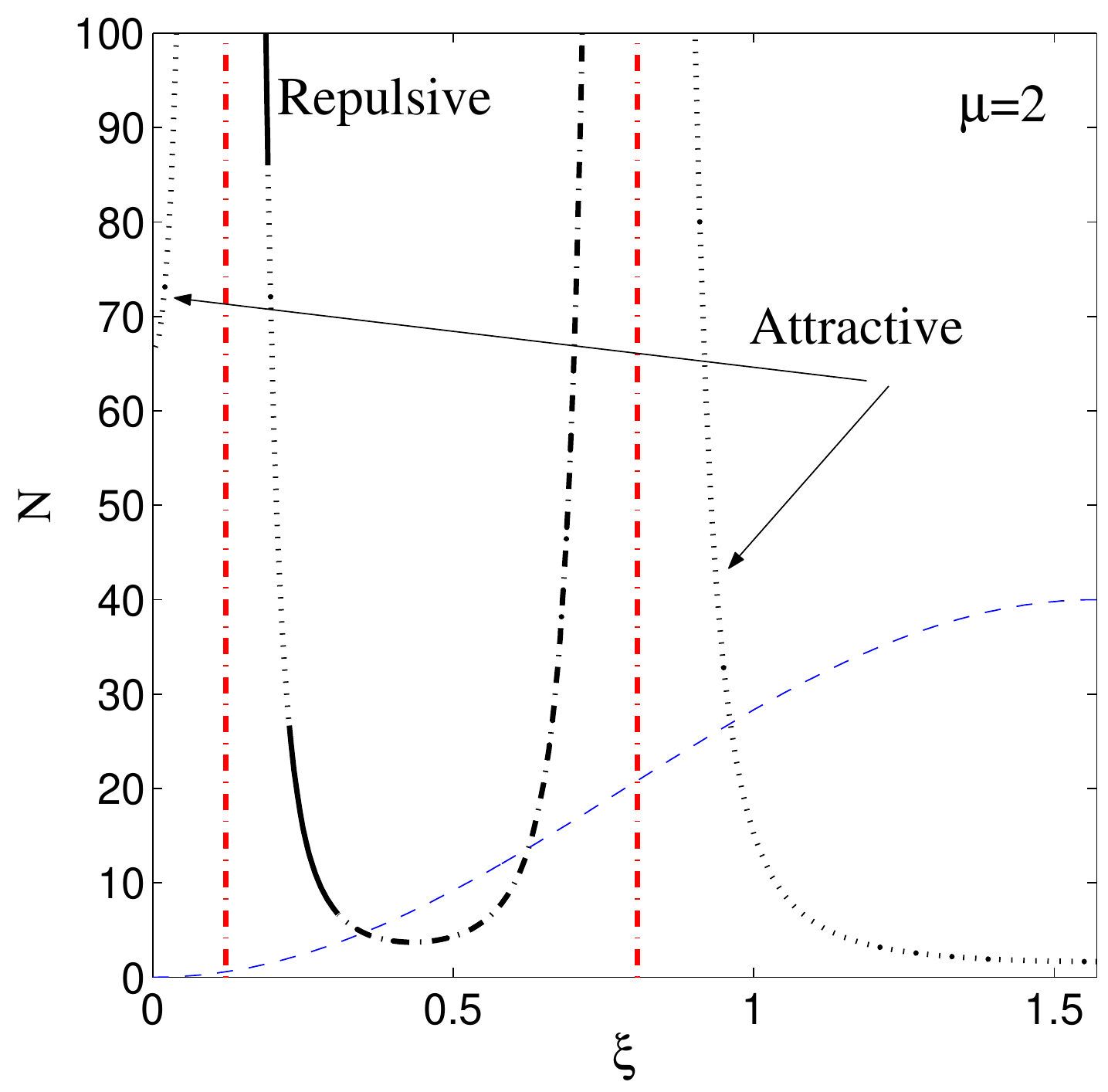}%
\label{NvsDeltax_SFG_Mup2}} \quad \subfigure[]{%
\includegraphics[width=2.7in]{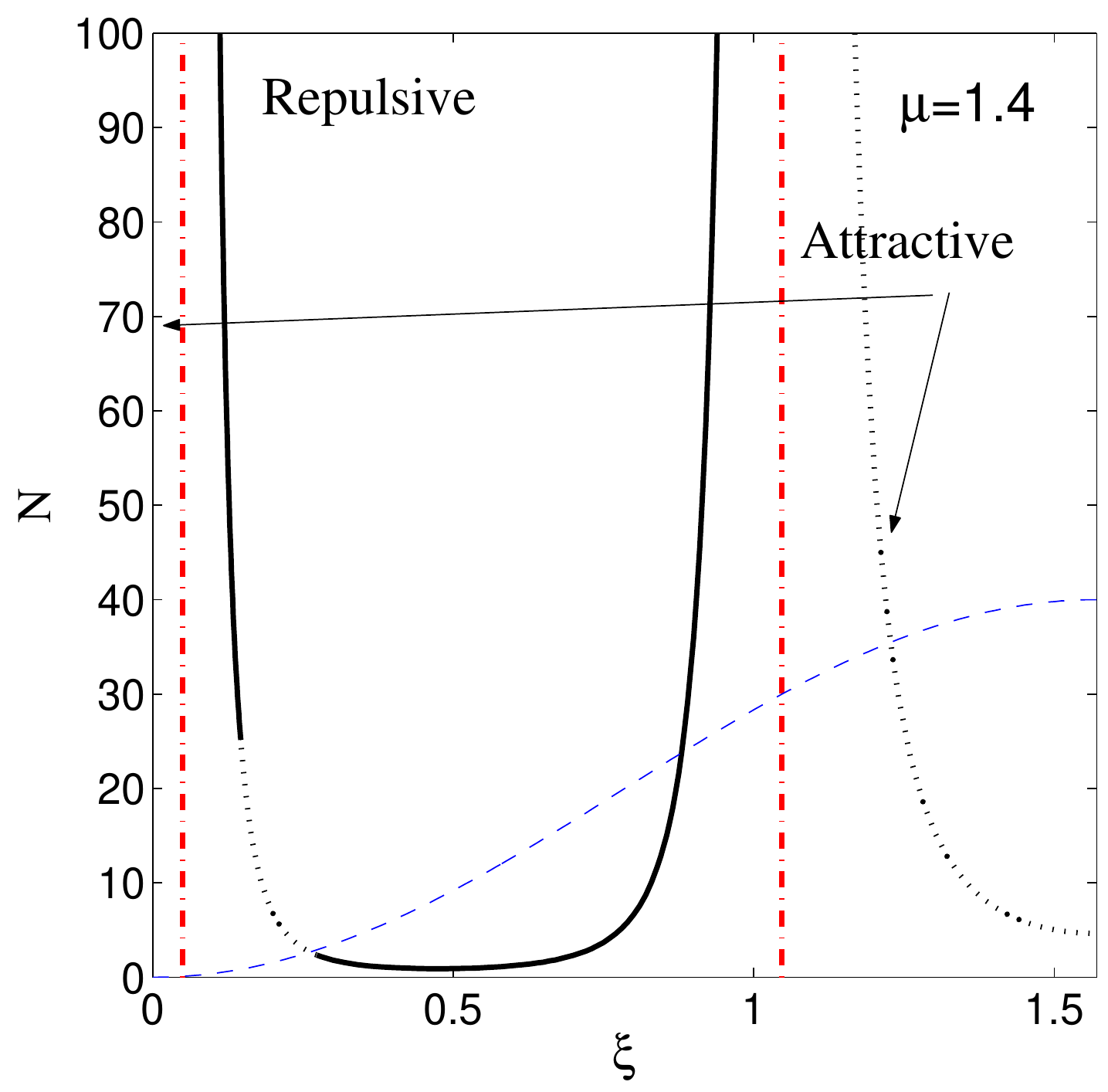}%
\label{NvsDeltax_SFG_Mup1p4}}
\caption{(Color online) The same as in Fig.~\protect\ref{NvsDeltax_FFG}, but
in the second finite bandgap, for $\protect\mu =2$ (a) and $\protect\mu =1.4$
(b). Note that, in very narrow regions of $\protect\xi $, such as the left
margin in (b), the norm of the (unstable) solitons, supported by the
attractive instability, is especially large, therefore it cannot be
displayed on the scale of this figure. The two thresholds (as explained in
the text) are represented, as usual, by dashed-dotted vertical lines.}
\label{NvsDeltax_SFG}
\end{figure}

\begin{figure}[tbp]
{\includegraphics[width=3.0in]{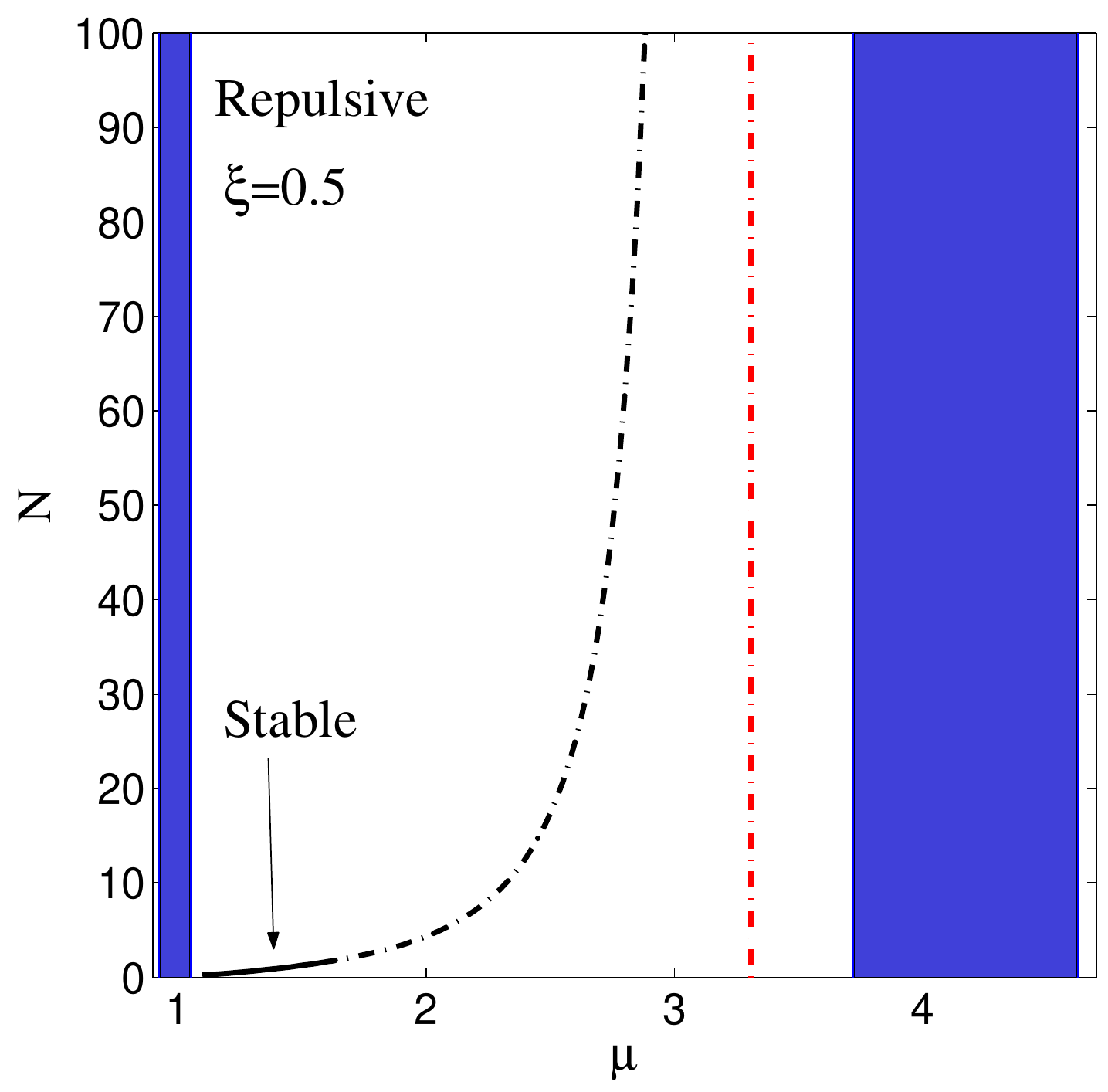}}
\caption{(Color online) The norm versus $\protect\mu $, in the second finite
bandgap for $\protect\varepsilon =5$, with the repulsive $\protect\delta $%
-function set at $\protect\xi =0.5$. The norm diverges at the threshold $%
\protect\mu=3.307$, marked by the vertical dashed-dotted curve.}
\label{NvsMu_SFG_Deltax05}
\end{figure}

With the attractive $\delta $-function, GSs are always unstable in the
second bandgap (featuring the localized instability), on the contrary to the
situation in the first bandgap, where a small stability region was found for
the case of attraction, see Figs. \ref{NvsDeltax_FFG} and \ref{NvsMu_FFG}.
On the other hand, in the case of the repulsive $\delta $-function, the
stability changes with the variation of parameters $\xi $, $\varepsilon $
and $\mu $. In particular, in the situation displayed in Fig.~\ref%
{NvsDeltax_SFG_Mup2} both localized and oscillatory instabilities can be
found, and \emph{two} stability regions are present. If the repulsive $%
\delta $-function is set at $\xi =0.5$, and $\varepsilon =5$, in which case
the second bandgap is $1.05<\mu <3.724$, stable solitons are found only at $%
\mu $ close to the lower edge of the gap, see Fig.~\ref{NvsMu_SFG_Deltax05}.
For instance, at $\mu =1.4$ the GSs are stable at almost all values of $\xi $%
, except for a narrow interval where the localized instability occurs, as
seen in Fig.~\ref{NvsDeltax_SFG_Mup1p4}. Similar results were found at other
values of $\varepsilon $.

Similar to the situation in the first bandgap, the GS families cover the
entire second bandgap only for $\xi =0$ and $\pi /2$. For different values
of $\xi $, there are regions of $\mu $ in which no solitons are present, see
Fig. \ref{NvsMu_SFG_Deltax05}.

\section{Numerical results for the model with the two symmetric $\protect%
\delta $-functions}

\label{sec:TwoDeltas}

Equation (\ref{Two_Dlata_NLSE}) presents a natural extension of the model,
which includes two $\delta $-functions symmetrically positioned around the
potential maxima or minima. We here consider the existence and stability
conditions for solitons in this model, in the semi-infinite and the first
finite gaps. In each case, two settings were explored, with either a
potential maximum or minimum located exactly at the midpoint between the two
$\delta $-functions. The stability analysis was carried out using the method
outlined in Sec. \ref{sec:SingleDelta}, with the nonlinearity coefficient $%
\delta (x-\xi )$ in Eq.(\ref{Eigenvalue_problem}) replaced by $\delta (x-\xi
)+\delta (x+\xi )$.

\subsection{Solitons in the semi-infinite gap}

As in the models with the single $\delta $-function, solitons in the
semi-infinite gap exist only for the attractive nonlinearity. First, we
examined the changes that the solitons undergo with the increase of distance
$\xi $ of each $\delta $-function from the potential \emph{minimum} located
between them, which corresponds to $\varepsilon >0$ in Eq. (\ref%
{Two_Dlata_NLSE}).

\begin{figure}[tbp]
\subfigure[]{%
\includegraphics[width=1.7in]{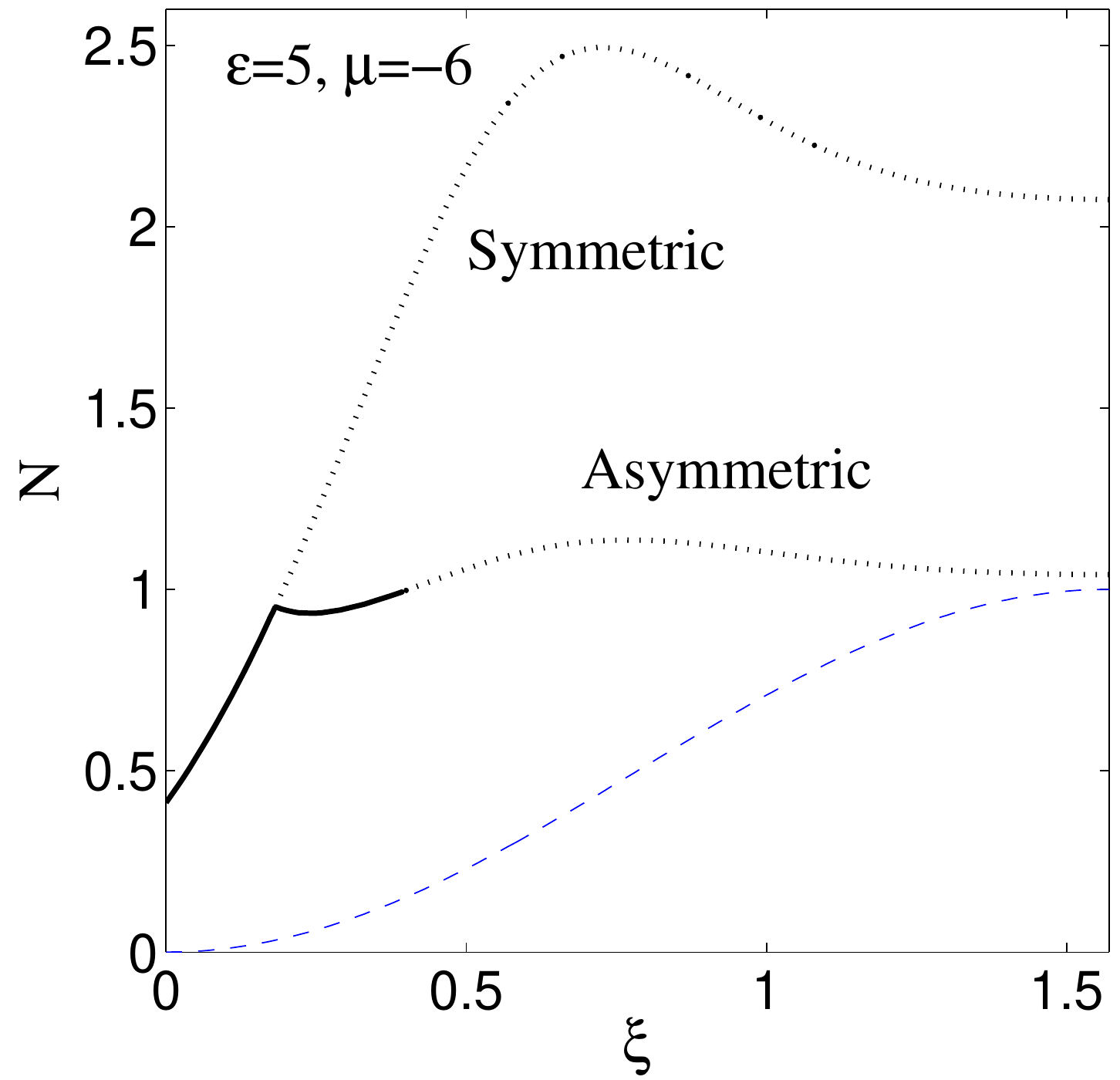}%
\label{NvsDeltax_2Delta_SIG_eps5_mum6}} \subfigure[]{%
\includegraphics[width=1.7in]{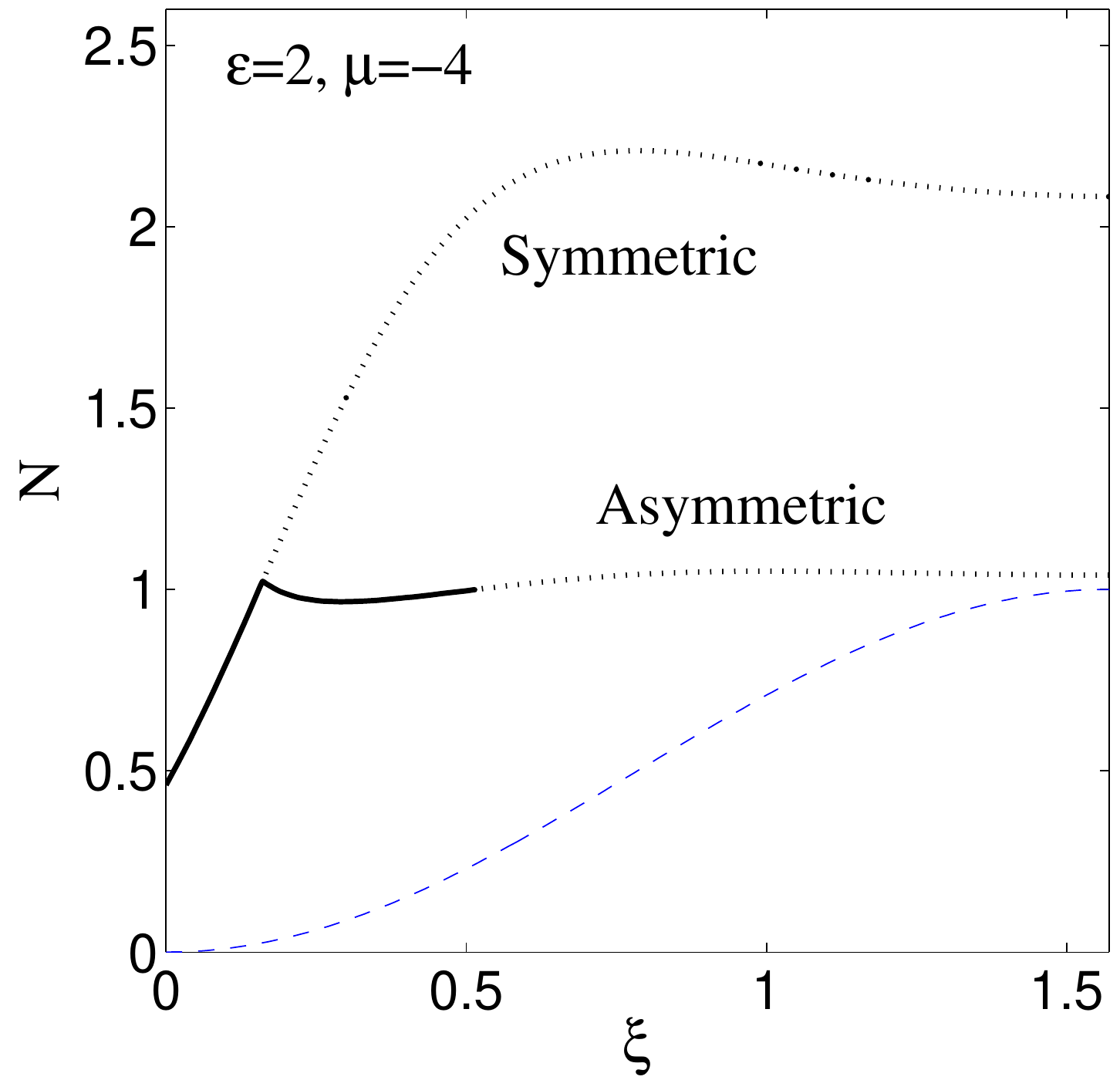}%
\label{NvsDeltax_2Delta_SIG_eps2_mum4}} \subfigure[]{%
\includegraphics[width=1.7in]{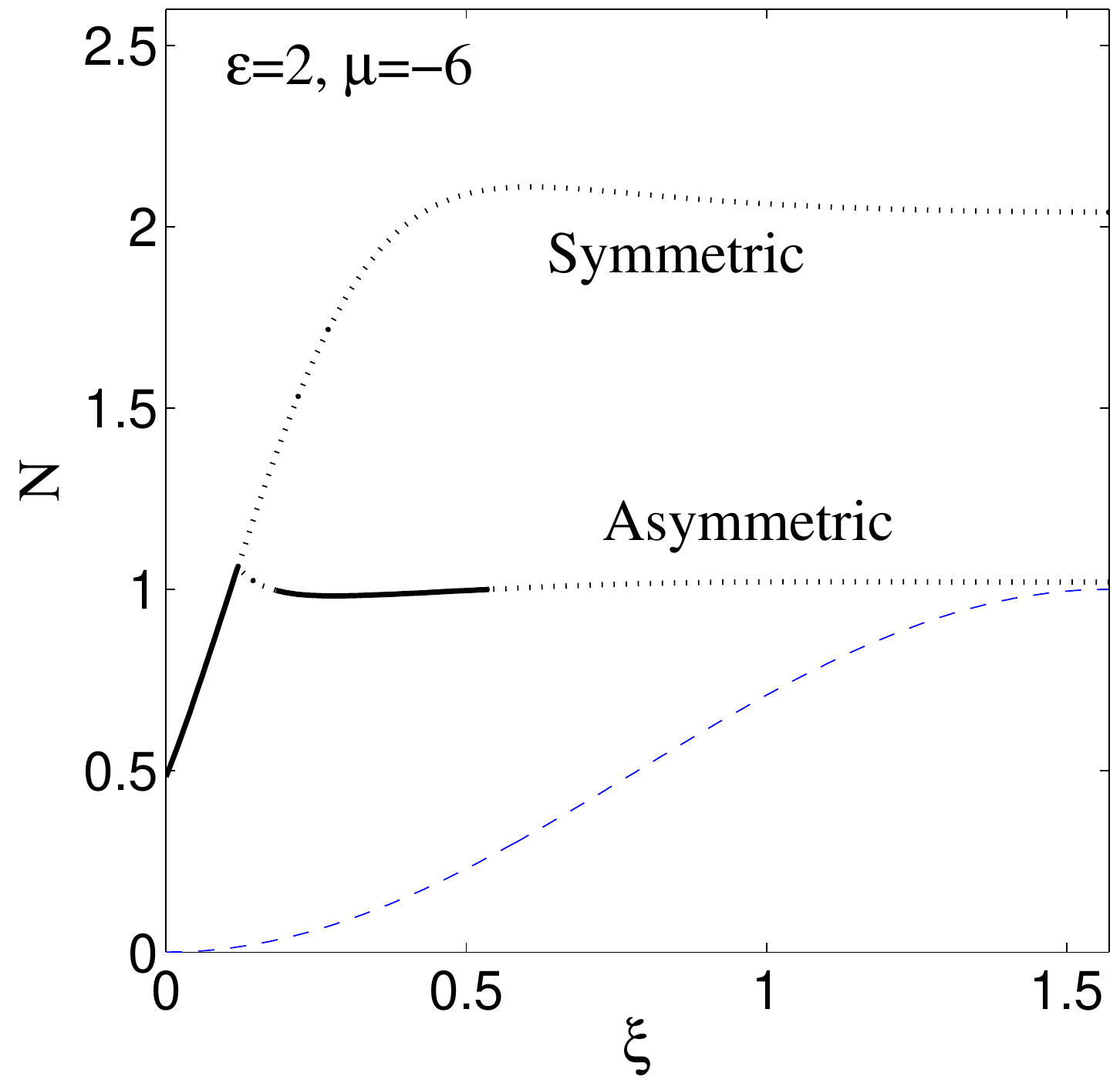}%
\label{NvsDeltax_2Delta_SIG_eps2_mum6}} \subfigure[]{%
\includegraphics[width=1.7in]{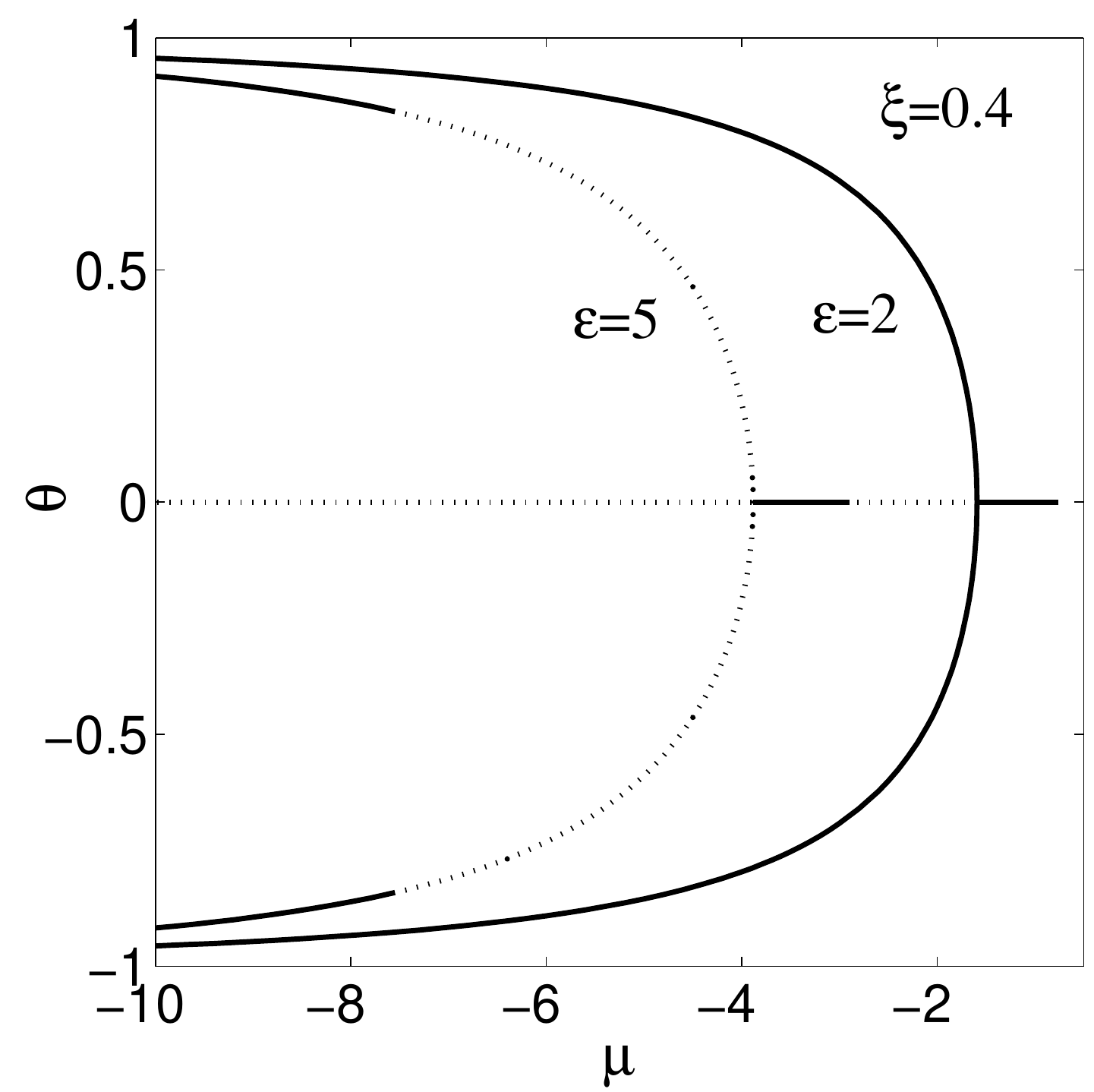}%
\label{ThetavsMu_deltax04_SIG}}
\caption{(a)-(c) Branches of the symmetric and asymmetric states in the $(%
\protect\xi ,N)$ plane near the bifurcation point, in the semi-infinite gap
of the model with two $\protect\delta $-functions, for (a) $\protect%
\varepsilon =5$, $\protect\mu =-6$, (b) $\protect\varepsilon =2$, $\protect%
\mu =-4$, and (c) $\protect\varepsilon =2$, $\protect\mu =-6$. (d) The
bifurcation diagrams in the $(\protect\mu ,\protect\theta )$ plane, for $%
\protect\xi =0.4$ and $\protect\varepsilon =2$ and $5$.}
\label{NvsDeltax_2Delta_SIG}
\end{figure}

For small values of $\xi $, a region of stable symmetric solitons always
exists. Increasing $\xi $, we reach a bifurcation point, after which the
symmetric solutions lose their stability (against non-oscillatory
perturbations) and a new \emph{asymmetric} branch emerges, that may be
partially stable. An example is shown in Fig.~\ref%
{NvsDeltax_2Delta_SIG_SymAsym}, for $\varepsilon =5$ and $\mu =-4$. In this
case, there is a tiny region of stable asymmetric solitons, abutting on the
bifurcation point, which is too small to be visible in the figure. Closer to
the edge of the semi-infinite gap, the bifurcation occurs at higher values
of $\xi $. For example, $\xi _{\mathrm{bif}}(\varepsilon =2,\mu =-1)=0.777$,
$\xi _{\mathrm{bif}}(\varepsilon =5,\mu =-3)=0.935$, $\xi _{\mathrm{bif}%
}(\varepsilon =6,\mu =-4)=0.683$, and $\xi _{\mathrm{bif}}(\varepsilon
=8,\mu =-6)=0.491$, and in all these cases, the asymmetric modes are
completed unstable. On the other hand, the bifurcation occurs at smaller
values of $\xi $ for sets of $(\varepsilon ,\mu )$ taken deeper inside gap,
i.e., when $\varepsilon $ is smaller and/or $\mu $ is more negative, the
bifurcation giving rise to a conspicuous stability region for asymmetric
states in such cases. Typical examples are presented in Fig.~\ref%
{NvsDeltax_2Delta_SIG_eps5_mum6}-\subref{NvsDeltax_2Delta_SIG_eps2_mum4},
for $(\varepsilon ,\mu )=$ $(2,-4)$ and $(5,-6)$. Going still deeper into
the semi-infinite gap, the value of $\xi $ at the bifurcation point keeps
decreasing. The corresponding domain of stable asymmetric solutions may not
necessarily emerge exactly at the bifurcation point, but slightly later. An
example is displayed in Fig.~\ref{NvsDeltax_2Delta_SIG_eps2_mum6} for $%
(\varepsilon ,\mu )=(2,-6)$.

Bifurcation diagrams in the $(\mu ,\theta )$ plane, displayed in Fig.~\ref%
{ThetavsMu_deltax04_SIG} for $\xi =0.4$ and $\varepsilon =2$ and $5$,
demonstrate the SSB (spontaneous symmetry breaking) of supercritical type.
For instance, at $\varepsilon =2$, stable branches of the asymmetric modes
emerge at the bifurcation point, where the symmetric branch loses its
stability. On the other hand, for $\varepsilon =5$ the asymmetric branches
are not immediately stable after the bifurcation point, as the VK criterion
is satisfied for them only when further decreasing $\mu $.

\begin{figure}[tbp]
\subfigure[]{\includegraphics[width=2.7in]{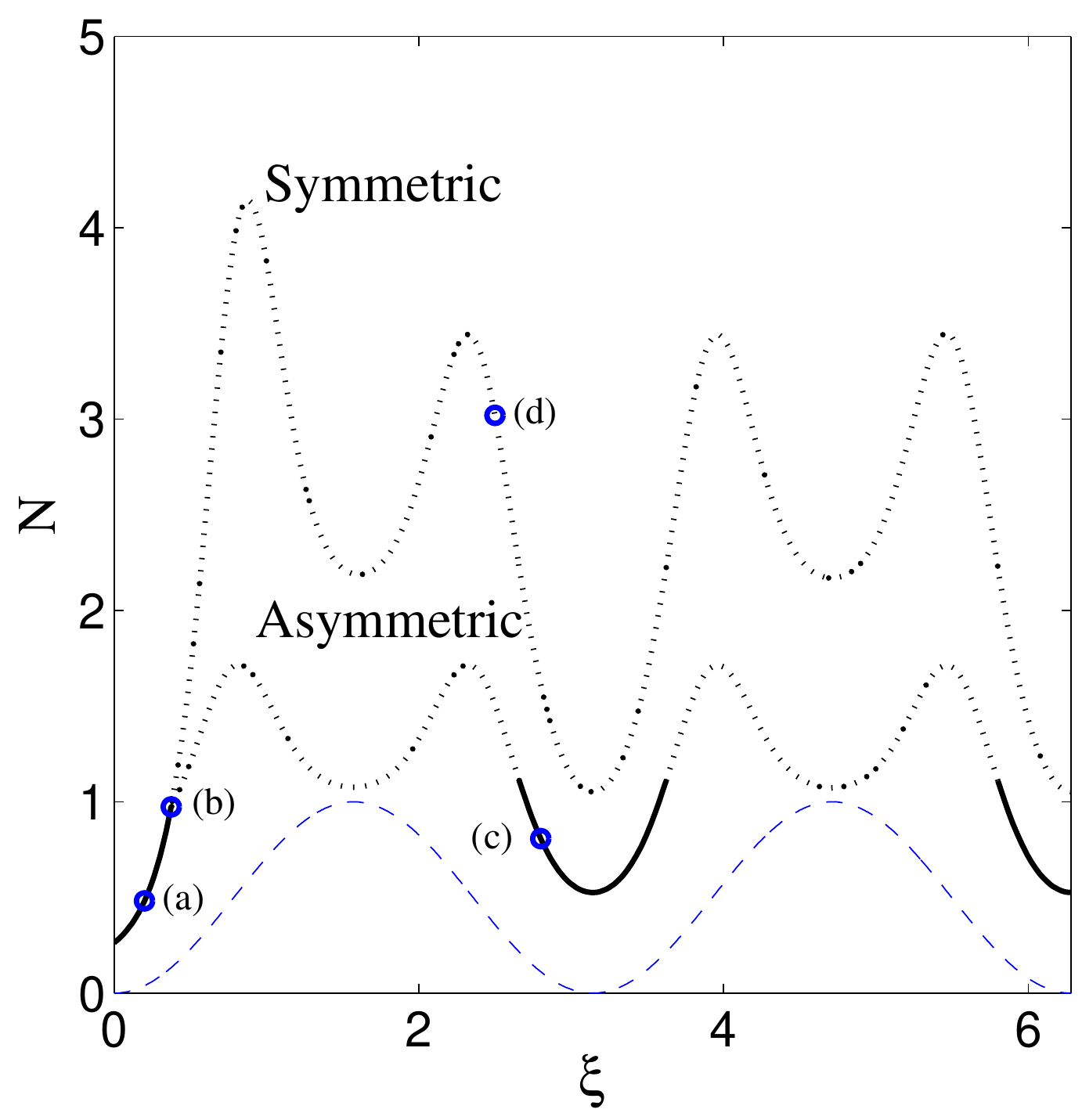}%
\label{NvsDeltax_2Delta_SIG_SymAsym}} \subfigure[]{%
\includegraphics[width=2.7in]{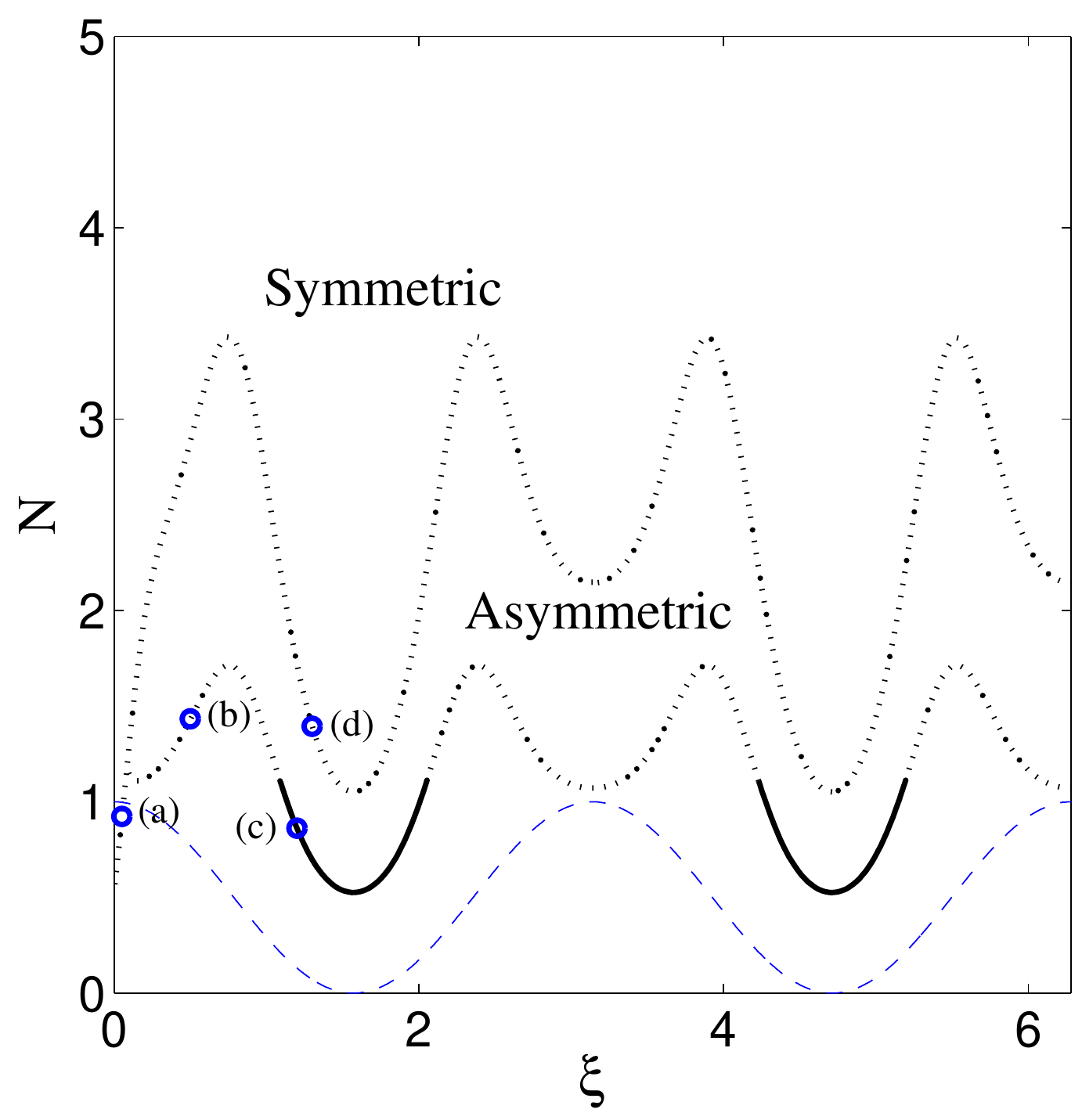}%
\label{NvsDeltax_2Delta_SIG_SymAsym_epsm}}
\caption{Branches of symmetric and asymmetric soliton modes in the
semi-infinite gap of the model with two $\protect\delta $-functions, for $%
\protect\mu =-4 $ and $\protect\varepsilon =5$ (a) or $-5$ (b), which
correspond, respectively, to the local minimum or maximum of the periodic
potential located between the $\protect\delta $-functions. As before, the
continuous and dotted lines represent stable and unstable portions of the
soliton families. Circles correspond to representative examples of solitons
that are shown in panels (a)-(d) of Figs.~\protect\ref%
{Profile_2Delta_mum4_epsp5} and~\protect\ref{Profile_2Delta_mum4_epsm5} .}
\label{NvsDeltax_2Delta_SIG_SymAsym_epspm}
\end{figure}

At large values of $\xi $, the asymmetric soliton gradually transforms into
a fundamental one, pinned to either of the two $\delta $-functions. As
concerns the symmetric modes, with the increase of the distance between the $%
\delta $-functions ($2\xi $), they transform into two-soliton bound states
which never regain the stability they had prior to the bifurcation.
Specifically, near the OL minimum points, where the fundamental soliton is
stable (see Fig.~\ref{NvsDeltax_2Delta_SIG_SymAsym}), eigenvalues accounting
for the instability of the bound state decrease as its two constituents are
pulled farther apart with the increase of $\xi $ (but still, never fall
exactly to zero).

\begin{figure}[tbp]
\subfigure[]{%
\includegraphics[width=1.8in]{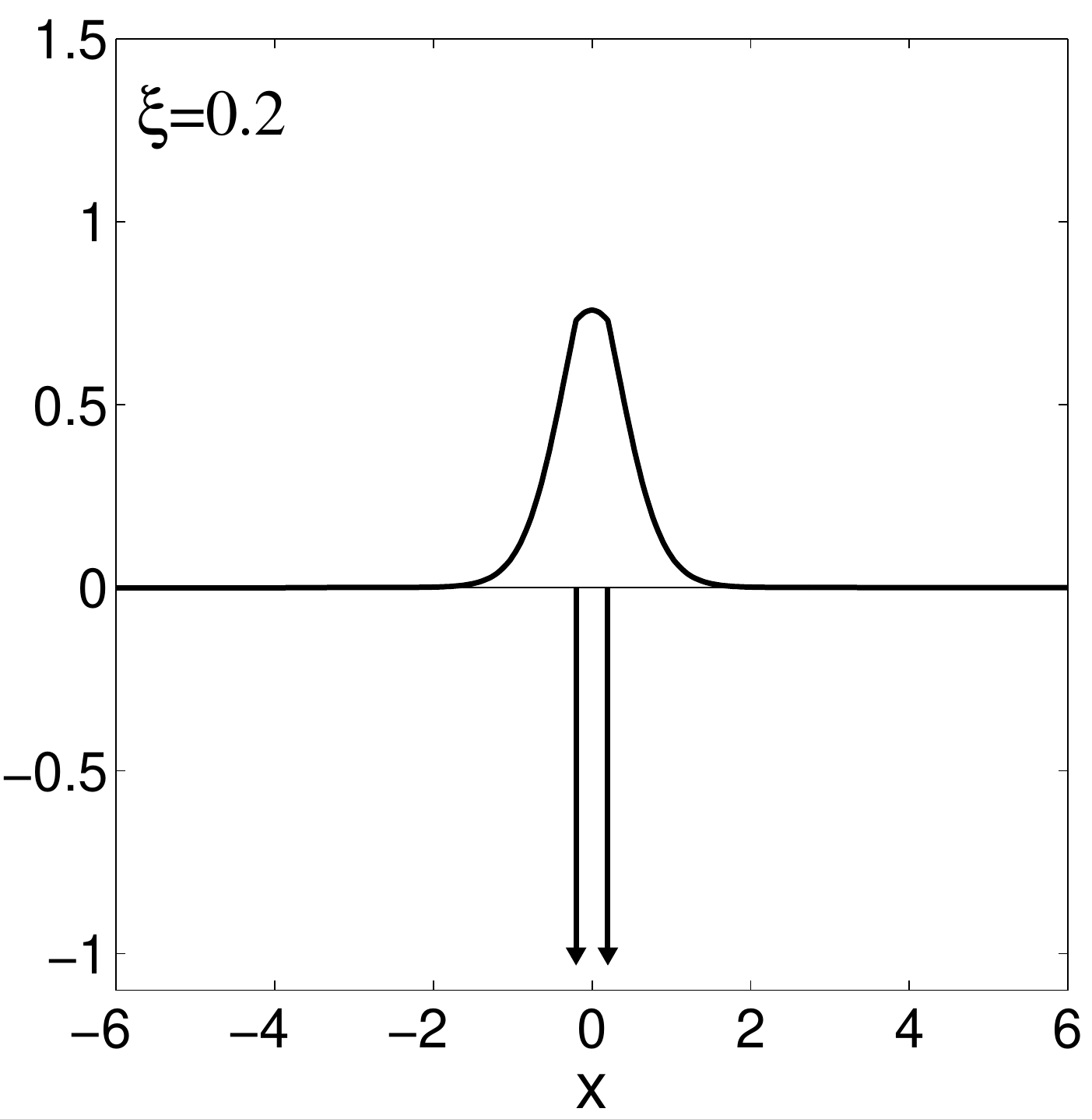}%
\label{Profile_2Delta_deltax02_mum4_sym}} \subfigure[]{%
\includegraphics[width=1.8in]{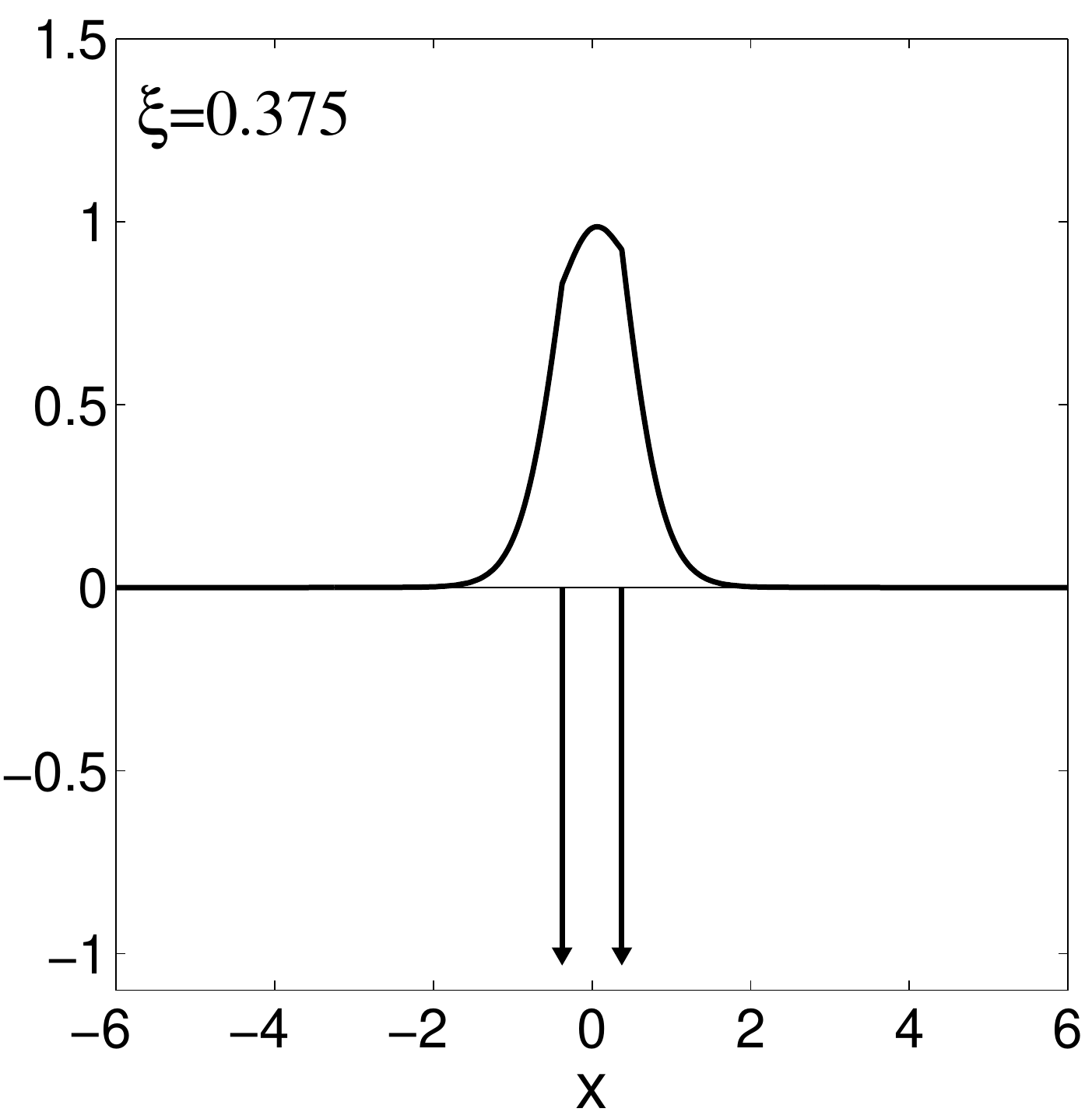}%
\label{Profile_2Delta_deltax0375_mum4_asym}} \subfigure[]{%
\includegraphics[width=1.8in]{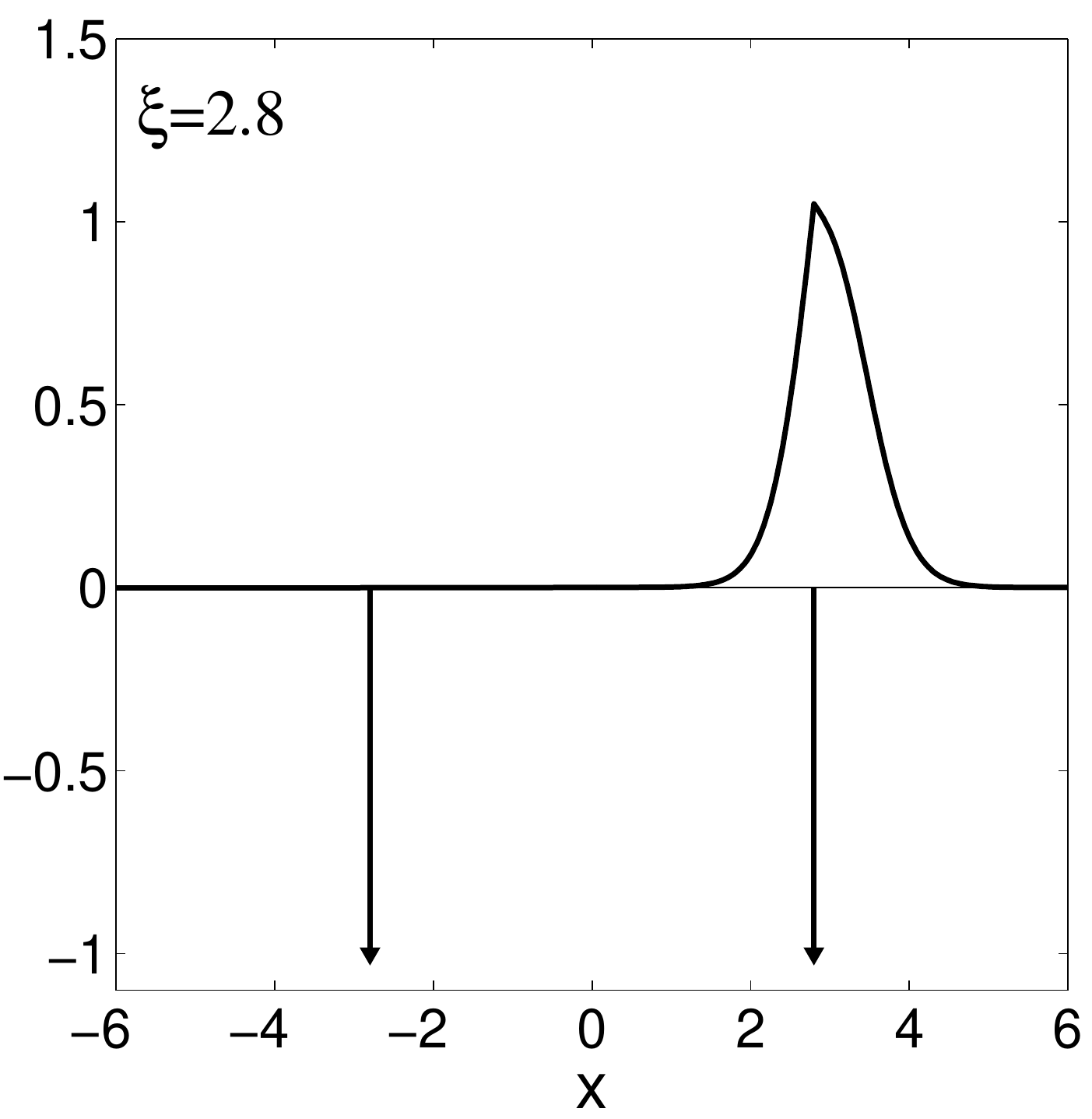}%
\label{Profile_2Delta_deltax28_mum4_asym}} \\ 
\subfigure[]{%
\includegraphics[width=1.8in]{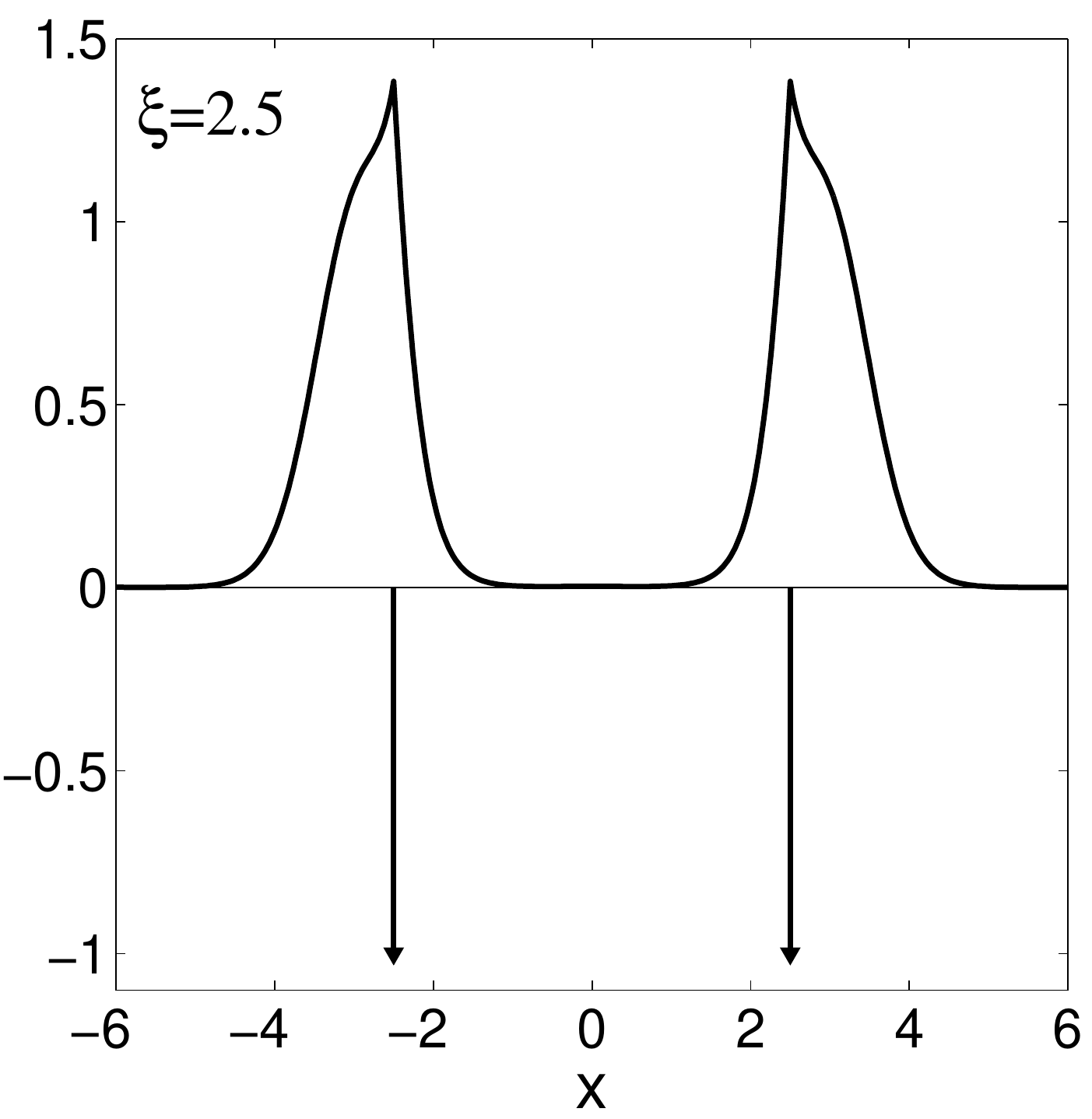}%
\label{Profile_2Delta_deltax25_mum4_sym}} \subfigure[]{%
\includegraphics[width=1.8in]{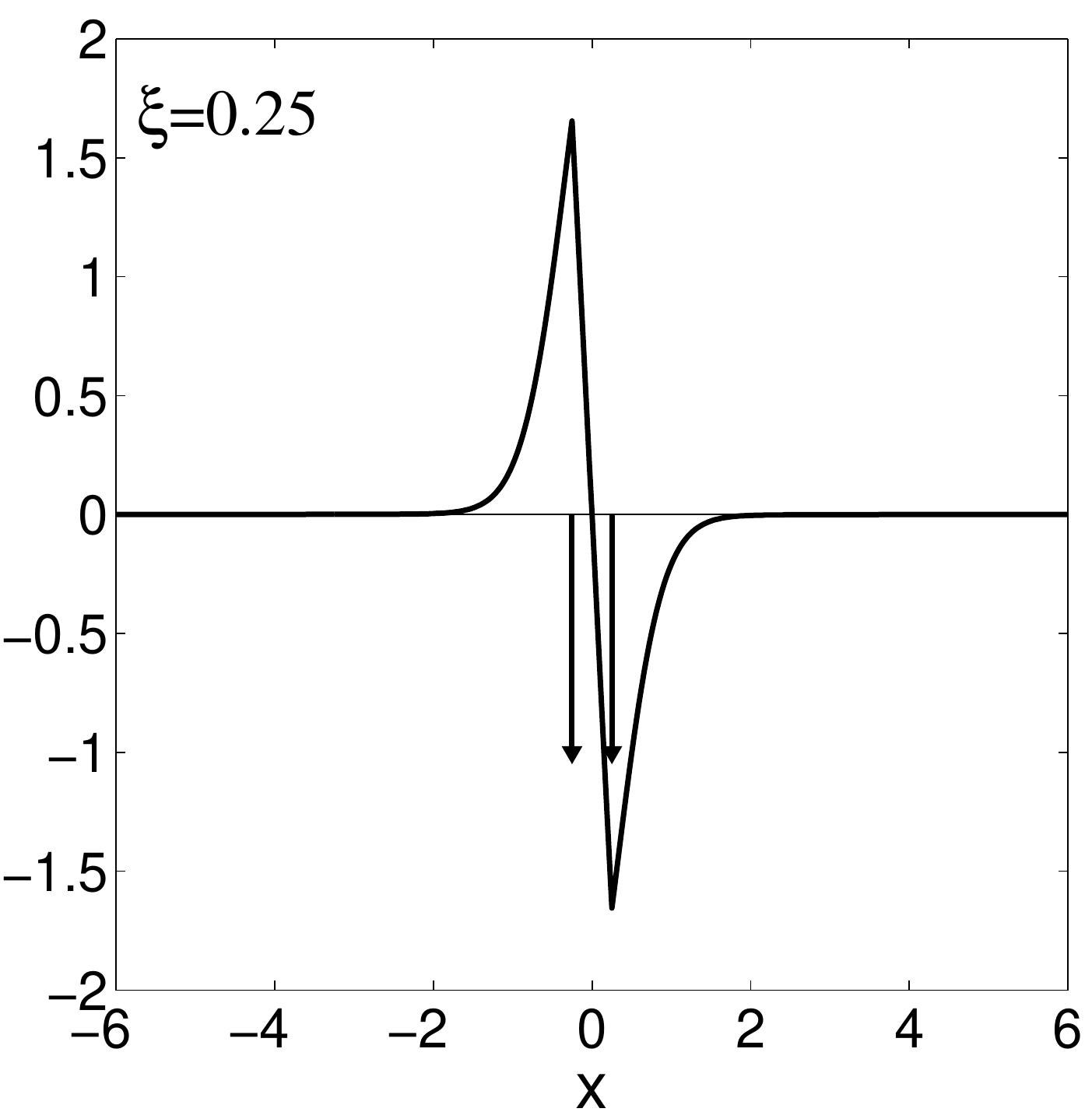}%
\label{Profile_2Delta_deltax025_mum4_antisym}} \subfigure[]{%
\includegraphics[width=1.8in]{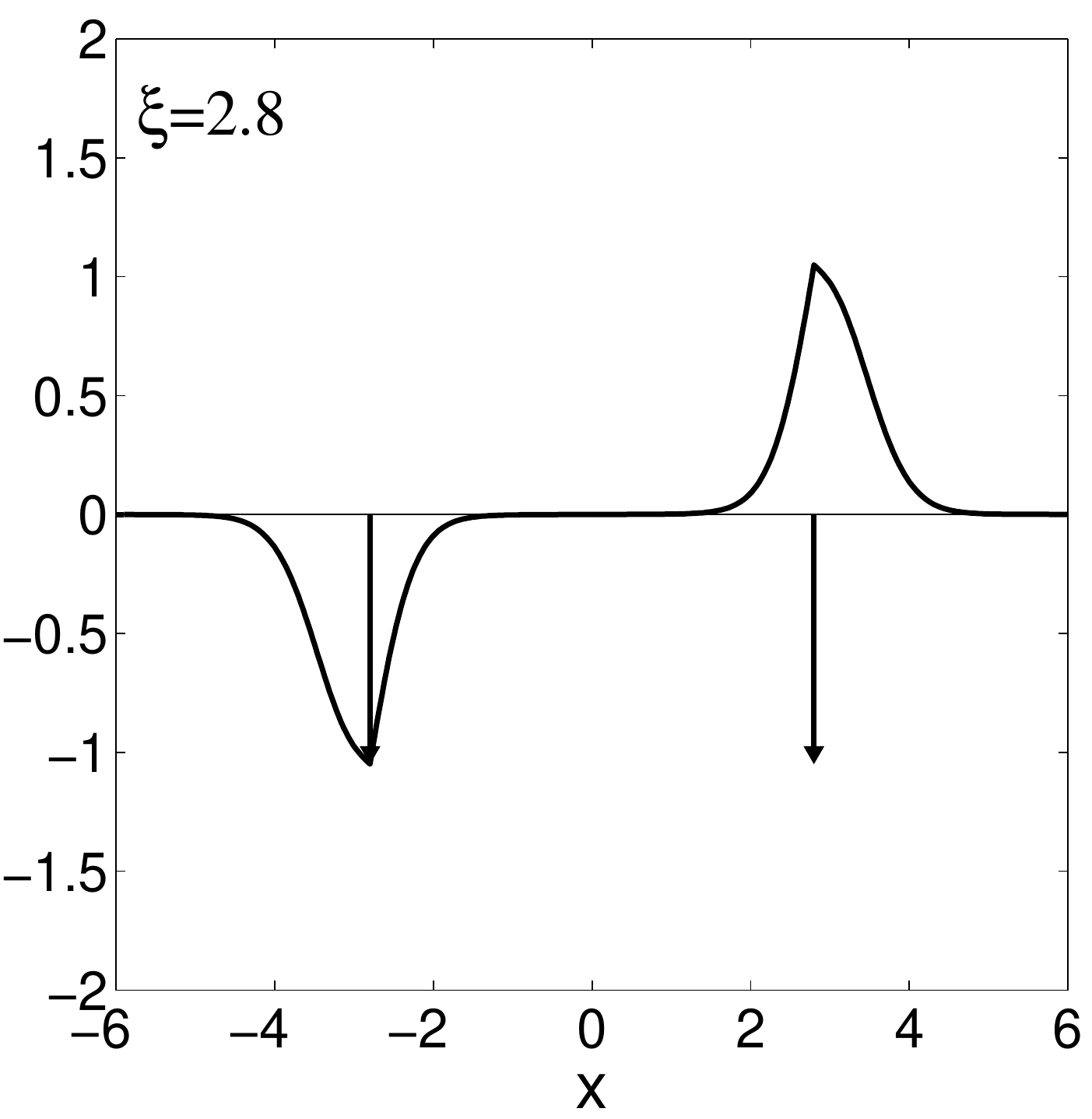}%
\label{Profile_2Delta_deltax28_mum4_antisym}}
\caption{Typical profiles of modes generated by two attractive $\protect%
\delta $-functions in the semi-infinite gap. Panels (a)-(d) correspond to
the points marked by circles in Fig.~\protect\ref%
{NvsDeltax_2Delta_SIG_SymAsym}. The arrows indicate the position of the $%
\protect\delta $-functions. Stable symmetric (a) and asymmetric (b)-(c)
solitons are demonstrated, as well as an unstable symmetric bound state (d).
In addition, panels (e) and (f) show, respectively, an unstable
antisymmetric soliton, and an antisymmetric bound state which is unstable
with respect to asymmetric perturbations.}
\label{Profile_2Delta_mum4_epsp5}
\end{figure}

As in the two-$\delta $-functions model without the lattice ($\varepsilon =0$%
) \cite{DeltaSSB}, antisymmetric states exist too and are unstable at small
values of $\xi $. As $\xi $ increases, they develop into antisymmetric
two-soliton bound states, for which unstable eigenmodes could not be found
(featuring instead zero eigenvalues) exactly in the region where the stable
fundamental soliton is supported by the single $\delta $-function. Direct
simulations have shown that, while such bound states are stable against
symmetric disturbances, asymmetric perturbations break them into mutually
incoherent fundamental solitons. Figure~\ref{Profile_2Delta_mum4_epsp5}
presents several examples of the solitons of all the aforementioned types,
in the case corresponding to Fig.~\ref{NvsDeltax_2Delta_SIG_SymAsym}.

\begin{figure}[tbp]
\subfigure[]{%
\includegraphics[width=1.8in]{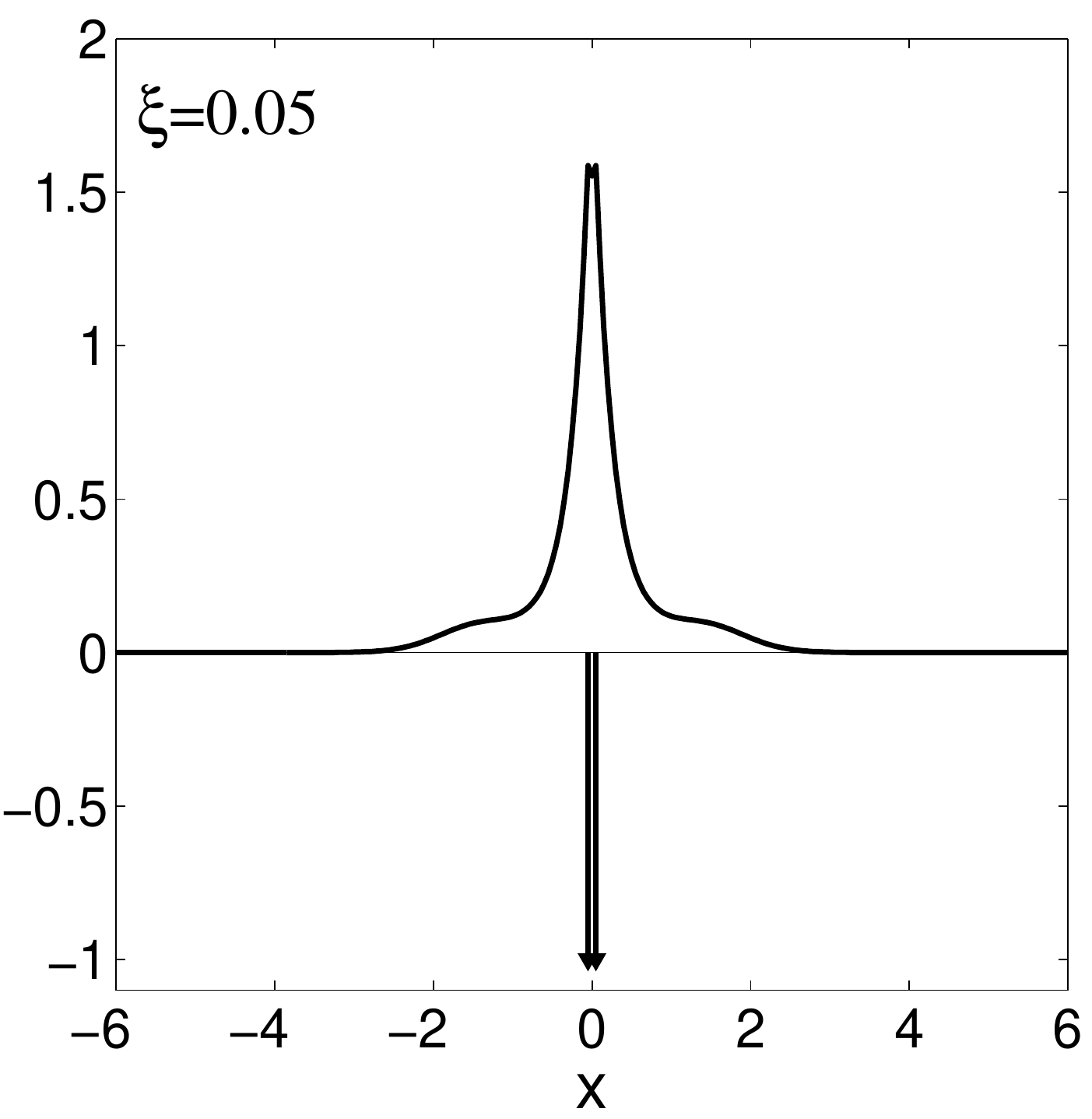}%
\label{Profile_2Delta_deltax005_mum4_epsm5_sym}} \subfigure[]{%
\includegraphics[width=1.8in]{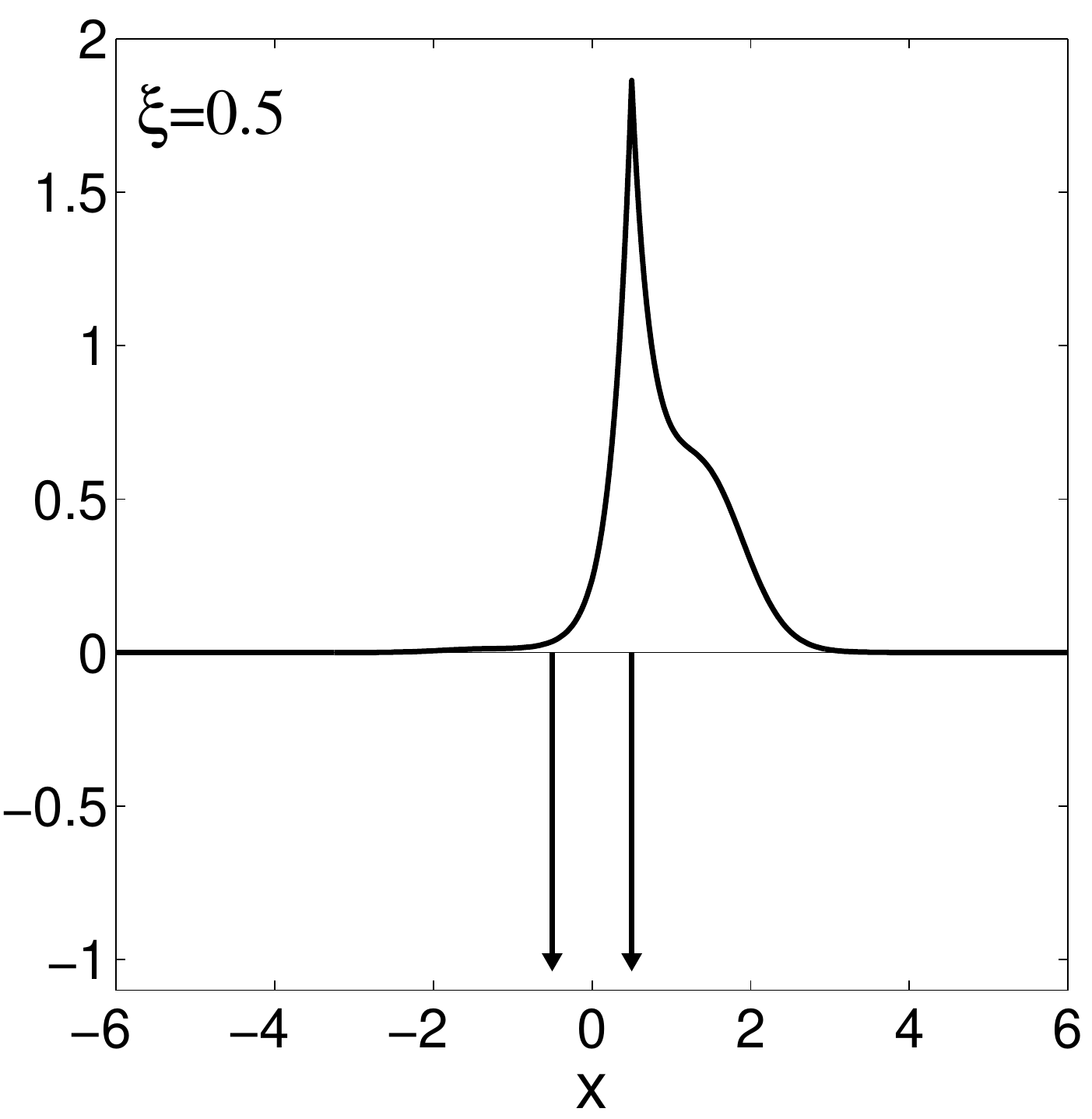}%
\label{Profile_2Delta_deltax05_mum4_epsm5_asym}} \subfigure[]{%
\includegraphics[width=1.8in]{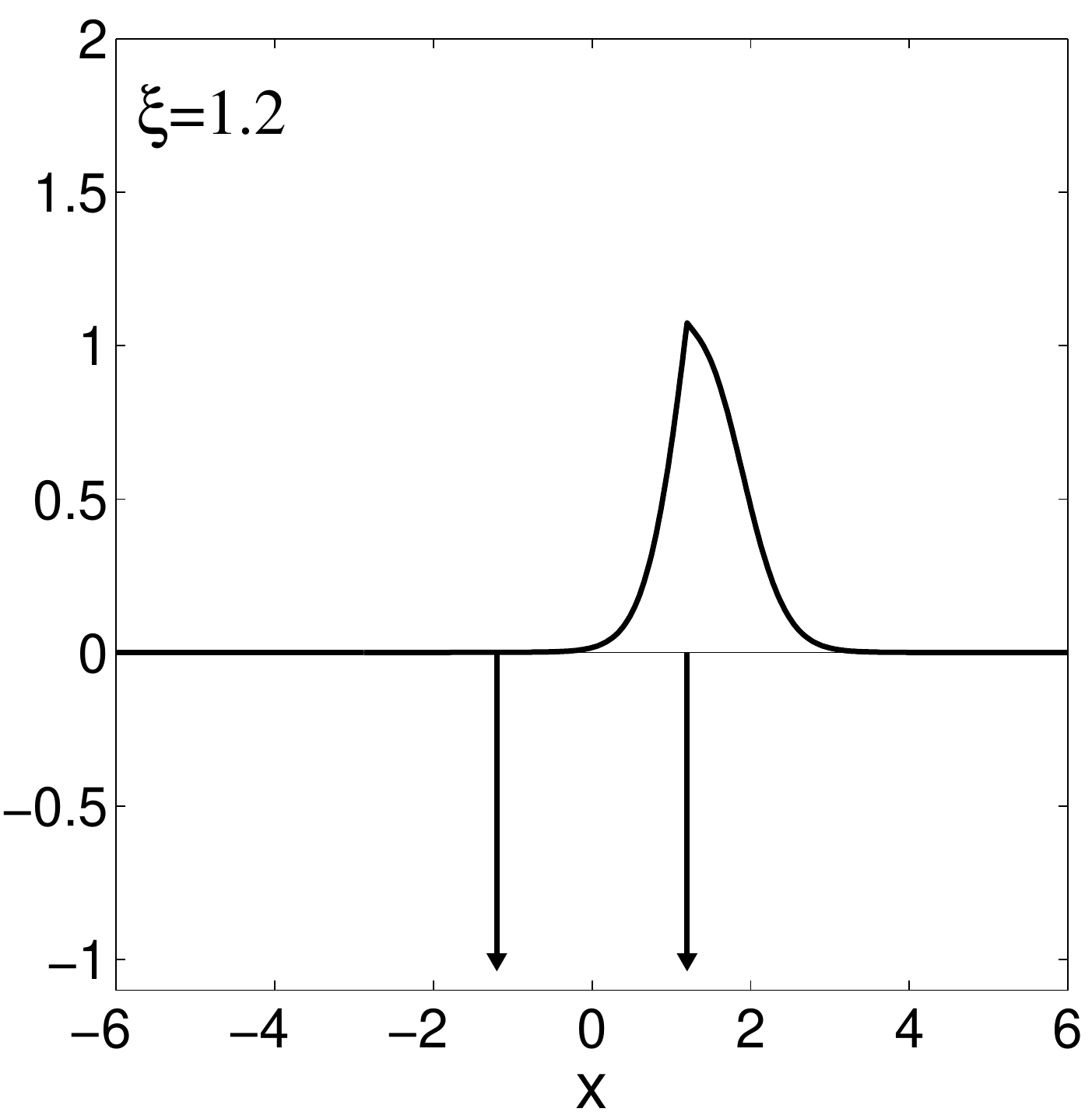}%
\label{Profile_2Delta_deltax12_mum4_epsm5_asym}} \\ 
\subfigure[]{%
\includegraphics[width=1.8in]{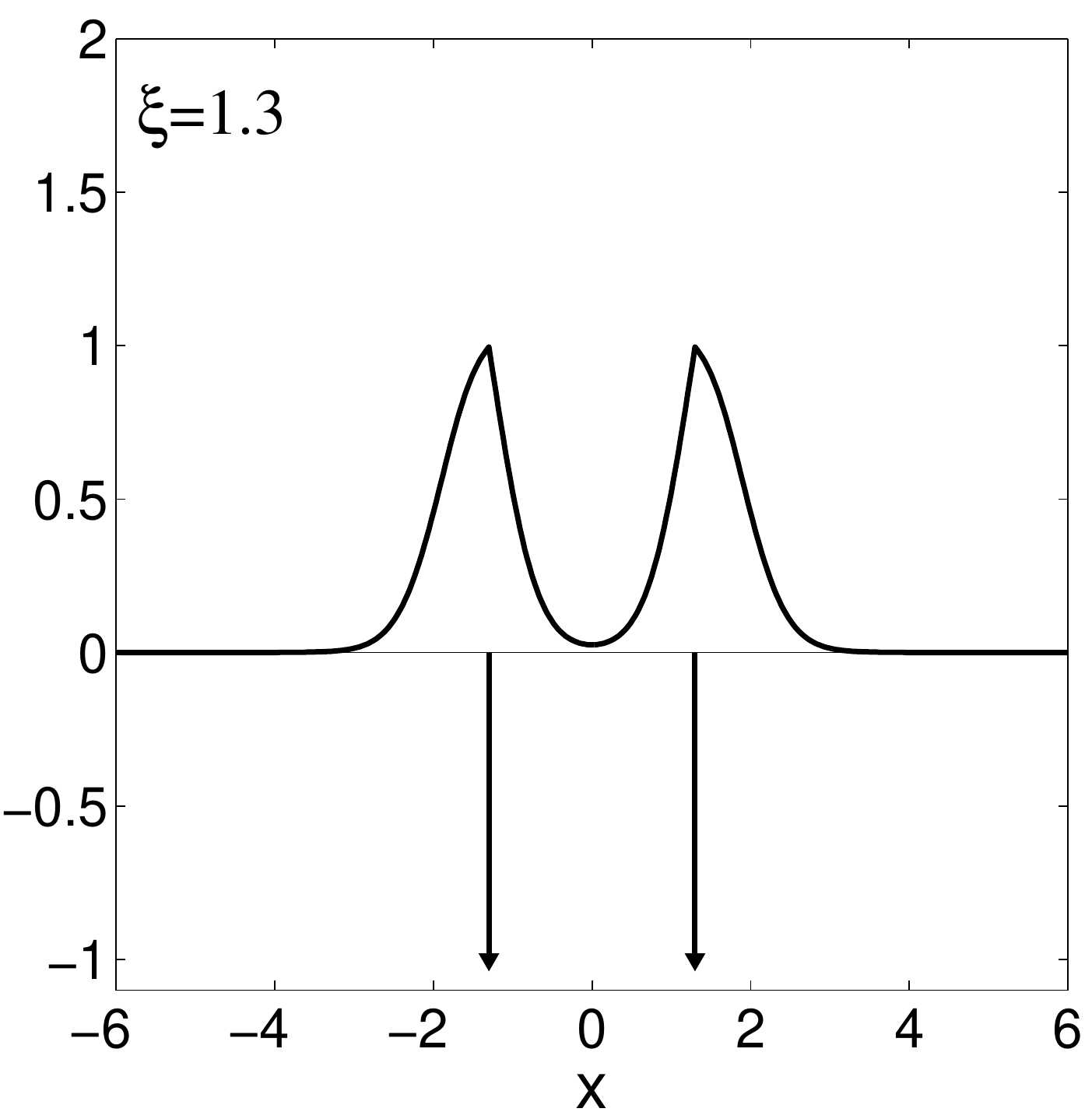}%
\label{Profile_2Delta_deltax13_mum4_epsm5_sym}}
\subfigure[]{\includegraphics[width=1.8in]
{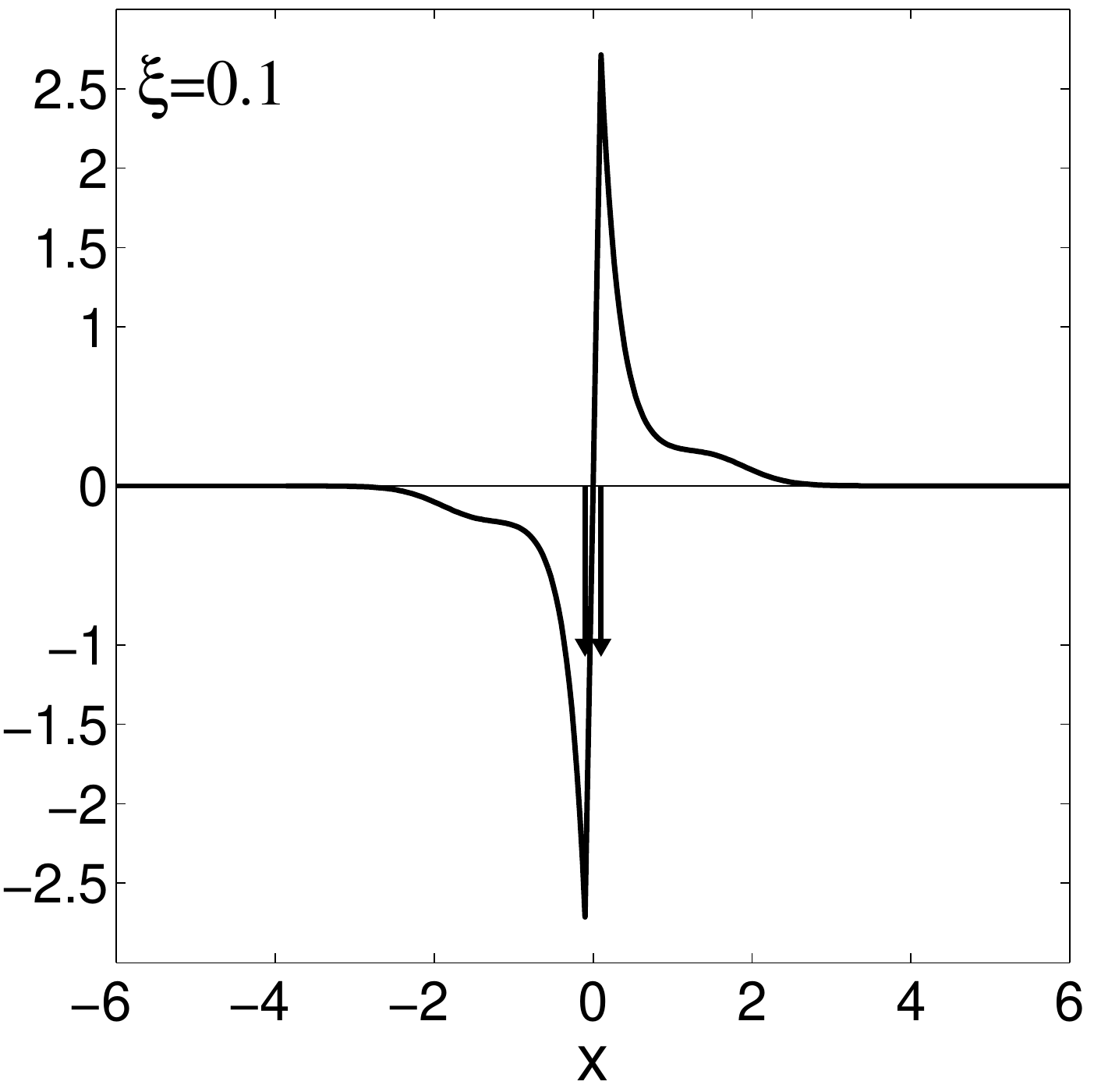}\label{Profile_2Delta_deltax01_mum4_epsm5_antisym}}
\subfigure[]{\includegraphics[width=1.8in]{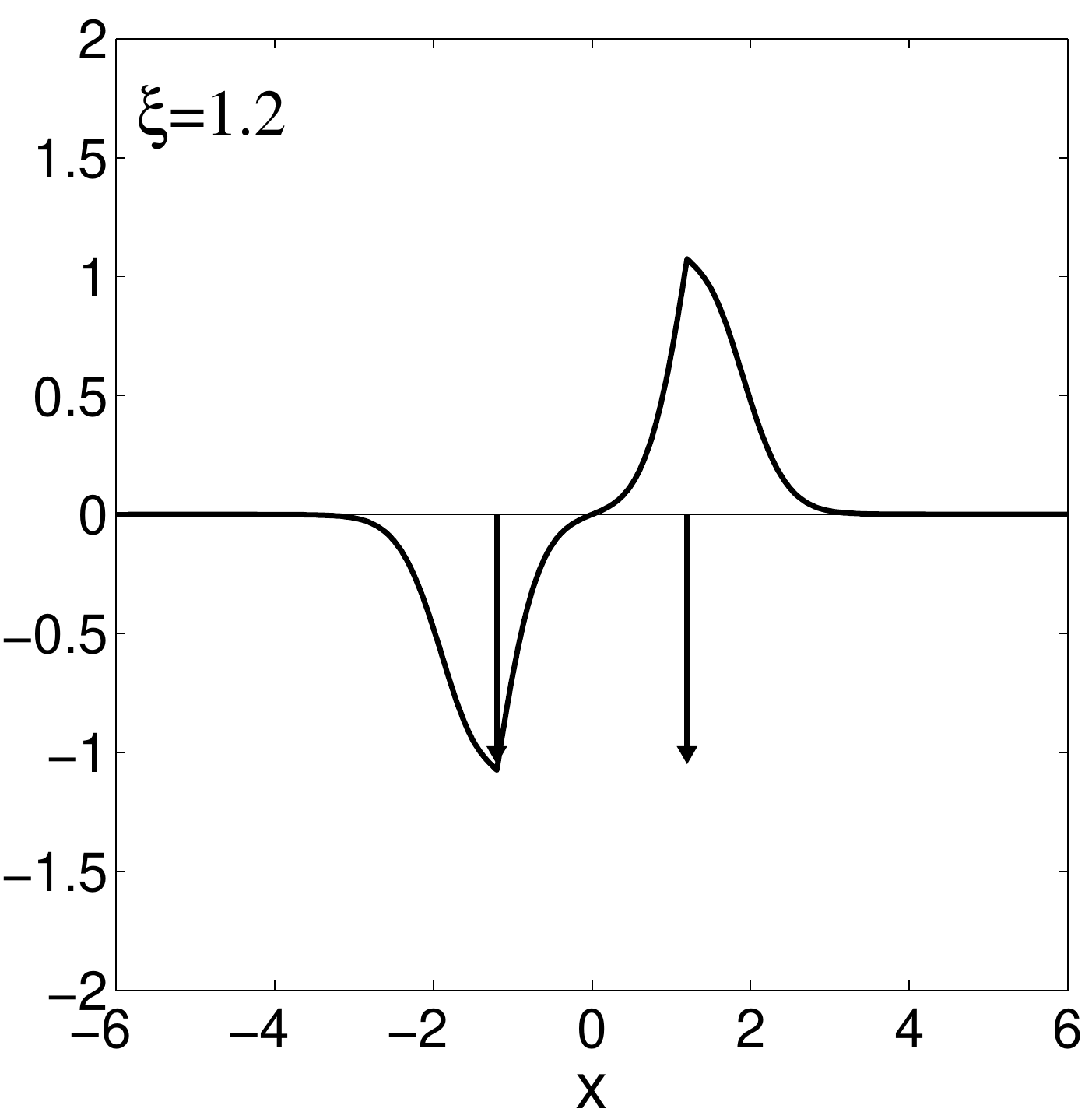}
\label{Profile_2Delta_deltax12_mum4_epsm5_antisym}}
\caption{Examples of solitons corresponding to the points marked by circles
in Fig.~\protect\ref{NvsDeltax_2Delta_SIG_SymAsym_epsm}. Panels (a) and (b)
demonstrate unstable symmetric and asymmetric solitons. An example for a
stable asymmetric soliton is displayed in (c). Panel (d) shows an unstable
symmetric bound state. In addition, examples of an antisymmetric soliton,
that features the strong local instability, and antisymmetric bound state,
which is unstable against asymmetric perturbations, are shown in panels (e)
and (f), respectively.}
\label{Profile_2Delta_mum4_epsm5}
\end{figure}

Similar analysis was carried out for two attractive $\delta $-functions
placed on both sides of a local maximum of the periodic potential, which
implies $\varepsilon <0$ in Eq. (\ref{Two_Dlata_NLSE}). Figures~\ref%
{NvsDeltax_2Delta_SIG_SymAsym_epsm} and~\ref{Profile_2Delta_mum4_epsm5}
present a typical example of the results in the $(\xi ,N)$ plane, for $%
\varepsilon =-5$ and $\mu =-4$. In this case, both the symmetric and
asymmetric solitons are unstable at small values of $\xi $. In other
aspects, the results resemble those reported above for the case of two $%
\delta $-functions placed symmetrically around a potential minimum.

It is relevant to compare these results with those reported in Ref. \cite%
{DeltaSSB} for two attractive $\delta $-functions in the absence of the
periodic potential ($\varepsilon =0$). In that case, exact analytical
solutions are available for the pinned states of all the types---symmetric,
asymmetric, and antisymmetric. The symmetric states are stable before the
symmetry-breaking bifurcation, and unstable after it, terminating at final $%
N $. The bifurcation is of an ``extreme subcritical" type, with branches of
the asymmetric states going backwards and never turning forward, hence they
are completely unstable. In fact, these results, obtained with the ideal $%
\delta $-functions, are degenerate. The numerical analysis with regularized $%
\delta $-functions lifts the degeneracy, demonstrating that the branches of
the asymmetric modes eventually turn forward, stabilizing themselves.
Simultaneously, the family of the symmetric modes (unstable past the
bifurcation point) extends to $N\rightarrow \infty $. Antisymmetric states
are completely unstable in that model (they become stable if the regularized
$\delta $-functions are made broad enough). The comparison with the present
findings suggests that the addition of the periodic potential also lifts the
degeneracy of the system, even if the $\delta $-functions are kept in the
ideal form. In fact, this is similar to how the inclusion of the weak
periodic potential lifts the degeneracy of the soliton family (\ref{simple})
supported by the single attractive $\delta $-function and stabilizes the
family, see Eqs. (\ref{phi1}) and (\ref{N1}).

\subsection{Solitons in the first finite bandgap}

\begin{figure}[tbp]
\subfigure[]{%
\includegraphics[width=4.8in]{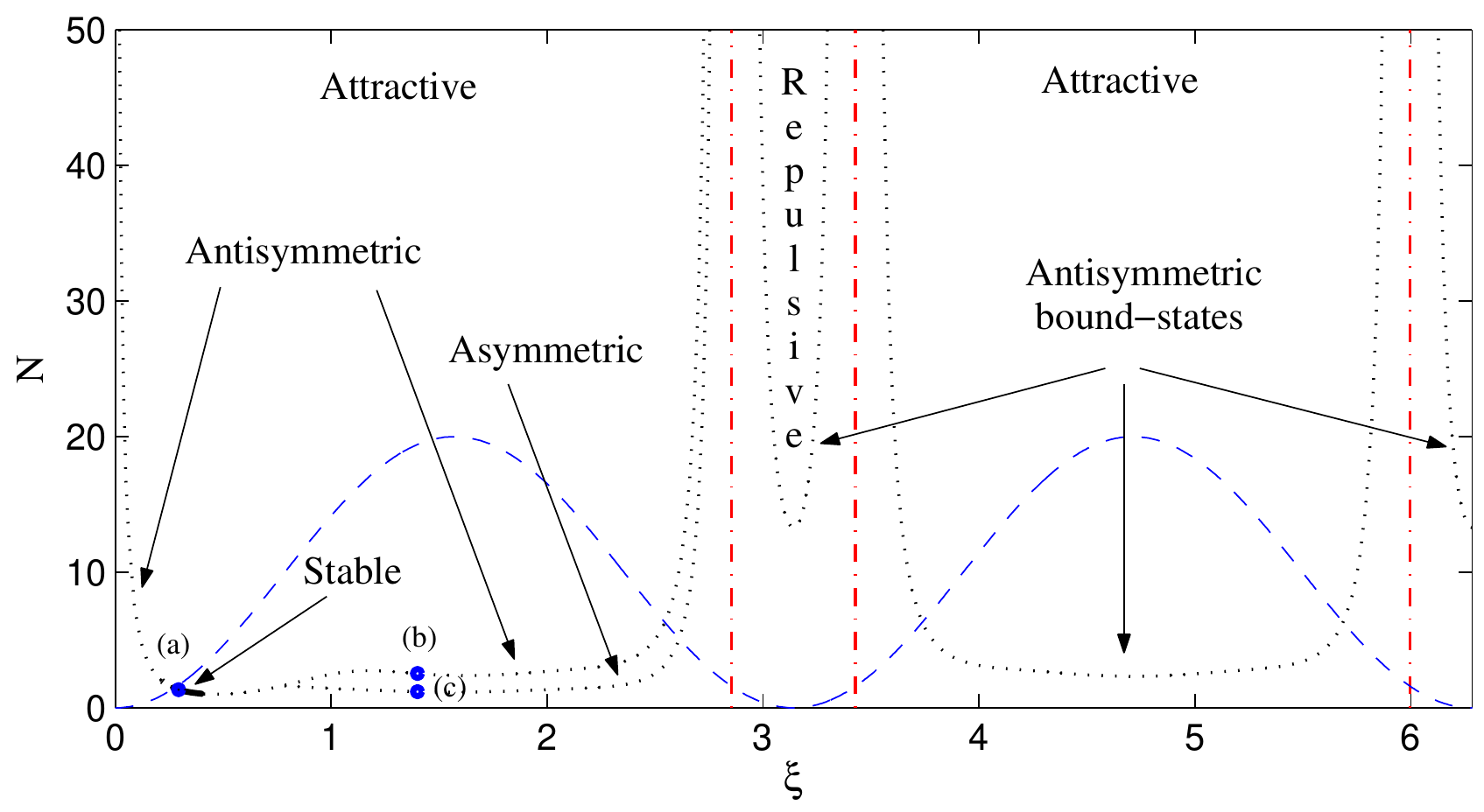}%
\label{NvsDeltax_2Delta_FFG_AntiAsymBoundStates}} \\ 
\subfigure[]{\includegraphics[width=2.4in]{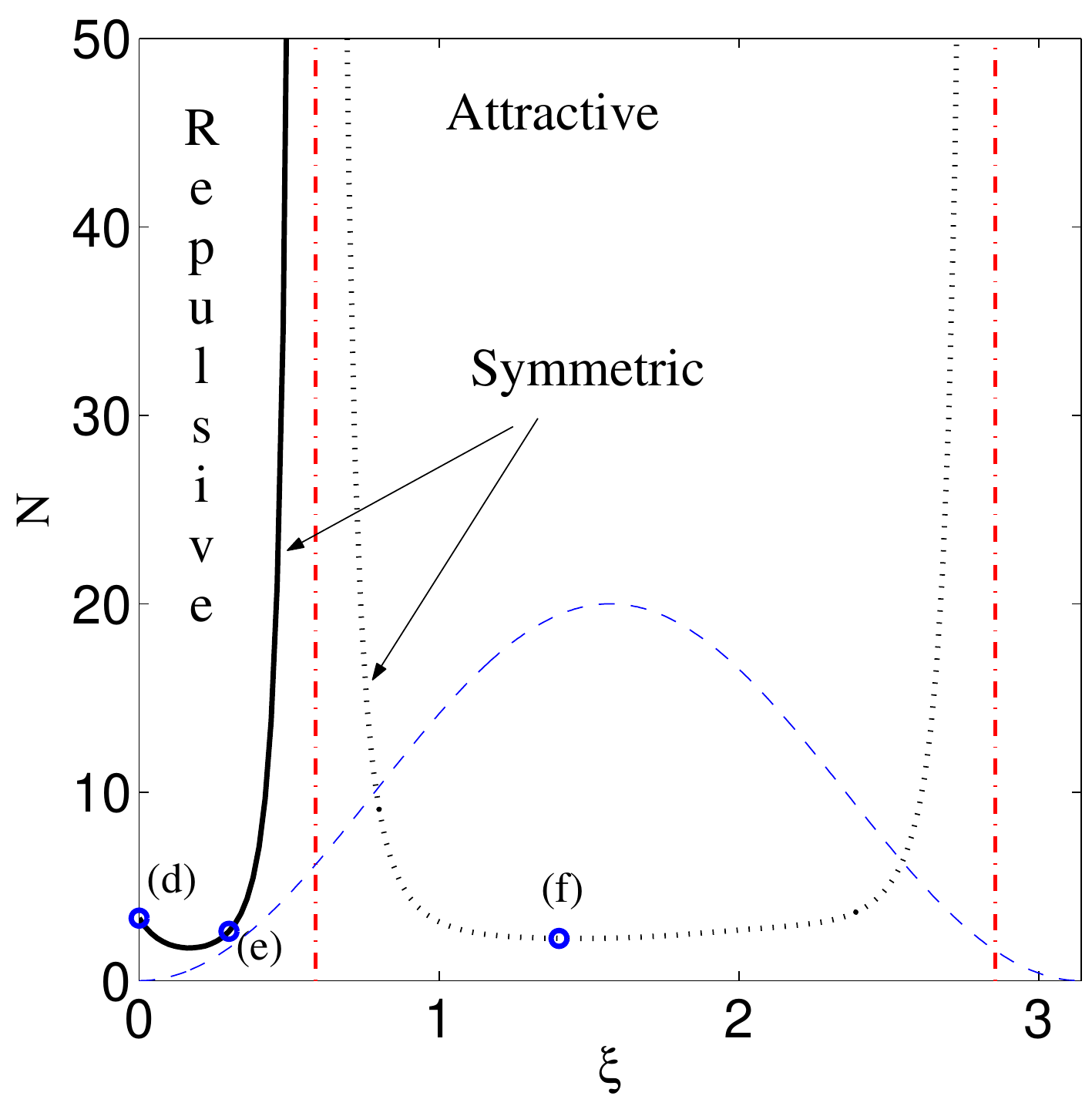}%
\label{NvsDeltax_2Delta_FFG_Sym}}\quad \subfigure[]{%
\includegraphics[width=2.4in]{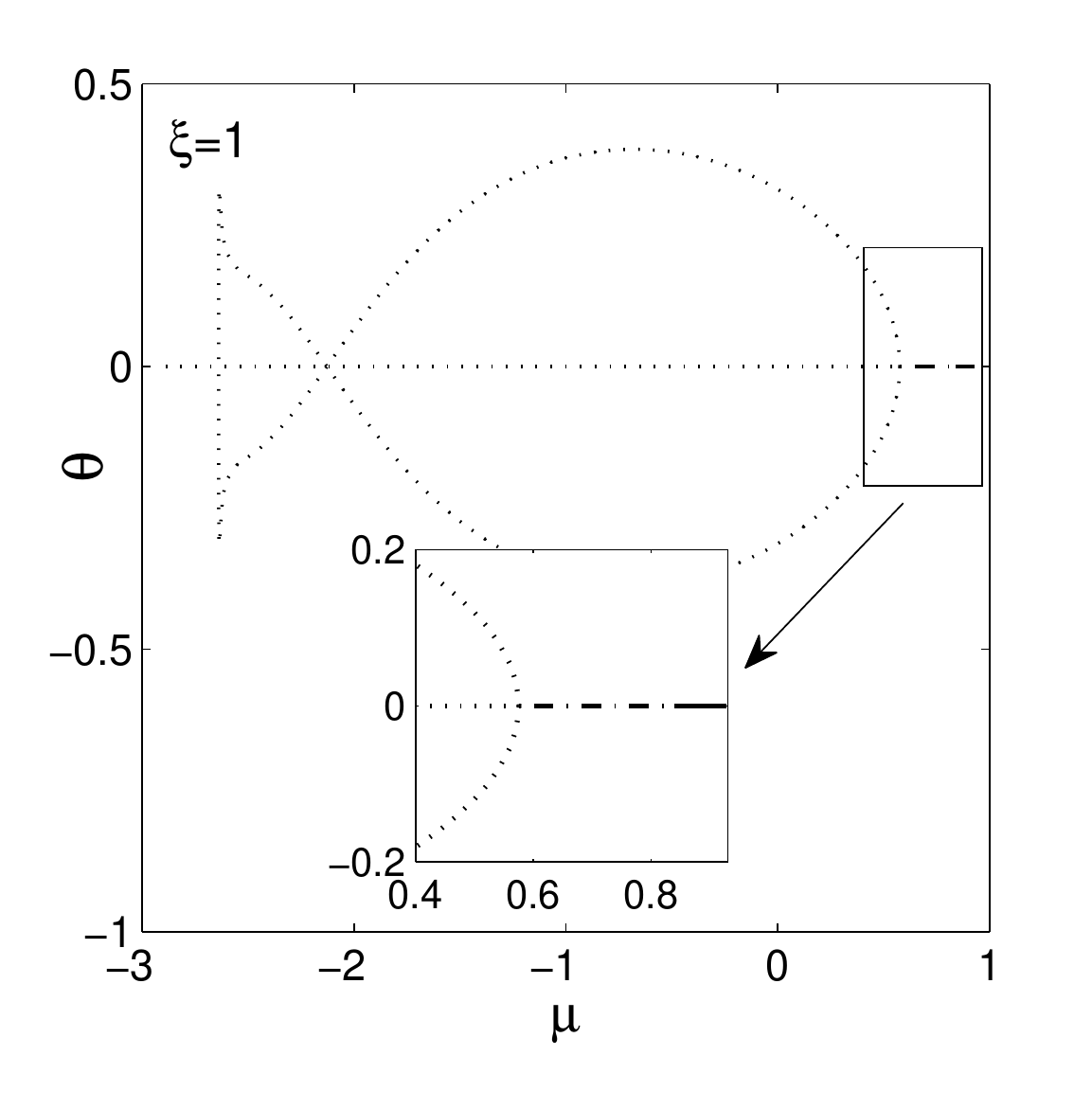}%
\label{ThetavsMu_deltax1_FFG}}
\caption{(Color online) (a)-(b) The soliton's norm versus half the distance
between the two $\protect\delta $-functions, $\protect\xi $, in the first
finite bandgap, for $\protect\varepsilon =5$ and $\protect\mu =-1$. In panel
(a), the asymmetric branch bifurcates from the antisymmetric one, in the
case of the attractive nonlinearity. At larger values of $\protect\xi $, the
soliton branches correspond to unstable antisymmetric bound states of two
solitons, for either repulsive or attractive nonlinearity. (b) Branches of
symmetric states are shown for both the repulsive and attractive
nonlinearities. Not shown in (b) are the unstable symmetric bound states,
for larger $\protect\xi $, as this part of the diagram is virtually
identical to the one in panel (a), for the antisymmetric bound states. The
dashed-dotted vertical lines represent the existence thresholds, $\protect%
\pi \pm \protect\xi _{\mathrm{thr}}=2.853,3.43$ (periodic with period $%
\protect\pi $) and $\protect\xi _{\mathrm{thr}}^{(2)}=0.588$ [only in (b)].
Circles indicate solitons whose profiles are displayed in Fig.~\protect\ref%
{Profile_2Delta_mum1}. (c) The bifurcation diagram for the antisymmetric and
asymmetric modes, in the $(\protect\mu ,\protect\theta )$ plane, at $\protect%
\varepsilon =5$ and $\protect\xi =1$.}
\label{NvsDeltax_2Delta_FFG}
\end{figure}

Similar to the case of the single $\delta $-function, GSs in the first
finite bandgap can be found for both the attractive and repulsive
nonlinearities. We again start by considering the pair of $\delta $%
-functions placed around a minimum of the potential. For the case of two
attractive $\delta $-functions, antisymmetric solitons exist in the region
of $0<\xi <\pi -\xi _{\mathrm{thr}}$, where $\xi _{\mathrm{thr}}$ is the
same threshold as in the model with the single $\delta $-function. Unlike
what was observed in the semi-infinite gap, in the first finite bandgap the
asymmetric solitons bifurcate from the antisymmetric branch, see an example
in Fig.~\ref{ThetavsMu_deltax1_FFG} for $\varepsilon =5$ and $\xi =1$. In
particular, this setting features a closed bifurcation loop, where both the
direct and the reverse bifurcations are of the supercritical type. In the
case presented in Fig.~\ref{ThetavsMu_deltax1_FFG}, there is a small region
of stable antisymmetric solitons, obtained for high values of $\mu $, near
the upper edge of the gap. Decreasing $\mu $, the soliton is destabilized by
an oscillatory instability, both the antisymmetric and asymmetric branches
being strongly unstable past the bifurcation point. A typical example,
plotted in the $(\xi ,N)$ plane for $\varepsilon =5$ and $\mu =-1$, is
displayed in Figs.~\ref{NvsDeltax_2Delta_FFG_AntiAsymBoundStates}-%
\subref{NvsDeltax_2Delta_FFG_Sym}, where a small section of the
antisymmetric branch is stable. The stability analysis, carried out for
different values of $\varepsilon $ and $\mu $, indicates a somewhat larger
stability region for larger $\varepsilon $, and for $\mu $ taken closer to
the upper edge of the bandgap. The stability region gradually disappears at
smaller values of $\varepsilon $, or close to the lower edge of the bandgap.
A weak oscillatory instability also occurs in the present case. In
particular, in the situation corresponding to Fig.~\ref{NvsDeltax_2Delta_FFG}%
, a very small section (too small to be visible in Fig.~\ref%
{NvsDeltax_2Delta_FFG_AntiAsymBoundStates}) of the corresponding weakly
unstable antisymmetric solitons exists, starting from the edge of the stable
region and extending to slightly smaller values of $\xi $.

\begin{figure}[tbp]
\subfigure[]{%
\includegraphics[width=1.8in]{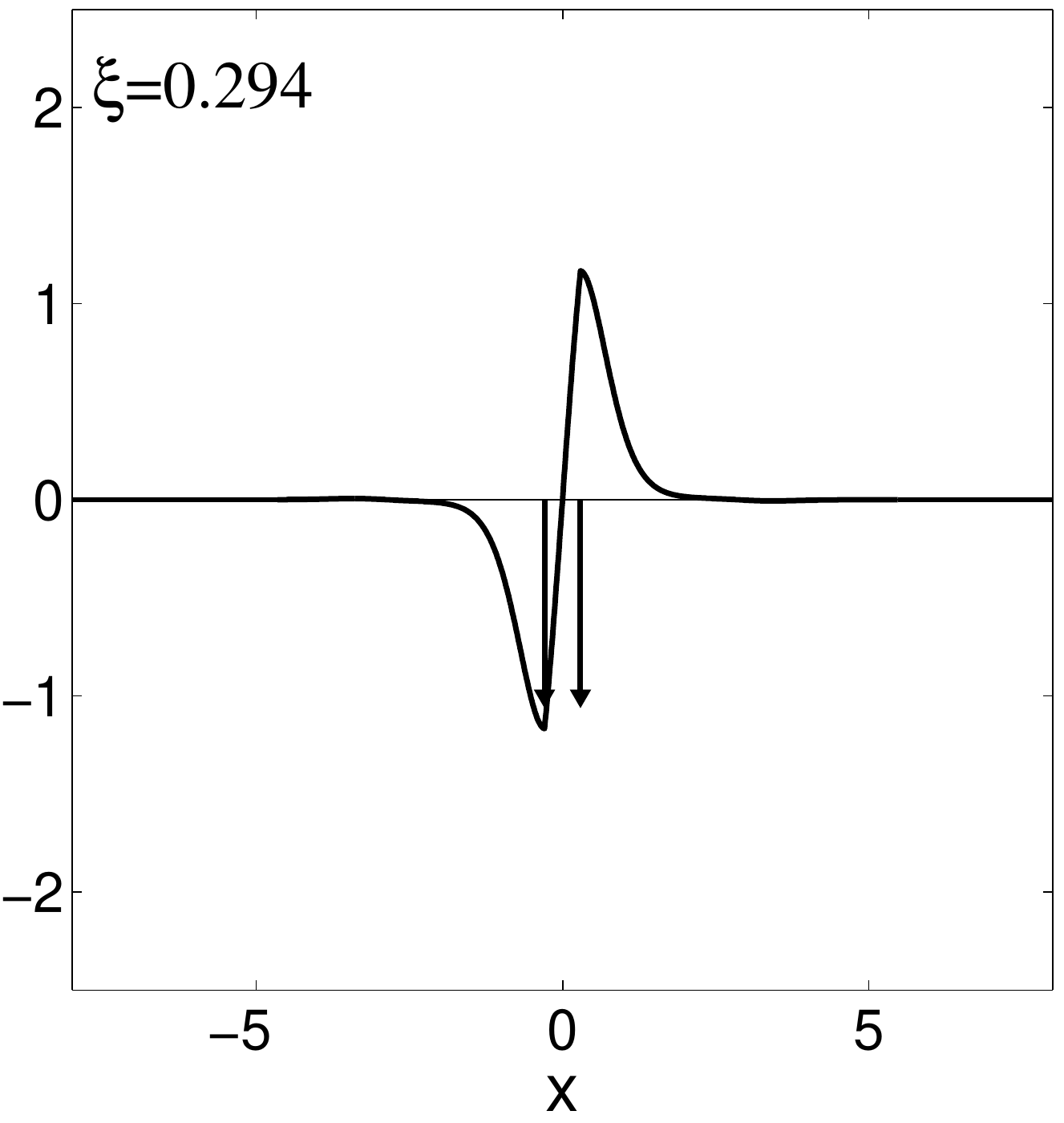}%
\label{Profile_2Delta_deltax0294_mum1_anti}} \subfigure[]{%
\includegraphics[width=1.8in]{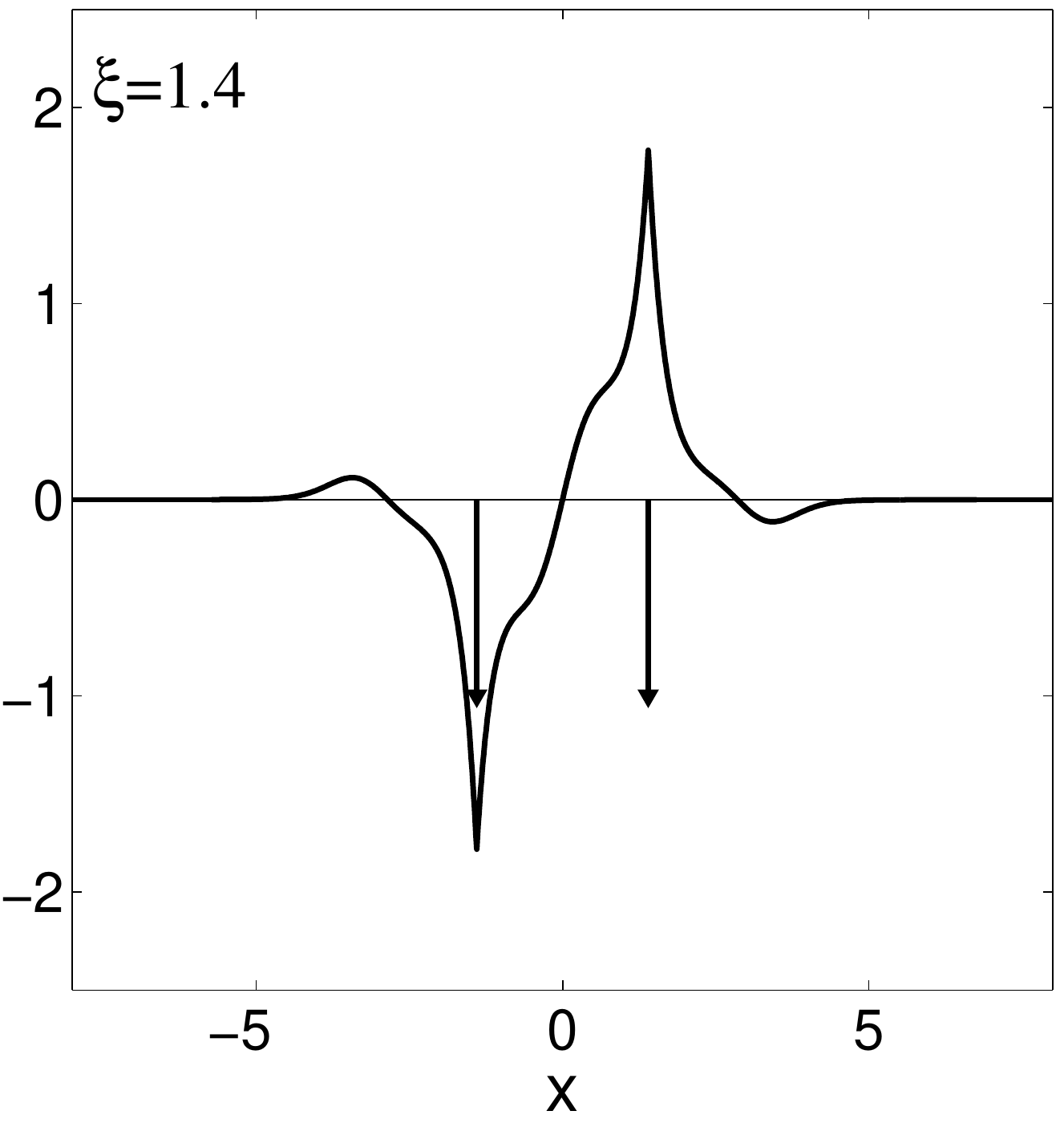}%
\label{Profile_2Delta_deltax14_mum1_anti}} \subfigure[]{%
\includegraphics[width=1.8in]{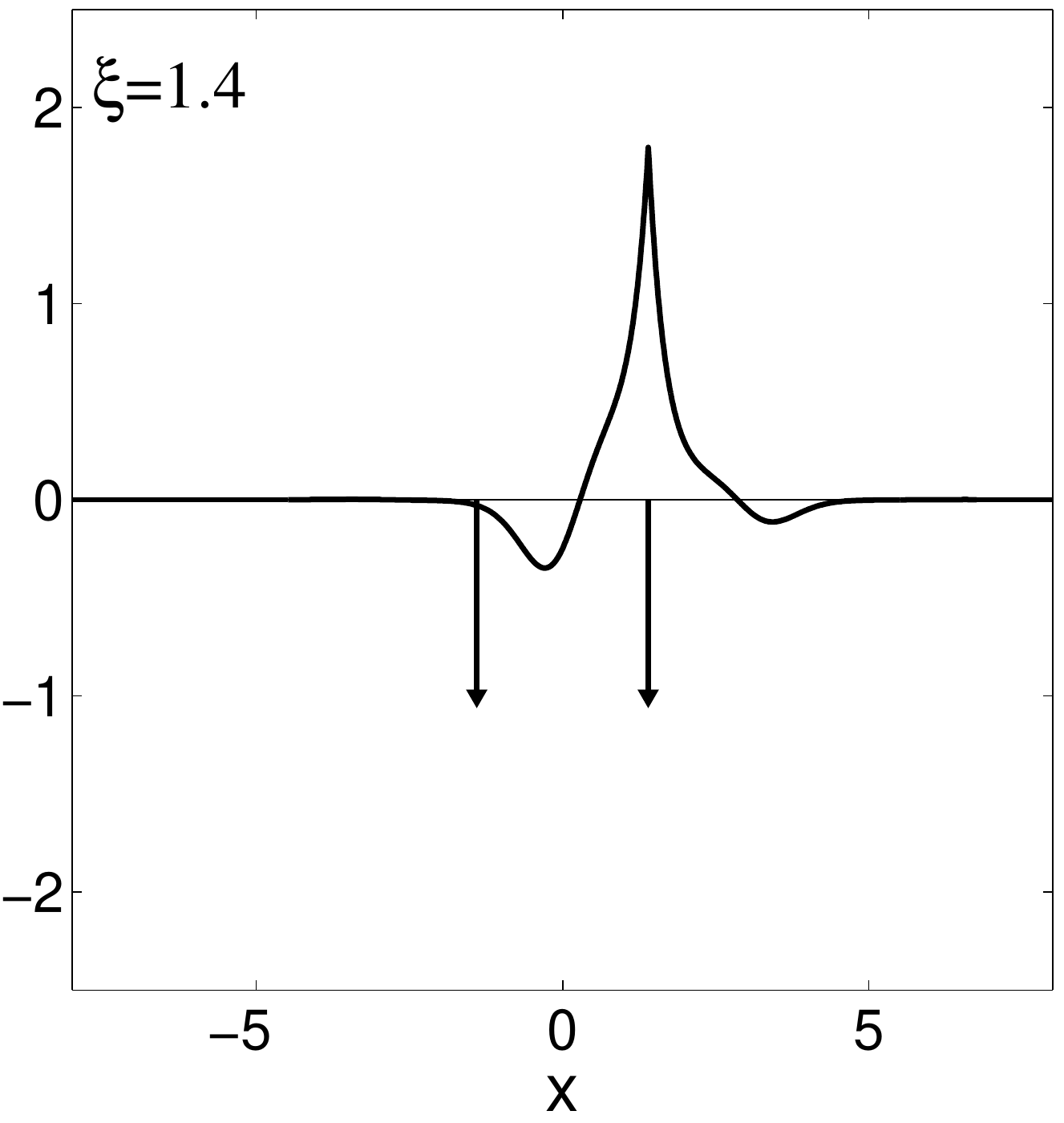}%
\label{Profile_2Delta_deltax14_mum1_asym}} \\ 
\subfigure[]{%
\includegraphics[width=1.8in]{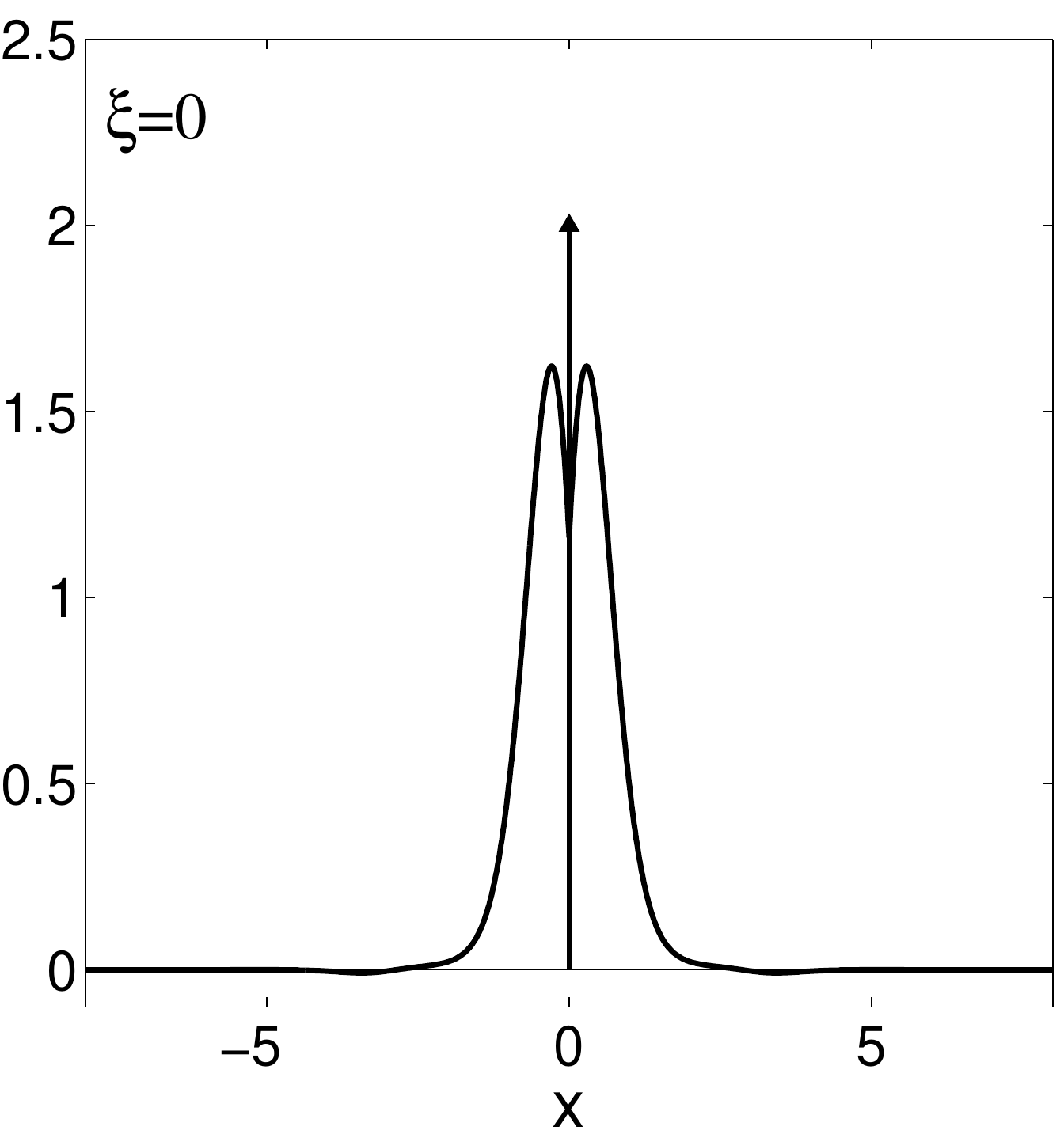}%
\label{Profile_2Delta_deltax0_mum1_sym}} \subfigure[]{%
\includegraphics[width=1.8in]{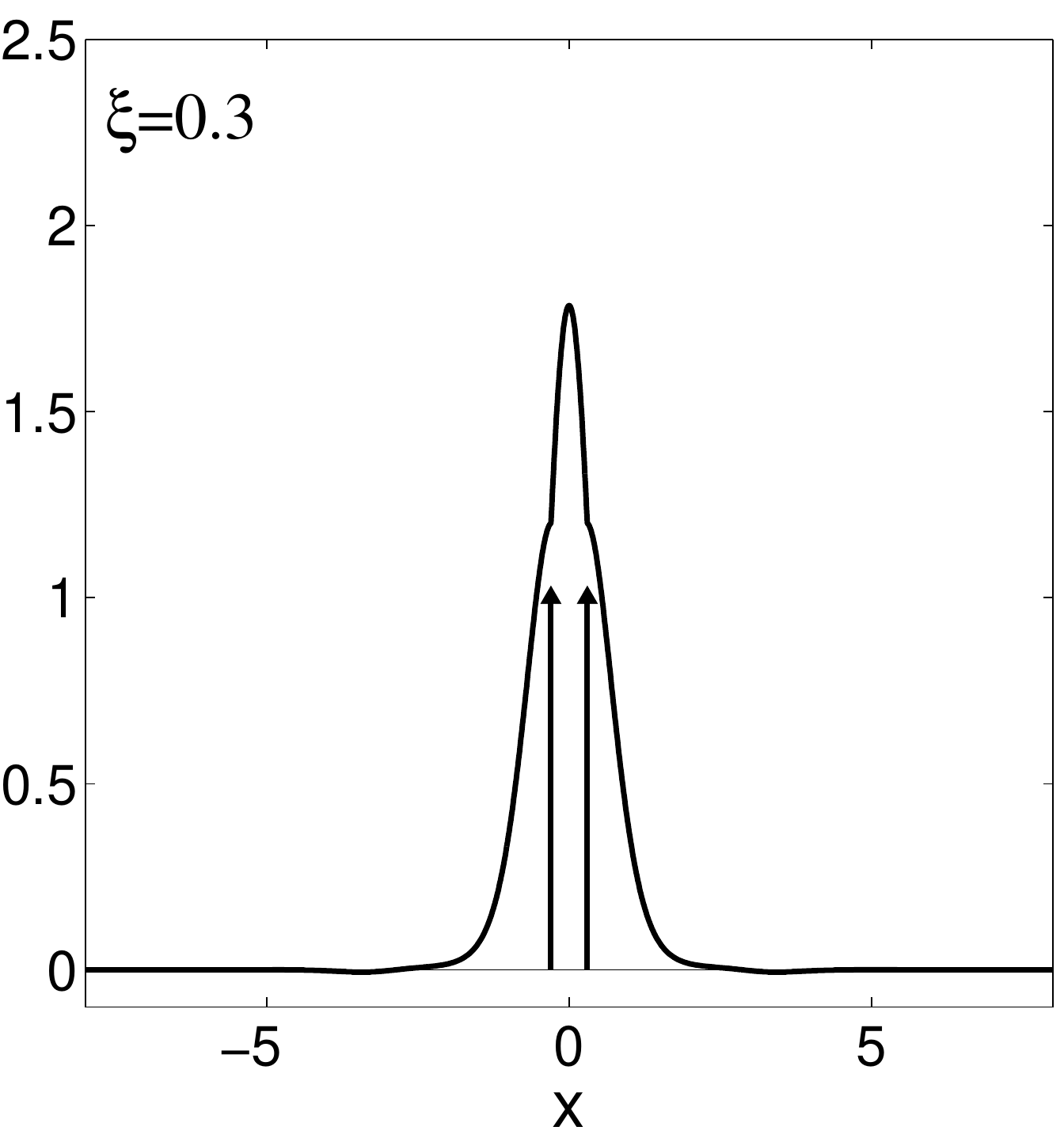}%
\label{Profile_2Delta_deltax03_mum1_sym}} \subfigure[]{%
\includegraphics[width=1.85in]{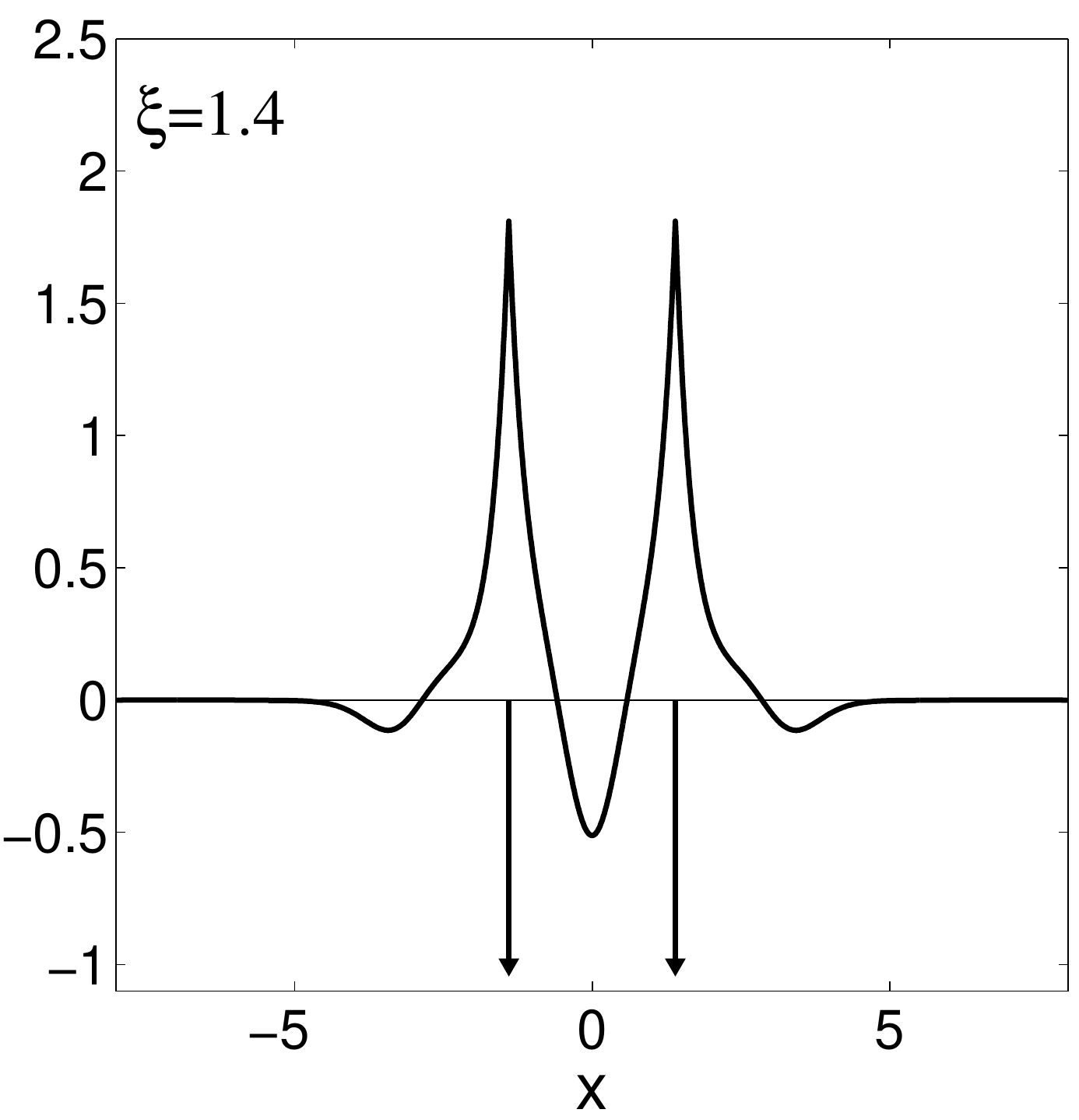}%
\label{Profile_2Delta_deltax14_mum1_sym}}
\caption{Examples of solitons generated by two $\protect\delta $-functions
in the first finite bandgap, which correspond to the marked points in Figs.
\protect\ref{NvsDeltax_2Delta_FFG_AntiAsymBoundStates},
\subref{NvsDeltax_2Delta_FFG_Sym}. As in Fig.~\protect\ref%
{Profile_1Delta_mum1}, the arrows indicate the position and sign of the $%
\protect\delta $-functions. Stable and unstable antisymmetric solitons, as
well as an unstable asymmetric one, are shown in (a), (b) and (c),
respectively, in the case of two attractive $\protect\delta $-functions.
Symmetric solitons, stable and unstable, are shown in panels (d),(e) and
(f), for the repulsive and attractive nonlinearity, respectively. }
\label{Profile_2Delta_mum1}
\end{figure}

In this setting, an additional threshold, $\xi _{\mathrm{thr}}^{(2)}$, was
found (which is not related to that in the single-$\delta $-function model, $%
\xi _{\mathrm{thr}}$, see Fig. \ref{Deltaxthresh_FFG}). The new threshold
serves as the upper boundary for \emph{always-stable} symmetric solitons
generated by a pair of closely placed \emph{repulsive} $\delta $-functions,
as shown in Fig.~\ref{NvsDeltax_2Delta_FFG_Sym}. Simultaneously, the same
threshold is a lower border for a branch of unstable symmetric states that
exists at $\xi _{\mathrm{thr}}^{(2)}<\xi <\pi -\xi _{\mathrm{thr}}$, in the
case of the attractive nonlinearity. Typical examples of the soliton
profiles of the symmetric, antisymmetric and asymmetric types are displayed
in Fig.~\ref{Profile_2Delta_mum1}. Further, Fig.~\ref%
{Deltaxthresh_FFG_2deltas} shows $\xi _{\mathrm{thr}}^{(2)}$ as a function
of $\varepsilon $ and $\mu $ (within the first finite bandgap). All the
symmetric GSs are stable in the case of repulsion, and unstable under
attraction.

Modes found at $\xi >\pi -\xi _{\mathrm{thr}}$ may be considered as bound
states of two fundamental GSs. For both signs of the nonlinearity, symmetric
and antisymmetric bound states are found precisely where their fundamental
counterparts exist. This can be seen in Fig.~\ref%
{NvsDeltax_2Delta_FFG_AntiAsymBoundStates}, comparing the location of the
bound-state branches with those obtained in the model with the single $%
\delta $-function, cf. Fig.~\ref{NvsDeltax_FFG_Mum1}. It is not surprising
that both symmetric and antisymmetric bound states are strongly unstable
close to the potential maximum, in the case of the attractive nonlinearity.
For the repulsive nonlinearity, the symmetric bound states experience weak
local instability in the regions near the minimum of the OL. On the other
hand, their antisymmetric counterparts may appear to be stable in terms of
the eigenvalues, and in direct simulations with respect to symmetric
perturbations, but they split into their fundamental constituents when
asymmetric disturbances are imposed (similar to the antisymmetric bound
states in the semi-infinite gap.)

When the two $\delta $-functions are placed around a maximum of the periodic
potential, the picture is somewhat simpler. In this case, the secondary
threshold, $\xi _{\mathrm{thr}}^{(2)}$, does not exist, while, for the
attractive nonlinearity, all types of the soliton families---symmetric,
asymmetric and antisymmetric ones---exist in the region of $0<\xi <\pi
/2-\xi _{\mathrm{thr}}$ (not shown here). Also, on the contrary to the
previous setting, the asymmetric solitons bifurcate from the symmetric
branch (not from the antisymmetric one), all the solitons being unstable in
the case of the attractive nonlinearity. Results for the bound states are
qualitatively similar to those reported for the minimum-centered
configuration, with the difference that, in the present case, the weak
localized instability occurs for the antisymmetric solutions, while the
symmetric bound states are the ones which are unstable only against
asymmetric perturbations.
\begin{figure}[tbp]
\subfigure[]{\includegraphics[width=2.7in]{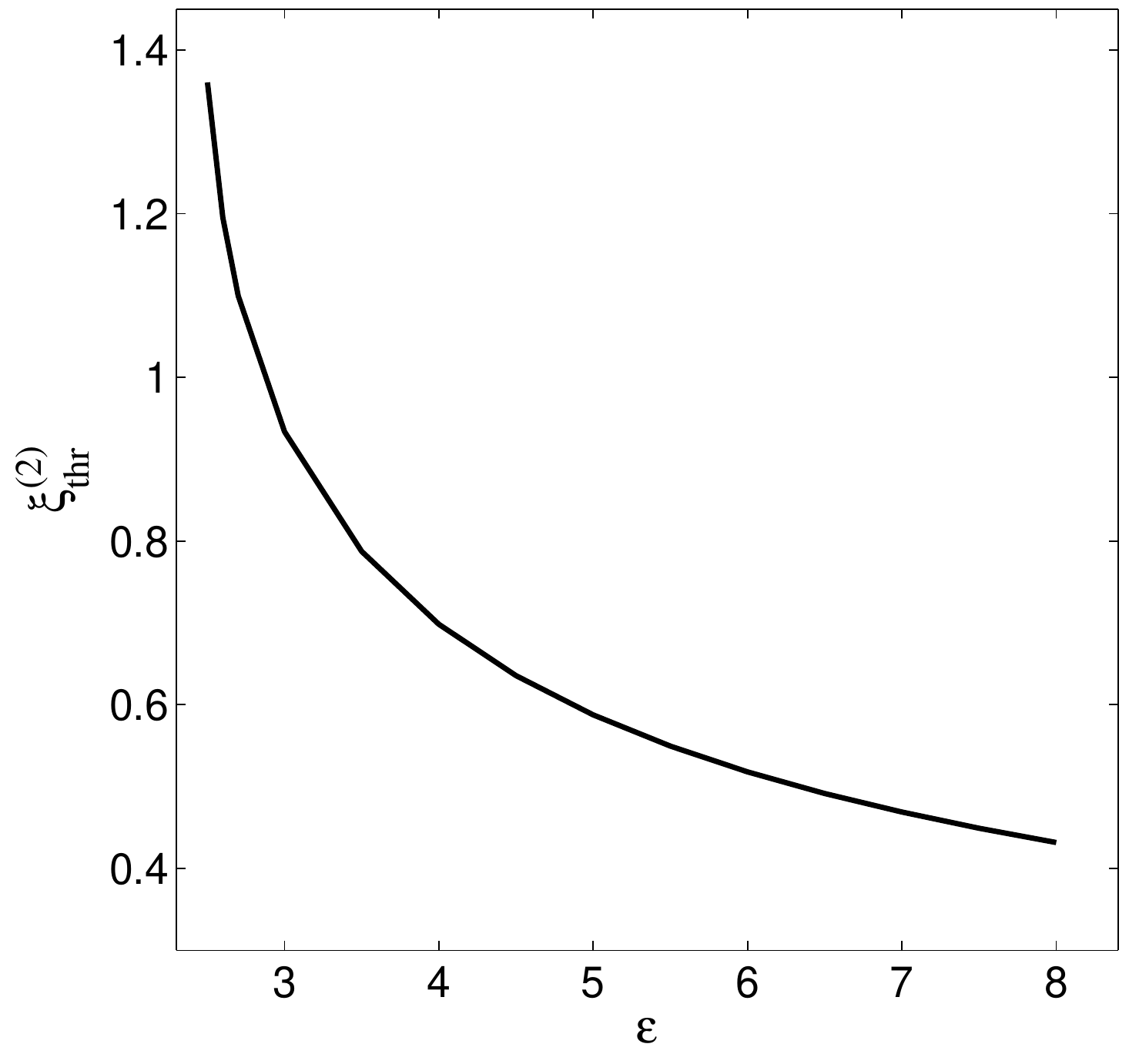}%
\label{Deltaxthresh_vs_eps_2deltas}} \quad \subfigure[]{%
\includegraphics[width=2.7in]{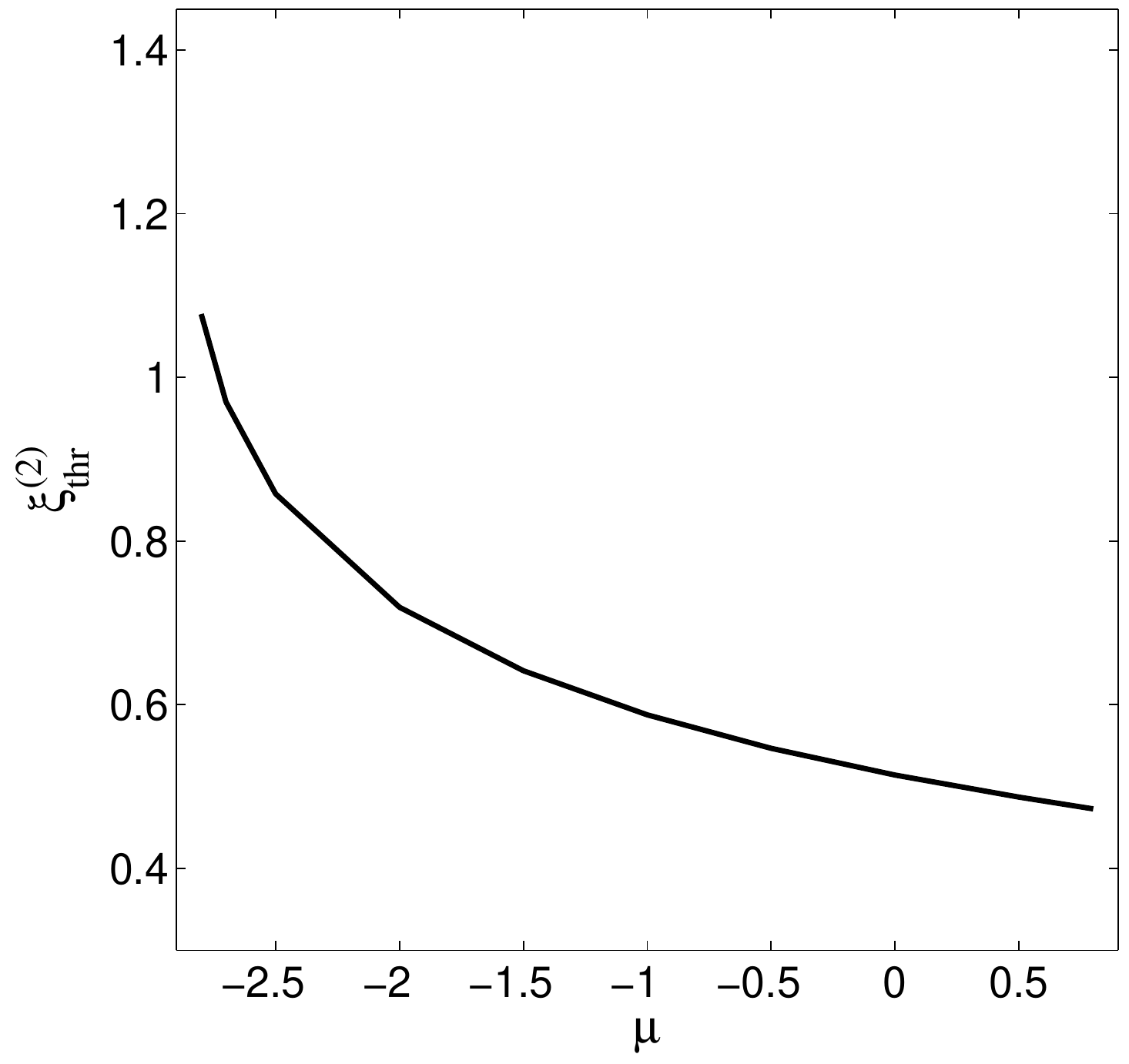}%
\label{Deltaxthresh_vs_mu_2deltas}}
\caption{(a) The secondary threshold, $\protect\xi _{\mathrm{thr}}^{(2)}$,
for the pair of two closely set $\protect\delta $-functions in the first
finite band gap, versus $\protect\varepsilon $, for $\protect\mu =-1$. (b)
The same as a function of $\protect\mu $, for $\protect\varepsilon =5$.}
\label{Deltaxthresh_FFG_2deltas}
\end{figure}

\section{Conclusions}

\label{sec:Conclusion}

In this work, we have introduced two settings that combine the attractive or
repulsive nonlinearity, concentrated in one or two points, and the linear
periodic potential. For the model with the single $\delta $-function, we
have found stable solutions in the semi-infinite and two lowest finite gaps.
In particular, in the case of the attractive nonlinearity ($\sigma =-1$),
the degenerate family of exact soliton solutions exists in the absence of
the periodic potential, being fully unstable. Even a weak potential lifts
the degeneracy and stabilizes the entire family in the semi-infinite gap,
provided that the $\delta $-function is set in a finite region around a
local minimum of the periodic potential. The stability of this soliton
family agrees with the Vakhitov-Kolokolov criterion.

In the first finite bandgap, GSs (gap solitons) have been found for both
attractive and repulsive nonlinearities, although they do not coexist: if
the $\delta $-function is placed in a finite area around a local minimum of
the periodic potential, the GS exists only under the repulsion, and in the
remainder of the period of the potential, GS can be supported solely by
attractive nonlinearity. In the first bandgap, all the GSs are stable in the
case of the repulsion, while the soliton pinned to the attractive $\delta $%
-function is stable only in a small region, if any.

In the second bandgap, two soliton branches, centered around the attractive $%
\delta $-function set at either the minimum or maximum of the periodic
potential, were found. While none of them is stable, stability regions were
produced for the soliton branch supported by the repulsive $\delta $%
a-function. It exists exactly for those values of shift $\xi $ of the $%
\delta $-function relative to the underlying potential at which both
aforementioned branches are absent in the case of the attraction.

In addition to the numerical results, analytical predictions, based on the
perturbation theory for the Mathieu equation, were presented for the case
when the $\delta $-function is positioned exactly at the maximum or minimum
of the periodic potential. This approximation was developed for the solitons
in the semi-infinite and two lowest finite gaps, showing a good accuracy in
comparison with the numerical findings, in all the cases.

In the model with two separated local nonlinearities, represented by the
symmetric pair of the $\delta $-functions, the numerical analysis was
carried out for the semi-infinite and the first finite gaps. Families of
symmetric, antisymmetric and asymmetric solitons were found, with respect to
the symmetric set of the two $\delta $-functions. In the semi-infinite gap,
with the $\delta $-functions set symmetrically around a minimum of the
potential, the symmetric solitons are stable up to the symmetry-breaking
bifurcation point, from which asymmetric branches emerge, that turn out to
be partially stable. Antisymmetric modes were found too in the semi-infinite
gap, being completely unstable. With the two $\delta $-functions placed
symmetrically around a point of the potential maximum, all the soliton
families obtained at small separations between the $\delta $-functions, as
well as all the two-soliton bound states, are unstable.

In the first finite bandgap, asymmetric solitons bifurcate from the
antisymmetric branch, if the potential minimum is set between the attractive
$\delta $-functions. This antisymmetric branch has a very short stability
segment, while the asymmetric GSs are completely unstable. Under the
repulsion, there is a stable branch of symmetric modes centered around the
minimum of the potential. At a certain threshold value of the separation $%
2\xi $ between the $\delta $-functions (which is different from the
threshold found in the single-$\delta $-function setting), the symmetric
mode switches from the repulsive to attractive nonlinearity, simultaneously
loosing its stability. Finally, in the configuration with the local maximum
of the periodic potential fixed at the midpoint, no stable solitons were
found for sufficiently small separations $2\xi $. For both configurations,
with the midpoint coinciding with either the maximum or minimum of the
underlying potential, all the families of two-soliton bound states are
unstable against various perturbation modes.

This work may be naturally extended in other directions. In particular, it
may be interesting to consider the nonlinearity represented by a periodic
array of $\delta $-functions. A challenging issue is to analyze similar
models in two dimensions.

\end{document}